\documentclass[10pt,final,journal,a4paper,oneside,twocolumn]{IEEEtran}
\usepackage{amssymb}
\usepackage[tbtags,sumlimits,nointlimits,reqno]{amsmath}

\usepackage{bbm}
\usepackage{float}
\usepackage{srcltx}
\usepackage{psfrag}
\usepackage{epsfig}
\usepackage{color}
\usepackage{mathdots}
\usepackage[export]{adjustbox}
\usepackage{graphicx}

\usepackage{color}
\usepackage{float}
\usepackage{mathtools, cuted}
\usepackage[makeroom]{cancel}
\usepackage{relsize}
\usepackage{bbm}
\usepackage{ulem}
\usepackage{cite}

\makeatletter
\let\MYcaption\@makecaption
\makeatother

\usepackage[font=footnotesize]{subcaption}

\makeatletter
\let\@makecaption\MYcaption
\makeatother

\newcommand{\bbmatrix}{\begin{bmatrix}}
\newcommand{\ebmatrix}{\end{bmatrix}}

\newcommand{\E}{\mathbf{E}}

\newcommand{\C}{\mathbf{C}}
\renewcommand{\H}{\mathbf{H}}
\newcommand{\N}{\mathbf{\hat n}}
\newcommand{\NN}{\mathbb{\bar N}}
\newcommand{\J}{\mathbf{J}}
\newcommand{\K}{\mathbf{K}}

\newcommand{\M}{\mathbf{M}}
\renewcommand{\P}{\mathbf{P}}
\renewcommand{\r}{\mathbf{r}}

\newcommand{\F}{\mathbf{F}}

\renewcommand{\P}{\mathbf{P}}

\newcommand{\rbb}{\mathbbm{r}}

\renewcommand{\L}{\mathbf{\mathcal{L}}}
\newcommand{\R}{\mathbf{\mathcal{R} }}

\newcommand{\Nh}{\mathbf{\hat n}}

\newcommand{\0}{\varnothing}
\newcommand{\I}{\mathbb{I}}

\newcommand{\Cv}{\mathbb{C}}
\newcommand{\Qv}{\mathbb{\bar Q}}
\newcommand{\Av}{\mathbb{A}}
\newcommand{\Bv}{\mathbb{B}}
\newcommand{\Pv}{\overline{\mathbb{P}}}

\newcommand{\Sv}{\mathbb{S}}
\newcommand{\Fv}{\mathbb{F}}

\newcommand{\Iv}{\mathbb{I}}
\newcommand{\Rv}{\mathbb{R}}
\newcommand{\Xv}{\mathbb{X}}

\newcommand{\Lv}{\mathbb{L}}

\newcommand{\Ev}{\mathbb{E}}

\newcommand{\Dv}{\mathbb{\bar D}}
\newcommand{\Gv}{\mathbb{\bar G}}

\newcommand{\Jv}{\mathbb{J}}
\newcommand{\Hv}{\mathbb{H}}
\newcommand{\Kv}{\mathbb{K}}

\newcommand{\Nv}{\overline{\mathbb{N}}}

\definecolor{myGreen}{rgb}{0.0, 0.5, 0.0}
\definecolor{amber}{rgb}{0.8, 0.33, 0.0}

\newcommand{\tjs}[1] {#1}
\newcommand{\sg}[1] {#1}

\usepackage{flushend}

\begin{document}

\normalem

\title{Surface Susceptibility Synthesis of Metasurface Holograms for creating Electromagnetic Illusions}

\author{%
        Tom. J. Smy, Scott A. Stewart, and Shulabh~Gupta
\thanks{This work was supported and funded by the Department of National Defence's Innovation for Defence Excellence and Security (IDEaS) Program.}
\thanks{\sg{T. J. Smy, S. A. Stewart, and S.~Gupta, are with the Department of Electronics, Carleton University, Ottawa, Ontario, Canada. Email: TomSmy@cunet.carleton.ca}}
}

 %The paper headers
\markboth{MANUSCRIPT DRAFT}
{Shell \MakeLowercase{\textit{et al.}}: Bare Demo of IEEEtran.cls for Journals}

% make the title area
\maketitle
\begin{abstract}
A systematic approach is presented to exploit the rich field transformation capabilities of Electromagnetic (EM) metasurfaces for creating a variety of illusions using the concept of metasurface holograms. A system level approach to metasurface hologram synthesis is presented here, in which the hologram is co-designed with the desired object to be projected. A structured approach for the classification of the creation of EM illusions is proposed for better organization and tractability of the overall synthesis problem. The deliniation is in terms of the initial incident (reference) illumination of the object to be recreated (front/back-lit), the position of illusion (posterior/anterior), and the illumination used to create the illusion (front/back). Therefore the classification is based on the specific relationship between the reference object to be recreated, the observer measuring the object, the orientation and placement of the reference and illumination field, and the desired placement of the metasurface hologram creating a virtual image. In the paper a general design procedure to synthesize metasurface holograms is presented based on Integral Equations (IE)  and  Generalized Sheet Transition Conditions (GSTCs), where the metasurface hologram is described as zero thickness sheet with tensorial surface susceptibility densities. Several selected configurations are chosen to illustrate various aspects of the hologram creation in 2D, along with a novel numerical technique to artificially reverse-propagate the scattered fields, required in the synthesis process. Finally, the impact of the metasurface size and the illumination field strength on the quality of the reconstructed scattered fields is also discussed.
\end{abstract}


%\tableofcontents

\section{Introduction}

Electromagnetic (EM) metasurfaces are 2D arrays of sub-wavelength resonating particles, where control of the spatial distribution and EM properties of the individual particles allows the scattered fields to be engineered with unprecedented control of both reflection and transmission, and with complete polarization control \cite{meta2,MS_review_Yu}. Metasurfaces have been proposed for achieving a variety of wave transformation functionalities including -- cloaking, waveform generation, and lensing\cite{MetaHolo, MetaCloak, MetaFieldTransformation, ReconfgMSoptics, ShaltoutSTMetasurface}, This has resulted in many proposed thin sub-wavelength surface devices across the EM spectrum, with a large number of structural designs and topologies using metals, dielectrics and other exotic materials \cite{Cylindrical_DMS, meta3}. There has been also recent development in the application of metasurface concepts to system-level applications, such as next generation wireless networks for 5G/6G, where such surfaces act as smart \sg{reflectors} in large Radio Frequency environments to achieve intelligent wave propagation \cite{Zhang_Smart_MS_WirelessComm, Basr_MIMO6G, Fink_AI_Metasurface}. In such cases, the metasurfaces act as part of an overall system and therefore must be co-designed and integrated  with the rest of the system components. 

One such application is an \emph{Hologram}. Holograms are well-known in optics where the spatial (and possibly temporal) information of an arbitrary object is encoded onto the surface (typically photographic plates) \cite{Goodman_Fourier_Optics, Saleh_Teich_FP}. This is a two step process, where the information about scattering properties of an object of interest is first recorded using a given reference beam and modulated onto a given surface. Once the information is recorded, the encoded surface, when illuminated with a reconstructing beam, projects an illusion of the object. With increasing sophistication of encoding capability, more complex illusions can naturally be created. This process essentially exploits the wave-transformation capability of a surface -- manipulating the reference beam electromagnetically. Metasurfaces therefore naturally represent a powerful platform to create sophisticated EM illusions with complete control over the scattered fields with respect to both complex amplitude and polarization. Consequently, these surfaces can also be used to create \emph{metasurface holograms}, due to their advanced information encoding capability \cite{MSHologram_Review, MShologram_Review2}.

In this work, a systematic description of metasurface holograms is presented and rigorous procedures are defined to design and synthesize these metasurfaces for achieving a desired EM illusion. To enable a general treatment of this problem, practical metasurfaces can conveniently be modeled as zero thickness sheets characterized using frequency dependent electromagnetic surface susceptibility tensors $\bar{\bar{\chi}}(\omega)$ \cite{Chi_Review, MS_Synthesis, Chi_extraction_Macrodmodel, TBC_vs_GSTC_Caloz}. The EM fields around the metasurface then can be described using Generalized Sheet Transition Conditions (GSTCs) \cite{KuesterGSTC}. The spatial distribution of surface susceptibilities of the metasurface $\bar{\bar{\chi}}(\mathbf{r})$ dictates the scattered (and thus total) fields produced by the metasurface when illuminated by an incident field. Therefore, the key design objective in creating metasurface based illusions is to synthesize the spatially varying surface susceptibilities, $\bar{\bar{\chi}}(\mathbf{r})$, to project the desired scattered fields at a given design frequency, $\omega$ to the observer location, identical to that of a real object.

Many metasurface synthesis and analysis problems using surface susceptibilities have been reported in the literature. In typical frequency-domain metasurface synthesis procedures, arbitrary incident and desired scattered fields may be specified, which in conjunction with the GSTCs, can be used to numerically solve, or optimize, for the required surface susceptibilities. For planar surfaces, surface susceptibilities can be analytically computed, for instance see \cite{MS_Synthesis, Caloz_EM_inversion}. \sg{When performing a synthesis the fields must be specified at the metasurface location and not anywhere in space}. While such methods are general in nature, their efficacy rests on a physically meaningful specification of the EM fields. On the other hand, metasurface analysis typically involves numerical computations where the GSTCs are coupled into bulk Maxwell's equations using a variety of standard numerical techniques based on Finite-Difference and Finite Element methods \cite{Caloz_MS_Siijm, Caloz_Spectral, Smy_Metasurface_Space_Time}, and Integral-Equation (IE) techniques \cite{stewart2019scattering, FE_BEM_Impedance, Caloz_MS_IE, AppBEMEM, Smy_EuCap_BEM_2020, Caloz_EM_inversion}. Given that the field scattering from a metasurface hologram may need to be evaluated for electrically large domains, IE-GSTC methods are  computationally efficient choices and, as will be shown later, are well suited for metasurface synthesis for holographic applications. 

Given this context, a general methodology of synthesizing and designing metasurface holograms is presented in this work using the IE-GSTC method from a system level perspective. The metasurface hologram design problem is systemically defined for various geometrical relationships between the desired object illusion, a reference beam, the observer location and the illuminating beam.
%, which may require individual treatment of the problem in some specific cases. 
The 2D IE-GSTC based numerical platform is further developed to synthesize metasurface susceptibilities with an integrated approach, where the desired fields, \sg{specified anywhere in space and not necessarily at the metasurface}, are generated using a system level description and fed-back into the metasurface design. A novel wave-propagation technique (a numerical inverse propagation) is further proposed for accurate determination of the desired scattered fields in certain specific scenarios. This ensures a physically meaningful field specification for arbitrary shaped metasurface designs and avoid ill-posed metasurface synthesis problems. Consequently, the proposed IE-GSTC based numerical platform computes spatially varying surface susceptibilities, $\bar{\bar{\chi}}(\mathbf{r})$ and is demonstrated to project complex EM illusions in a variety of physical scenarios.

The paper is structured as follows. Sec.~\ref{Sec:MH} defines the overall problem of metasurface holograms from a functional point-of-view in the context of observer and field illuminations, along with basic properties of the general EM metasurfaces. Sec.~\ref{Sec:MI} presents the IE-GTSC approach: developing the field equations to be applied to various hologram situations, the discretized form of the proposed IE-GSTC field solver, and the numerical procedure for surface susceptibility synthesis. Sec.~\ref{Sec:Res} shows several results demonstrating the metasurface hologram operation. Sec.~\ref{Sec:Prac} discusses some practical aspects of metasurface hologram designs such as the effect of finite-size surfaces and the field illuminations strengths on reconstructing the desired scattered fields. Finally, conclusions are provided in Sec.~\ref{Sec:Con}.

\begin{figure*}[htbp]
\begin{center}
\psfrag{J}[c][c][0.8]{$\psi_s(\mathbf{r}_0,\omega)$}
\psfrag{N}[c][c][0.8]{$\psi_s(\mathbf{r}_0,\omega) + \psi_0(\mathbf{r}_0,\omega)$}
\psfrag{K}[c][c][0.8]{$\psi_s(\mathbf{r},\omega)$}
\psfrag{C}[c][c][0.8]{$\psi_0(\mathbf{r},\omega)$}
\psfrag{D}[c][c][0.8]{$\bar{\bar{\chi}}_2(\mathbf{r}_m, \omega)$}
\psfrag{E}[c][c][0.8]{$\psi_i(\mathbf{r},\omega)$}
\psfrag{G}[c][c][0.8]{$\bar{\bar{\chi}}_1(\mathbf{r}_m, \omega)$}
\includegraphics[width=2\columnwidth]{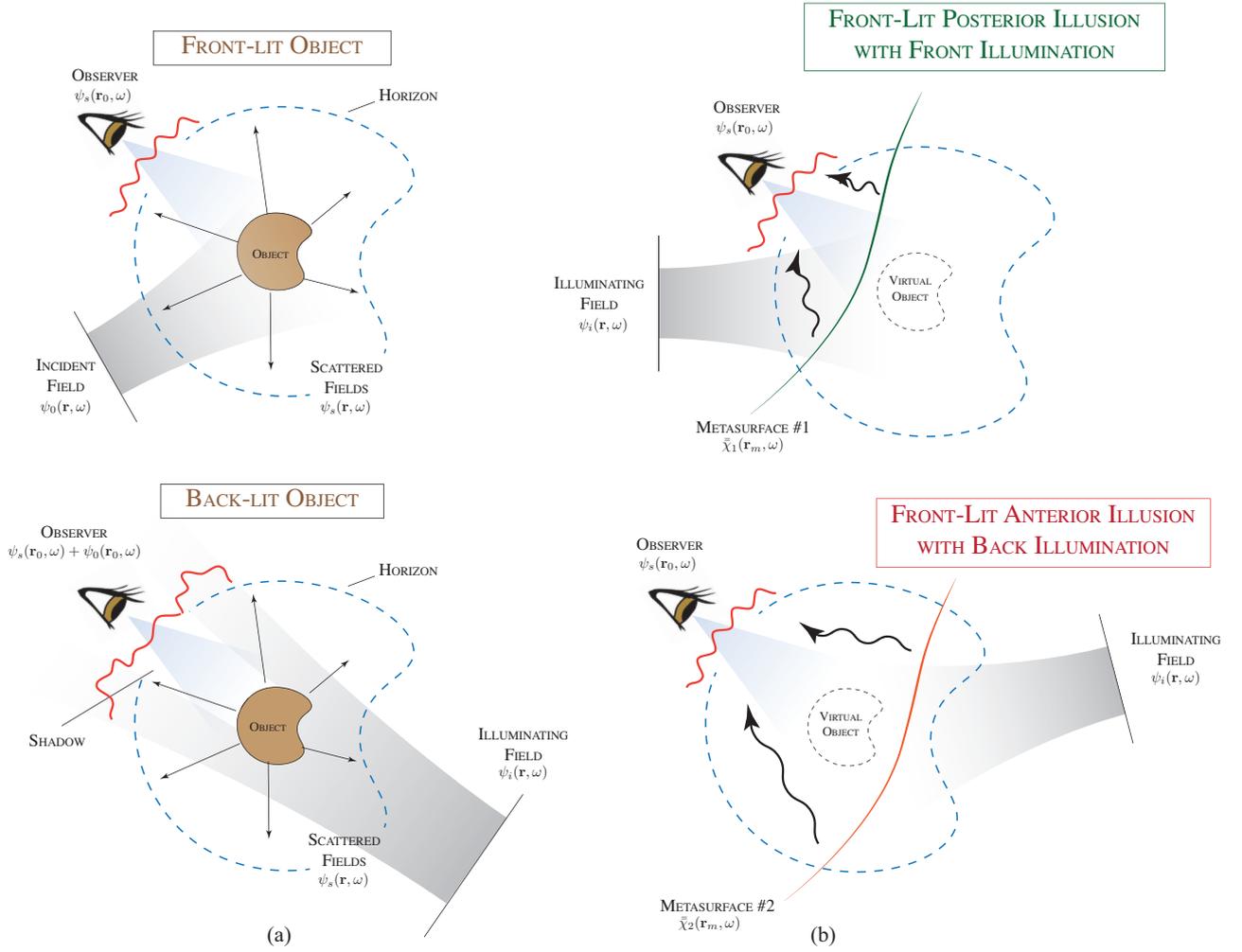}
\caption{Illustration of producing electromagnetic illusions using metasurfaces. a) Front-lit vs \tjs{back-lit objects} where the incident field and the observer are on the same side of the object. b) Anterior and posterior illusions depending on the relative position of the metasurface and the original object under front and back illumination cases; the two synthesized metasurfaces are different in these cases.}\label{Fig:Illusion_Cases}
\end{center}
\end{figure*}

\section{Metasurface Holograms}\label{Sec:MH}

\subsection{Principle of Creating Illusions}\label{Sec:II-A}

Consider an object of arbitrary shape and material composition (dielectric or metal), subject to an incident \emph{reference}  wave, $\psi_0(\mathbf{r},\omega)$, as shown in Fig.~\ref{Fig:Illusion_Cases}(a).\footnote{The field $\psi(\r)$ is a compact notation for a fully vectorial electric and magnetic field distribution.}
The reference wave interacts with the object producing propagating scattered fields, $\psi_s(\mathbf{r},\omega)$. We now place an observer located to the left of the object measuring the scattered fields $\psi_s(\mathbf{r}_0,\omega)$ within a certain field-of-view. Assume that the incident fields are propagating from left of the object, so that from the perspective of the observer, it appears as a \emph{Front-Lit Object}. By measuring and analyzing the incoming \sg{left} propagating scattered fields, the observer perceives (or detects) the presence and properties of the object, such as its geometrical shape and material characteristics. If, on the other hand,  the reference wave illuminates the object from the right side of the domain, the observer measures both the scattered and the reference fields (as both are \sg{left} propagating), including the shadow produced by an object. This is termed as a \emph{Back-Lit Object}.

Let us now remove the object, and introduce an artificial surface, i.e. a metasurface hologram, as shown in Fig.~\ref{Fig:Illusion_Cases}(b) at $\r =\r_m$. The metasurface is excited with an \emph{illumination} field, $\psi_i(\mathbf{r},\omega)$, which may or may not be the same as the reference field $\psi_0(\mathbf{r},\omega)$. Can this surface be engineered to rigorously recreate the scattered fields produced by the original object $\psi_s(\mathbf{r}_0,\omega)$ within the field-of-view of the observer? In such a case, the real physical object or a \emph{virtual object} produced by an artificial metasurface are indistinguishable from the perspective of the observer. This creation of a false perception by the metasurface hologram will be referred to as an \emph{Electromagnetic Illusion}.

There exist several possibilities in the placement of the metasurface hologram, which while irrelevant from the perspective of the observer, is important from the practical design point of view and impacts how the metasurface may later be synthesized. If the metasurface is located between the object and observer, we refer to it as \emph{a Posterior Illusion}, otherwise, when the metasurface is behind the illusion, we refer to it as an \emph{Anterior Illusion}. In case of a posterior illusion, the virtual object is formed behind the metasurface, while in case of anterior, it is formed in front of the metasurface. Another variable of importance is how the metasurface is excited, i.e. the relative location and direction of the \emph{Illuminating Field}. The illuminating field, in general, could be entirely different from the incident field that was used to determine the desired scattered fields from the real object. If it strikes the surface from the same side as the observer, it is referred to as \emph{Front Illumination}, else, as \emph{Back Illumination}.

With this description, the various illusion scenarios may be termed as: Front/Back-Lit Posterior/Anterior Illusions with Front/Back Illumination. The front/back-lit configuration determines which fields the observer detects -- either scattered fields or total fields. The Posterior or Anterior position of the metasurface determines how the desired fields are to be constructed \tjs{to form the virtual image} (with a finite physical separation from the surface) and are numerically propagated to the metasurface location. This is important since the metasurface synthesis requires fields infinitesimally close to the surface region. It will be shown later that unlike the posterior configuration anterior illusions require an unusual inverse field propagation as an initial step before metasurface synthesis can be performed. Finally, the front vs back illumination choice will determine whether a fully reflective metasurface is required or a transmissive one. This has important implications for practical realization of the synthesized metasurface, where compared to reflective ones, the transmission-type metasurface requires both reflection and transmission control.

\subsection{Metasurface Descriptions}\label{Sec:II-B}

In order to generate arbitrarily complex and fully-vectorial scattered fields $\psi_s(\r_0)$ from another equally arbitrary illuminating field $\psi_i(\r)$, the metasurface hologram located at $\r=\r_m\ne \r_0$ must be capable of general EM wave transformations with complete wave control. Also, it should be noted, that while the prescribed fields may be arbitrary and complex, they are completely physical, making the metasurface synthesis problem a well-posed physically meaningful problem. This is due to the fact that the desired scattered fields are first computed using physical objects under well-defined incident fields, and the metasurface is illuminated with a physical field description as well. 

The general wave transformation capability of physical EM metasurfaces can be described by expressing them as mathematical space discontinuities (zero thickness interfaces), owing to their their sub-wavelength thick nature and characterizing their EM wave interaction using electric and magnetic polarizabilities. To correctly model zero-thickness sheets, \cite{IdemenDiscont} introduced Generalized Sheet Transition Conditions (GSTCs) which were later applied to metasurface problems by \cite{GSTC_Holloway, KuesterGSTC}. GSTCs (frequency-domain) relate the tangential EM fields around the metasurface to its tangential and normal surface polarization response as,
\begin{subequations}\label{Eq:GSTC}
\begin{align}
	\Nh \times \Delta \E_T &= -j\omega\mu_0 \M_T - \Nh \times \nabla_{||}\left(\frac{\P_{n}}{\epsilon_0}\right)\\
	\Nh \times \Delta \H_T  &= j\omega \P_T - \Nh \times \nabla_{||}\M_{n},
\end{align}
\end{subequations}
\noindent where $\Nh$ is the normal vector to the surface, $\Delta\psi_T = (\psi_+- \psi_-)$ is the difference between the fields across the surface; $\{\M_T,~\P_T\}$, $\{\M_{n},~\P_{n}\}$ are the average tangential and normal magnetic and electric surface polarizability densities of the surface. The polarization densities can be seen as a response of the metasurface to the average fields, related through the surface susceptibility densities as \cite{Chi_Review},
\begin{subequations}\label{Eq:chis}
\begin{align}
    \P_T &= \epsilon \bar{\bar{\chi}}_\text{ee} \E_{T,av} + \bar{\bar{\chi}}_\text{em} \sqrt{\mu \epsilon}\; \H_{T,av}\\
    \M_T &= \epsilon \bar{\bar{\chi}}_\text{mm} \H_{T,av} + \bar{\bar{\chi}}_\text{me} \sqrt{\epsilon/\mu}\; \E_{T,av},
\end{align}
\end{subequations}
\noindent where $\psi_\text{av} = (\psi_+ + \psi_-)/2$ is the average field at the metasurface, and $\bar{\bar{\chi}}$ is a general $3\times3$ tensor accounting for various microscopic EM characteristics of the metasurface. Eqs.~\ref{Eq:GSTC} and \ref{Eq:chis} together rigorously model the EM interaction with the metasurface, while Eq.~\ref{Eq:chis} captures its field transformation capabilities via 36 variables inside the tensors.

Therefore, the metasurface hologram synthesis problem of Fig.~\ref{Fig:Illusion_Cases} can now be defined in terms of the determination of these unknown susceptibility tensors needed to produce the desired field configuration across the surface. This field configuration is, of course, determined by the various incident and scattered EM fields needed to create the illusion.

\section{Modelling Approach}\label{Sec:MI}

\subsection{Propagation and Problem Formulation}\label{Sec:III-A}

\begin{figure*}[htbp]
\begin{center}
\psfrag{a}[c][c][0.7]{$\J,~\K$}
\psfrag{K}[c][c][0.7]{\shortstack{\textsc{Scattered}\\$\F_s(\r)$} }
\psfrag{C}[c][c][0.7]{ \shortstack{\textsc{Incidence}\\$\F_0(\r)$} }
\psfrag{G}[c][c][0.7]{\shortstack{\textsc{Illumination}\\$\F_i(\r)$}  }
\psfrag{D}[c][c][0.7]{\color{myGreen}\shortstack{\textsc{Horizon}\\$\F_s(\r_h)$}}
\psfrag{J}[c][c][0.7]{\shortstack{\textsc{Observation}\\$\F_s(\r_0)$}}
\psfrag{E}[c][c][0.7]{\shortstack{\textsc{Metasurface}\\$\bar{\bar{\chi}}(\r_m)$} }
\psfrag{x}[c][c][0.7]{$x$}
\psfrag{y}[c][c][0.7]{$y$}
\psfrag{b}[c][c][0.7]{\color{blue}\shortstack{\textsc{Forward-propagated}\\$\F_s(\r_m^-)$} }
\psfrag{c}[c][c][0.7]{$\F_s(\r_m^+) = 0$}
\psfrag{e}[r][c][0.7]{\color{red}\shortstack{\textsc{Reverse-propagated}\\$\F_s(\r_m^-)$}}
\psfrag{P}[c][c][0.9]{ \textsc{\textbf{Reference} Simulation} }
\psfrag{Q}[c][c][0.9]{ \textsc{\textbf{Posterior} Illusion} }
\psfrag{R}[c][c][0.9]{ \textsc{\textbf{Anterior} Illusion} }
\includegraphics[width=2\columnwidth]{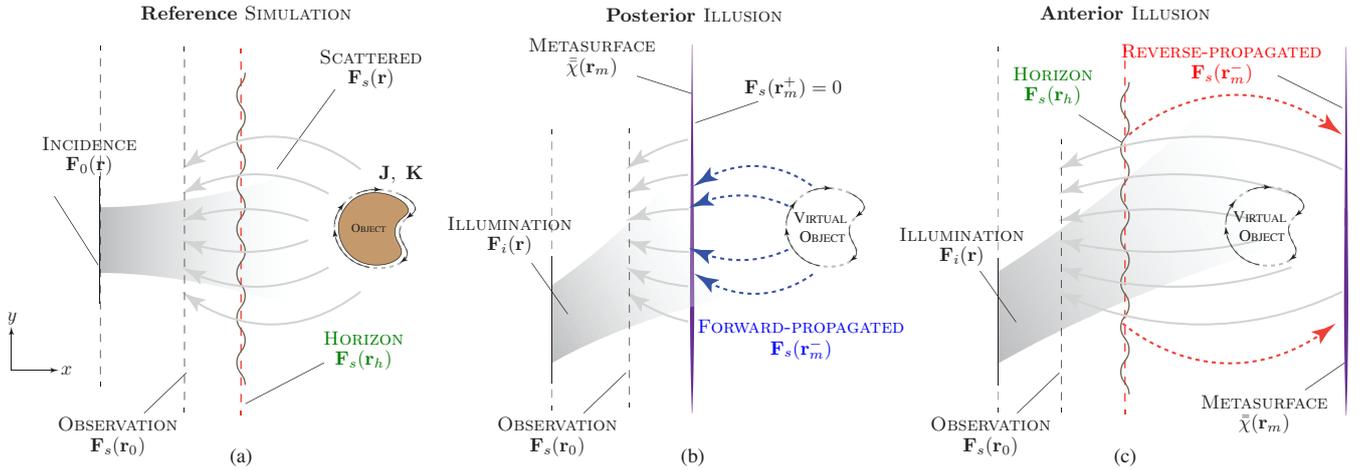}
\caption{Methodology to achieve various field transformations using a metasurface to achieve an EM illusion. a) Determination of the required scattered fields of the desired illusion object. b) Same scattered fields produced by inserting a metasurface in the absence the object, for the Posterior and Anterior (c) positions, for a given illumination field.}\label{Fig:Illusion_Setup}
\end{center}
\end{figure*}

To elucidate the metasurface hologram synthesis problem, consider, for simplicity, the 2D case of Fig.~\ref{Fig:Illusion_Setup}, where all field interactions with the metasurface happen in the $x-y$ plane and there is \tjs{no field variation along $z$}, and $\partial/\partial z = 0$. Consider a reference simulation first, where the first task is to compute the desired scattered fields from a given object for the specified reference wave, as shown in Fig.~\ref{Fig:Illusion_Setup}(a).  This is achieved by using the Integral Equation (IE) form of the Maxwell's equations and the creation of field propagators.

The EM fields radiated into free-space from electric and magnetic current sources, $\{\J,~\K\}$, can be generally expressed as \cite{chew2009integral, Method_Moments}:
\begin{subequations}\label{Eq:FieldProp}
\begin{align}
    \E_s(\r) &= -j\omega\mu(\L \J)(\r,\r') - (\R \K)(\r,\r')\\
    \H_s(\r) &= -j\omega\epsilon(\R \K)(\r,\r') + (\R \J)(\r,\r'),
\end{align}
\end{subequations}
\noindent where the various field operators are given by:
\begin{align*}
    (\L \J)(\r, \r') &= \int_{\ell}[1+\frac{1}{k^2}\nabla\nabla\cdotp] [G(\r,\r')\J(\r')] \,d\r'\\
    (\R \J)(\r, \r') &= \int_{\ell}\nabla \times [G(\r,\r')\J(\r')] \,d\r'\\
    (\L \K)(\r, \r') &= \int_{\ell}[1+\frac{1}{k^2}\nabla\nabla\cdotp] [G(\r,\r')\K(\r')] \,d\r'\\
    (\R \K)(\r, \r') &= \int_{\ell}\nabla \times [G(\r,\r')\K(\r')] \,d\r'.
\end{align*}
\noindent The operators $\L\{\cdot\}$ and $\R\{\cdot\}$ produce a field response at location $\r$ due to a distribution of current sources located at $\r'$. This can be conveniently expressed in a matrix form as:
\begin{align*}
    \bbmatrix \E_s(\r)\\ \H_s(\r) \ebmatrix
    &= \bbmatrix - j\omega \mu \mathbf{L}(\r, \r')  - \mathbf{R}(\r, \r')  \\ - j\omega \epsilon \mathbf{L}(\r, \r') + \mathbf{R}(\r, \r') \ebmatrix \bbmatrix \J(\r') \\ \K(\r') \ebmatrix 
\end{align*}
\noindent which can be further simplified as,
\begin{align}
  \F_s(\r) &= \mathbf{P}(\r, \r') \mathbf{C}(\r') \label{Eq:Propagators}
\end{align}
\noindent where,
\begin{align*}
    \mathbf{C}(\r') &= \bbmatrix \J(\r') & \K(\r') \ebmatrix^\top\\
    \F^s(\r) &= \bbmatrix \E^s(\r) & \H^s(\r) \ebmatrix^\top
\end{align*} 
are the current source and radiated field vectors respectively and $\{\cdot\}^\top$ denotes a matrix transpose. The $\mathbf{P}(\r, \r')$ propagator matrix thus relates the current sources at $\r'$ to the EM fields produced by them at an arbitrary location $\r$. Finally, $G(\r,\r')$ inside Eq.~\ref{Eq:FieldProp}, represents the Green's function, which for a 2D case is given by the \sg{2$^\text{nd}$} Hankel function, 
\begin{align}
G(\r, \r')=    H_0^{(2)}(\r) = J_0(\r,\r') - i Y_0(\r,\r'),
\end{align}
\noindent where $J_0$ and $Y_0$ are the Bessel functions of the 1st and 2nd kind and the function represents outwardly propagating radial waves.

Consider again the illumination of an object producing scattered fields as in Fig.~\ref{Fig:Illusion_Setup}a. These scattered fields can alternatively be produced by a set of equivalent currents $\mathbf{C} (\r_o)$ on the surface of the object, which are determined by Eq.~\ref{Eq:Propagators}. By specifying the appropriate boundary conditions at the known object surface, these unknown currents can be easily solved using standard IE solvers for a specifed reference field\cite{chew2009integral, Method_Moments}.

To obtain the desired scattered fields to be later reconstructed by the metasurface hologram, let us introduce a \emph{Horizon Plane} at $\r_h$, which is always located between the object ($\r_o$) and the observation plane ($\r_0$). Its utility will become clear shortly. The scattered fields at this horizon, $\F_s(\r_h)$ can be calculated by \emph{forward-propagating} the fields using Eq.~\ref{Eq:Propagators} giving,
\begin{align}\label{Eq:RefFieldHor}
\F_s^\text{ref.}(\r_h) = \mathbf{P}(\r_h, \r_o)\mathbf{C}(\r_o).
\end{align}
\noindent Thus, $\F_s^\text{ref.}(\r_h)$ represents the desired reference fields that our metasurface hologram must reconstruct in the absence of the object. This completes the first task.

Next, let us remove the object and introduce a metasurface described in terms of its surface susceptibilities $\bar{\bar{\chi}}(\r_m)$, which is excited with an arbitrary illumination field $\F_i(\r)$, as illustrated in Fig.~\ref{Fig:Illusion_Setup}(b). Dropping the perpendicular terms and assuming scalar susceptibilities, for simplicity, Eq.~\ref{Eq:GSTC} can be combined with Eq.~\ref{Eq:chis} to give,
\begin{subequations}\label{Eq:ExplicitGSTCS}
\begin{align}
	\Nh \times \Delta \E_T &= -j\omega\mu_0 (\epsilon \chi_\text{mm} \H_{T,av} + \chi_\text{me} \sqrt{\epsilon/\mu}\; \E_{T,av})\\
	\Nh \times \Delta \H_T  &= j\omega (\epsilon \chi_\text{ee} \E_{T,av} + \chi_\text{em} \sqrt{\mu \epsilon}\; \H_{T,av}),
\end{align}
\end{subequations}
\noindent where $\Delta \psi_T = \E(\r_{m+}) - \E(\r_{m-})$, and $\psi_\text{av} = \{\E(\r_{m+}) + \E(\r_{m-})\}/2$, are expressed in terms of total fields just before and after the metasurface. Since, the metasurface and the horizon are not in general co-located (i.e. $\r_m \ne\r_h$), the horizon fields $\F_s^\text{ref.}(\r_h)$ must now be used to determine the average fields around the metasurface, $\F(\r_m)$. The relationship between them depends on whether a Posterior or an Anterior illusion is desired.

\subsection{Anterior vs Posterior Illusions}\label{Sec:III-B}

Let us take the case of front-lit object and front illumination for explaining the procedures for relating the fields at the horizon and the metasurface.\footnote{In this paper, the observer is always assumed to be on the left of the object.} For the posterior illusion, the metasurface is located in front of the object, as shown in Fig.~\ref{Fig:Illusion_Setup}(b). In this case, a judicious choice is to place the metasurface directly at the horizon, so that $\r_m = \r_h$. This simplifies Eq.~\ref{Eq:ExplicitGSTCS}, so that the total fields around the surface are given by,
\begin{subequations}\label{Eq:MS_fields}
\begin{align}
\F(\r_{m-}) &=\F_i(\r_m) +  \F_s^\text{ref.}(\r_m) \\
\F(\r_{m+}) &=0,
\end{align}
\end{subequations}
\noindent where zero field is arbitrarily enforced on the right-half of the metasurface\footnote{This has no impact on reconstructing the desired fields at horizon, but impacts the required surface susceptibility distribution of the metasurface.}. All the fields in Eq.~\ref{Eq:ExplicitGSTCS} are now known, and the unknown surface susceptibilities can now be easily determined to complete the hologram synthesis. With the known susceptibilities and the illuminating field, the metasurface will correctly reconstruct $\F_s^\text{ref.}(\r_h)$ everywhere beyond the horizon, so that the observer (also located beyond the horizon) will perceive an illusion of the original object located at its original position. This completes the hologram field specification for this case and the synthesis of the surface can be performed as described in Sec. \ref{Sec:MSS}.

The relationship between $\F_s^\text{ref.}(\r_h)$ and $\F(\r_m)$ is however more complex for the anterior illusion case. See Fig.~\ref{Fig:Illusion_Setup}(c), where now the metasurface is located behind the object, i.e. $\r_m \ne \r_h$. In this case, the horizon fields must be \emph{reverse propagated} to the metasurface location. This reverse propagation is not the usual physical forward-propagation of the fields, but a purely numerical exercise, where horizon wave-fronts are mathematically propagated back to the metasurface location while maintaining the original flow of EM power towards the observer on the left of the horizon. This technique of reverse-propagation requires a few intermediate steps, before the metasurface is ready for synthesis.

\begin{figure*}[htbp]
\begin{center}
\psfrag{A}[c][c][0.9]{\textbf{\boxed{\textsc{\shortstack{Specify Object \\ \& Reference Field, $\F_0(\r)$}}}}}
\psfrag{B}[c][c][0.8]{\boxed{\textsc{\shortstack{Desired Scattered Fields \\ at Horizon, $\F_s^\text{ref.}(\r_h)$}}}}
\psfrag{D}[r][c][0.7]{\textsc{\shortstack{Compute Equivalent \\ Currents, $\C(\r_o)$}}}
\psfrag{E}[l][c][0.7]{\textsc{\shortstack{Field Propagator, \\Eq.~\ref{Eq:RefFieldHor}}}}
\psfrag{F}[l][c][0.9]{\color{blue}\boxed{\textsc{\shortstack{Posterior \\Illusion}}}}
\psfrag{C}[l][c][0.8]{\textsc{\shortstack{Metasurface \\ at Horizon \\$\r_m = \r_h$}}}
\psfrag{H}[l][c][0.9]{\color{blue}\boxed{\textsc{\shortstack{Anterior \\Illusion}}}}
\psfrag{G}[l][c][0.8]{\textsc{\shortstack{Metasurface \\ Behind Object \\$\r_m \ne \r_h$}}}
\psfrag{J}[c][c][0.8]{\color{blue}\boxed{\textsc{\shortstack{Front-Lit \\ $\F_s^\text{ref.}(\r_h)  \vcentcolon  = \F_s^\text{ref.}(\r_h) $}}}}
\psfrag{K}[c][c][0.8]{\color{blue}\boxed{\textsc{\shortstack{Back-Lit \\ $\F_s^\text{ref.}(\r_h)  \vcentcolon  = \F_0(\r_h) + \F_s^\text{ref.}(\r_h) $}}}}
\psfrag{L}[c][c][0.8]{\color{blue}\boxed{\shortstack{\textsc{Front Illumination} \\$\F_s(r_{m-}) = \F_i(\r_m) + \F_s^\text{ref.}(\r_m)$ \\ $\F_s(r_{m+}) = 0$}}}
\psfrag{M}[c][c][0.8]{\color{blue}\boxed{\shortstack{\textsc{Back Illumination} \\$\F_s(r_{m-}) = \F_s^\text{ref.}(\r_m)$ \\ $\F_s(r_{m+}) = \F_i(\r_m)$}}}
\psfrag{Z}[r][c][1.2]{\textbf{\color{amber}\boxed{\textsc{\shortstack{Compute $\bar{\bar{\chi}}$}}}}}
\psfrag{N}[c][c][0.8]{\textsc{\shortstack{GSTCs, \\Eq.~\ref{Eq:ExplicitGSTCS}}}}
\psfrag{O}[c][c][0.8]{\color{myGreen}\boxed{\shortstack{\textsc{Reverse-Propagation}, \\ $\F_s^\text{ref.}(\r_m)$, Eq.~\ref{Eq:ReverProp1},~\ref{Eq:ReverProp2}}}}
\psfrag{Y}[c][c][0.8]{\textsc{\shortstack{Reconstruct Fields \\ at Observer $\F(\r_0)$\\BEM Solver, Eq.~\ref{Eq:SysMat}\cite{stewart2019scattering, Smy_EuCap_BEM_2020}}}}
\psfrag{X}[r][c][0.7]{\textsc{\shortstack{Metasurface Hologram}}}
\includegraphics[width=2\columnwidth]{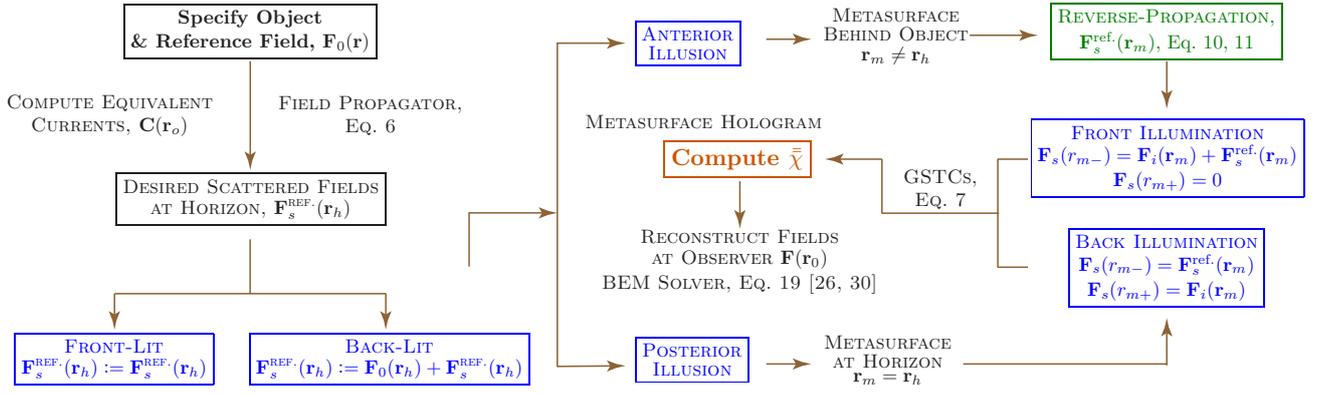}
\caption{Complete flowchart summarizing the design flow to synthesize a metasurface hologram for the general case of Front/Back-Lit Posterior/Anterior Illusion with Front/Back Illumination problem of Fig.~\ref{Fig:Illusion_Setup}, where the observer is assumed to be always located in the left-half region of the metasurface.}\label{Fig:FlowChart}
\end{center}
\end{figure*}

To reverse-propagate the horizon fields, the equivalence principle is invoked. The horizon is first represented as a surface with unknown equivalent currents $\mathbf{C}(\r_h)$, so that 
\begin{subequations}\label{Eq:Hor_C_E}
\begin{align}
\F_s(r_{h+}) &= \P(\r_{h+}, \r_h)\mathbf{C}(\r_h) = \F_s^\text{ref.}(\r_h)\\
\F_s(r_{h-}) &=  \P(\r_{h-}, \r_h)\mathbf{C}(\r_h) = 0,
\end{align}
\end{subequations}
\noindent where $\P(\r_{h\pm}, \r_h)$ are the self-propagator operators, and the fields on the left of the horizon are fixed to zero. In principle, the unknown equivalent horizon currents can be obtained by inverting Eq.~\ref{Eq:Hor_C_E}(a) giving $\mathbf{C}(\r_h) = \P(\r_{h+}, \r_h)^{-1}\F_s^\text{ref.}(\r_h)$. However, in practice, this is not a robust method and the related matrices are not well-behaved due to an ill-definition of the problem. To improve the robustness of the computation we formulate the problem as a set of unknown surface currents, $\C(\r_h)$, and scattered fields, $\F_s(\r_{h\pm})$.  Then in addition to Eq.~\ref{Eq:Hor_C_E}, the equivalent currents, $\C({\r_h})$ are enforced to be tangential to the horizon with zero normal components and the tangential field components across the horizon are set equal to the reference fields at the horizon, $\F_s^\text{ref.}(\r_h)$. These relations can be expressed conveniently in the matrix form as:
 \begin{align}\label{Eq:ReverProp1}
        \bbmatrix \P(\r_{h+}, \r_h)  & -\I &\0\\
         \P(\r_{h-}, \r_h)& \0  & -\I \\
                  \N\cdot& \0 & \0\\
                  \0 & \N_\times & - \N_\times
        \ebmatrix 
        \bbmatrix \mathbf{C}(\r_h) \\ \F_s(r_{h+}) \\\F_s(r_{h-}) \ebmatrix
        =
        \bbmatrix \0 \\ \0 \\ \0 \\ \N_\times\F_s^\text{ref.}(\r_h) \ebmatrix
\end{align}
\noindent  where $\N_\times\{\cdot\}$ extracts the tangential components of the argument vector. This solution of this system significantly improves the computation of $\C({\r_h})$ by ensuring that its normal components are zero and the interface boundary conditions are applied using the tangential fields only.

The fields produced by the equivalent sources, $\C({\r_h})$, must now be reverse-propagated to the metasurface location, $\r=\r_m$. To achieve this operation, the field propagator of Eq.~\ref{Eq:Propagators} is modified so that the propagation is reversed in a temporal sense,
\begin{align}\label{Eq:ReverProp2}
\F_s^\text{ref.}(r_m) = \P^r(\r_m, \r_h)\mathbf{C}(\r_h),
\end{align}
\noindent where the operator $\P^r(\r_m, \r_h)$ is formulated using an alternative Green's function involving a Henkel's function of the first kind:
\begin{align}
     G(\r, \r') = H_0^{(1)}(\r, \r') = J_0(\r, \r')+ j Y_0(\r, \r').
\end{align}

\sg{This function represents inwardly propagating radial waves and with respect to fields generated by surface currents is, of course, a non-physical time reversed solution to Maxwell's equations. However, in this case it is useful as a mathematical tool.}
\noindent Once the desired fields are reverse-propagated to the metasurface, the total fields around the metasurface can be formed as
\begin{subequations}\label{Eq:Fields_GSTC_A}
\begin{align}
\F_s(r_{m-}) &= \F_i(\r_m) + \P^r(\r_m, \r_h)\mathbf{C}(\r_h)\\
\F_s(r_{m+}) &= 0,
\end{align}
\end{subequations}
\noindent which when substituted inside the GSTCs of Eq.~\ref{Eq:ExplicitGSTCS} can now be solved for the unknown surface susceptibilities of the metasurface hologram (see Sec~\ref{Sec:MSS}). 

The complete design flow chart for a general Front/Back-Lit Posterior/Anterior Illusions with Front/Back Illumination problem is illustrated in Fig.~\ref{Fig:FlowChart}, summarizing various metasurface synthesis scenarios. This process also shows the final confirmation that the synthesized susceptibilities indeed produces desired fields beyond the horizon at the observer location $\r_0$ in which a BEM-GSTC framework is used to determine the output fields for the specified illuminating field used in the metasurface design  \cite{stewart2019scattering, Smy_EuCap_BEM_2020}. \tjs{The boundary element method (BEM) is a numerical computational method of solving linear partial differential equations which have been re-formulated as descritized integral equations (IE).}

% \section{Metasurface Susceptibility Synthesis}\label{Sec:MSS}

\subsection{Discretization - BEM Framework}

The field equations for synthesizing metasurface holograms have so far been expressed in terms of the continuous space variable $\r$. However, for numerical computation, each of these equations must be spatially discretized for incorporation into a BEM solver framework.

The surface position vector $\r_S$ and any equivalent currents ($\J$ and $\K$) on an arbitrary shape (such as the bounding region of a reference object) can be spatially discretized using $m$ meshing elements, such that $\r\rightarrow\rbb_S$ and $\J\rightarrow\Jv$, i.e. $\rbb_S = \bbmatrix \r_{s,1} & \r_{s,2} & \dots & \r_{s,m} \ebmatrix$ and $\Jv = \bbmatrix \J_1 & \J_2 & \dots & \J_m \ebmatrix$, for instance. Similarly, the field propagation operators $\L\{\cdot\}$ and $\R\{\cdot\}$, can also be discretized. For example, 
\begin{align*}
\Lv &= \bbmatrix 
\L_{1}(\r) & \L_{2}(\r) & \dots &\L_{m}(\r) \\
\ebmatrix
\end{align*}
with 
\begin{align*}
    \L_{m_i}(\r) \J_i &= \int_{\Delta\ell_i}[1+\frac{1}{k^2}\nabla\nabla\cdotp] [G(\r,\r_{s,i})\J_i] \,d\r_{s,i}.
\end{align*}
\noindent The resulting fields anywhere in space, produced by these discretized current sources, are obtained using the field propagation operators of Eq.~\ref{Eq:FieldProp}. If the region where the fields are measured is also discretized, $\rbb_p = \bbmatrix \r_{p,1} & \r_{p,2} & \dots & \r_{p,n} \ebmatrix$, we get
\begin{align*}
    \Fv_s(\rbb_p, \rbb_S) &= 
    \bbmatrix - j\omega \mu \Lv(\rbb_p, \rbb_S) \Jv(\rbb_S) - \Rv(\rbb_p, \rbb_S) \Kv(\rbb_S) \\ - j\omega \epsilon \Lv(\rbb_p, \rbb_S) \Kv(\rbb_S) + \Rv(\rbb_p, \rbb_S) \Jv(\rbb_S) \ebmatrix 
\end{align*}
\noindent which further can be written in compact form as
\begin{align}
    \Fv_s(\rbb_p, \rbb_S) = \Pv(\rbb_p, \rbb_S) \Cv(\rbb_S),~\text{with}~ \Cv = \bbmatrix \Jv & \Kv \ebmatrix^\top.
\end{align}
\noindent If the observation fields are on the surface itself (required when solving for equivalent currents of the reference object or during reverse-propagation), then $\rbb_p = \rbb_S$, then we have for the two sides of the surface (+/-),
\begin{align*}
    \Fv_s(\rbb_p =\rbb_{S+}) &= \Fv^s_{S+} = \Pv(\rbb_S, \rbb_{S+}) \Cv (\rbb_S)= \Pv_{S+} \Cv(\rbb_S)\\
    \Fv_s(\rbb_p =\rbb_{S-}) &= \Fv^s_{S-} = \Pv(\rbb_S, \rbb_{S-}) \Cv (\rbb_S)= \Pv_{S-} \Cv(\rbb_S).
\end{align*}
Defining a surface field configuration $\Sv_F = \bbmatrix \Fv^s_{S+} & \Fv^s_{S-} \ebmatrix^T$ and a surface propagator $\Pv_S = \bbmatrix \Pv_{S+} & \Pv_{S-} \ebmatrix$ we have,
\begin{align} \label{eq:SurfPropBoth}
    \Sv_F = \Pv_S \Cv
\end{align}

The metasurface GSTCs of Eq.~\ref{Eq:ExplicitGSTCS} can be explicitly written in terms of total E- and H-fields on the left ($\{\cdot_1\}$) and right ($\{\cdot_2\}$) half of the surface as,
\begin{align*}
&\left[ \begin{array}{cccc}
     \NN_{T\times} & \0 & -\NN_{T\times}  & \0\\
     \0 &\NN_{T\times} &\0 & -\NN_{T\times}
\end{array}\right] \left[ \begin{array}{c}\Ev_+\\\Hv_+\\\Ev_-\\\Hv_-\end{array}\right]
 = \\
 &\left[ \begin{array}{cccc}
     \gamma_\text{me}\NN_{T}  & \gamma_\text{mm}\NN_{T} & \gamma_\text{me}\NN_{T} & \gamma_\text{mm}\NN_{T}\\
     \gamma_\text{ee}\NN_{T}  & \gamma_\text{em}\NN_{T} & \gamma_\text{ee}\NN_{T} & \gamma_\text{em}\NN_{T}\\
\end{array}\right]
\left[ \begin{array}{c}\Ev_+\\\Hv_+\\\Ev_-\\\Hv_-\end{array}\right]
\end{align*}
\noindent where the surface susceptibility terms are expressed using auxiliary variables as,
\begin{align*}
    \gamma_\text{ee} = \frac{j\chi_\text{ee}\omega\epsilon}{2},~\gamma_\text{me/em} = \mp \frac{j\chi_\text{me/em}\omega\sqrt{\mu\epsilon}}{2},~\gamma_\text{mm} = -\frac{j\chi_\text{mm}\omega\mu}{2}.
\end{align*}
The matrix operator $\NN_T$ performs the operation of extracting the two tangential fields at the surface (one in the $x-y$ plane and the other with respect to $\hat z$) obtaining $\E_T$ from $\E$ for every surface element. The operator $\NN_{T\times}$ extracts the total tangent field and then rotates these two fields to implement the $\Nh \times\{\cdot\}_T$ operation on every element. 
The discretized GSTC matrices can now be expressed compactly as
\begin{subequations}\label{Eq:DTF}
\begin{align} 
&   \Dv_{TF} \Sv_F = \Gv_{TF} \Sv_F~\text{with}~\\
& \Gv_{TF} = \left[ \begin{array}{cccc}
     \gamma_\text{me}\NN_{T} & \gamma_\text{mm}\NN_{T} & \gamma_\text{me}\NN_{T} & \gamma_\text{mm}\NN_{T}\\
     \gamma_\text{ee}\NN_{T} & \gamma_\text{em}\NN_{T} & \gamma_\text{ee}\NN_{T} & \gamma_\text{em}\NN_{T}\\
\end{array}\right]\\
& \Dv_{TF} = \left[ \begin{array}{cccc}
     \NN_{T\times} & \0 & -\NN_{T\times}  & \0\\
     \0 &\NN_{T\times} &\0 & -\NN_{T\times}
\end{array}\right].
\end{align}
\end{subequations}

\subsection{Susceptibility Synthesis}\label{Sec:MSS}

The metasurface surface susceptibility synthesis rests on solving the GSTCs matrix equation of Eq.~\ref{Eq:DTF}(a), once all the desired scattered fields $\Sv_F^s$ are known and the illumination fields, $\Sv_F^i$, are specified. To extract the unknown surface susceptibilities, let us consider scalar susceptibilities, for simplicity. This extraction can be conveniently performed by rearranging $\Gv_{TF}$ of Eq.~\ref{Eq:DTF}(b), as

\sg{\begin{align*}
&\Gv_{TF} = \left[ \begin{array}{cccc}
     \chi_\text{me}\Av_\text{me} & \chi_\text{mm}\Av_\text{mm}  & \chi_\text{me}\Av_\text{me} & \chi_\text{mm}\Av_\text{mm}\\
     \chi_\text{ee}\Av_\text{ee} & \chi_\text{em}\Av_\text{em} & \chi_\text{ee}\Av_\text{ee} & \chi_\text{em}\Av_\text{em}\\
\end{array}\right]~\text{with}\\
  &  \Av_\text{ee} =  \frac{j\Nv_{T}\omega\epsilon}{2},~ \Av_\text{em/me} = \pm \frac{j\Nv_{T}\omega\sqrt{\mu\epsilon}}{2},~\Av_\text{mm} = -\frac{j\Nv_{T}\omega\mu}{2} 
\end{align*}}

Considering that the metasurface susceptibilities are discretized over the surface as $\chi(\r_m)$ and thus become vectors of localized susceptibilities ($\Xv$), we can express the right hand side of Eq.~\ref{Eq:DTF}(a), as

\begin{align*}
    \Gv_{TF}\Sv_{F} 
    % &= \bbmatrix
    %   \chi_\text{me}\Av_{me} & \chi_\text{mm}\Av_{mm}  & \chi_\text{me}\Av_{me} & \chi_\text{mm}\Av_{mm}\\
    %  \chi_\text{ee}\Av_{ee} & \chi_\text{em}\Av_{em} & \chi_\text{ee}\Av_{ee} & \chi_\text{em}\Av_{em}\\
    % \ebmatrix 
    % \bbmatrix \Ev_1\\\Hv_1\\\Ev_2\\\Hv_2\ebmatrix\\
    % &= \bbmatrix
    %   \chi_\text{me}\Av_{me}\Ev_1 + \chi_\text{mm}\Av_{mm}\Hv_1  + \chi_\text{me}\Av_{me}\Ev_2 + \chi_\text{mm}\Av_{mm}\Hv_2\\
    %  \chi_\text{ee}\Av_{ee}\Ev_1 + \chi_\text{em}\Av_{em}\Hv_1 + \chi_\text{ee}\Av_{ee}\Ev_2 + \chi_\text{em}\Av_{em}\Hv_2
    % \ebmatrix \\
    &= \left[ \begin{array}{l}
      \Xv_\text{me}\circ\left(\Av_\text{me}(\Ev_1 + \Ev_2)\right) \; + \\ \quad \quad \Xv_\text{mm}\circ\left(\Av_\text{mm}(\Hv_1 + \Hv_2)\right)\\
     \Xv_\text{ee}\circ\left(\Av_\text{ee}(\Ev_1 + \Ev_2)\right)  \; + \\ \quad \quad \Xv_\text{em}\circ\left(\Av_\text{em}(\Hv_1 +  \Hv_2)\right)
    \end{array}\right],
\end{align*}

\noindent where $\circ$ is the point-wise Hagamard product. Each of the terms,
\begin{align*}
    \Av_\text{me}(\Ev_1 + \Ev_2) 
    % = \bbmatrix \E_{Gme,1}\\\vdots\\ \E_{Gme,N}\\\hline \E_{Gme,N+1} \\ \vdots \\ \E_{Gme,2N} 
    % \ebmatrix 
    = \bbmatrix \Bv_{\text{me},xy}\\\Bv_{\text{me},z} \ebmatrix = \Bv_\text{me},
\end{align*}
\noindent is a column vector of one component of tangent fields ($xy$ and $z$). If we wish to create a distributed $\chi$ vector we can form,
\begin{align*}
    \Gv_\text{me} \Xv_\text{me} &= \Xv_\text{me} \circ \Av_\text{me}(\Ev_1 + \Ev_2)   
\end{align*}
\noindent where we define a diagonal matrix,
\begin{align*}
    \Gv_\text{me} = \bbmatrix 
    B_\text{me,1} & 0 & \dots & 0\\ 
    0  &B_\text{me,2} & \dots & 0 \\
    0  & 0 & \ddots & 0 \\
    0  & 0 & 0 & B_\text{me,2N}
    \ebmatrix
\end{align*}
\noindent This form allows a very convenient expression of RHS of Eq.~\ref{Eq:DTF}(a), where the susceptiblity matrix term is explicitly extracted as
\begin{align}
    \Gv_{TF}\Sv_{F} 
    % &= \bbmatrix
    %   \chi_\text{me}\Av_{me}(\Ev_1 + \Ev_2) + \chi_\text{mm}\Av_{mm}(\Hv_1 + \Hv_2)\\
    %  \chi_\text{ee}\Av_{ee}(\Ev_1 + \Ev_2)  + \chi_\text{em}\Av_{em}(\Hv_1 +  \Hv_2)
    % \ebmatrix \\
    % &\bbmatrix
    %   \Bv_{me} \Xv_{em} +\Bv_{mm} \Xv_{mm}\\
    %   \Hv_{Gee} \Xv_{ee} +\Hv_{Gme} \Xv_{me}
    % \ebmatrix 
    &= 
    \bbmatrix
       \Gv_{me} & \Gv_{mm} &\0 & \0\\
      \0 & \0 &  \Gv_{ee} & \Gv_{me} 
    \ebmatrix
    \bbmatrix\Xv_{em}\\\Xv_{mm}\\\Xv_{ee} \\ \Xv_{me}\ebmatrix = \Qv \Xv.
\end{align}

 Finally using Eq.~\ref{Eq:DTF}, we now have the explicit relationship for the spatially varying surface susceptibility matrix as

\sg{\begin{align}
    \Xv = \Qv^{-1} \Dv_{TF}\Sv_{F},
\end{align}}

\noindent which can be used directly for metasurface synthesis for a given $\Sv_{F}$.

\subsection{Solution of the Final Configuration}

To confirm the synthesis of the MS a full simulation of the surface with appropriate illumination is needed. For a single surface subject to an illumination we use Eq. \ref{eq:SurfPropBoth} to determine the surface fields, force the normal components of the currents on the metasurface to be zero and enforce the interface conditions prescribed by Eq. \ref{Eq:DTF}a. These equations can be assembled into a final matrix equation,
\begin{align} \label{Eq:SysMat}
    \left[\begin{array}{ccc} 
    \Pv_{S} & -\Iv \\
    \Nv_{DC} & \0 \\
    \0 & (\Dv_{TF} -\Gv_{TF}) 
    \end{array}\right]
    \left[\begin{array}{c}\raisebox{-.5\normalbaselineskip}{$\Cv$} \\ \\ \raisebox{.5\normalbaselineskip}{$\Sv_F^s$} 
    \end{array}\right]
    = 
    \left[\begin{array}{c} \0 \\ \0 \\ -(\Dv_{TF} -\Gv_{TF}) \Sv_F^i 
    \end{array}\right]
\end{align}
where $\Nv_{DC}$ takes the dot product of the currents for all elements and enforces $\Nv_{DC}\Cv = \0$ -- setting the normal component of the currents to zero. The surface fields $\Sv_F$ have been split into two components: 1) the unknown scattered fields $\Sv_F^s$, and 2) the known applied field on the metasurface (reference or the illumination fields, for instance) $\Sv_F^i$, so that $\Sv_F = \Sv_F^s + \Sv_F^i$. The solution of this equation provides $\Cv$ and $\Sv^s_F$ and using $\Cv$, the fields at any point in the simulation domain can be obtained using a propagation matrix. 

\begin{figure*}[htbp]
\begin{center}
     \begin{subfigure}[b]{0.3\textwidth}
         \centering
         \psfrag{a}[c][c][0.7]{$y$~(m)}
         \psfrag{b}[c][c][0.7]{$x$~(m)}
         \psfrag{I}[c][c][0.6]{\shortstack{\textsc{Horizon}\\ $x=0$}}
         \psfrag{O}[c][c][0.6]{\shortstack{\textsc{Observation}\\ $x=x_0$}}
         \psfrag{R}[c][c][0.6]{\color{white}\textsc{\shortstack{Object}}}         
	\psfrag{T}[c][c][0.7]{\color{white}$\theta_\text{in}$}
	\includegraphics[width=\columnwidth]{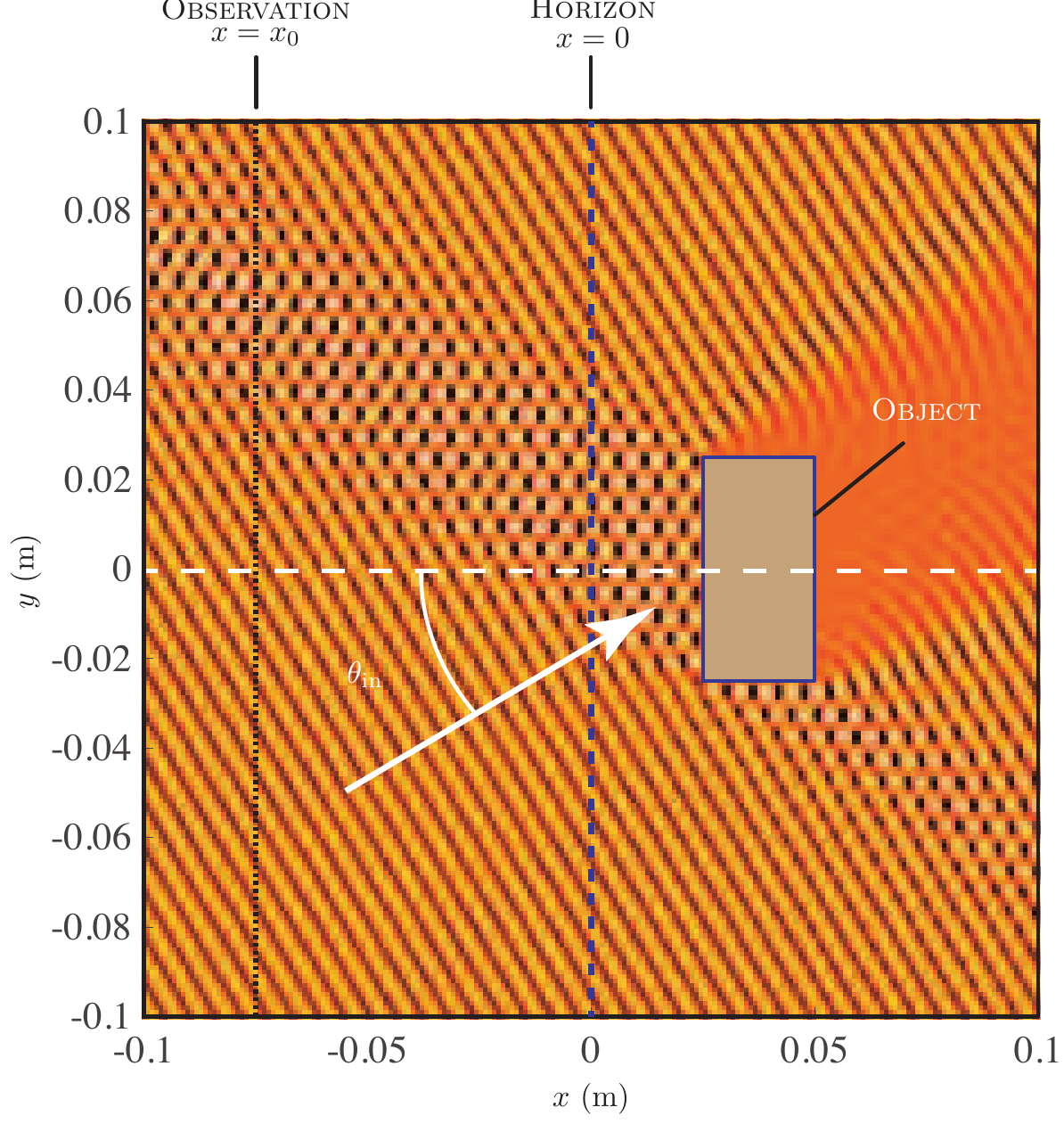}\caption{Re$\{E_{z,t}^\text{ref.}(x,y)\}$ - Object only}
     \end{subfigure}
     \begin{subfigure}[b]{0.3\textwidth}
         \centering
         \psfrag{a}[c][c][0.7]{$y$~(m)}
         \psfrag{b}[c][c][0.7]{$x$~(m)}
         \psfrag{I}[c][c][0.6]{\shortstack{\textsc{Horizon}\\ $x=0$}}
         \psfrag{O}[c][c][0.6]{\shortstack{\textsc{Observation}\\ $x=x_0$}}
         \psfrag{R}[c][c][0.6]{\color{white}\textsc{\shortstack{Object}}}                
         \psfrag{T}[c][c][0.7]{\color{white}$\theta_\text{in}$}
         	\includegraphics[width=\columnwidth]{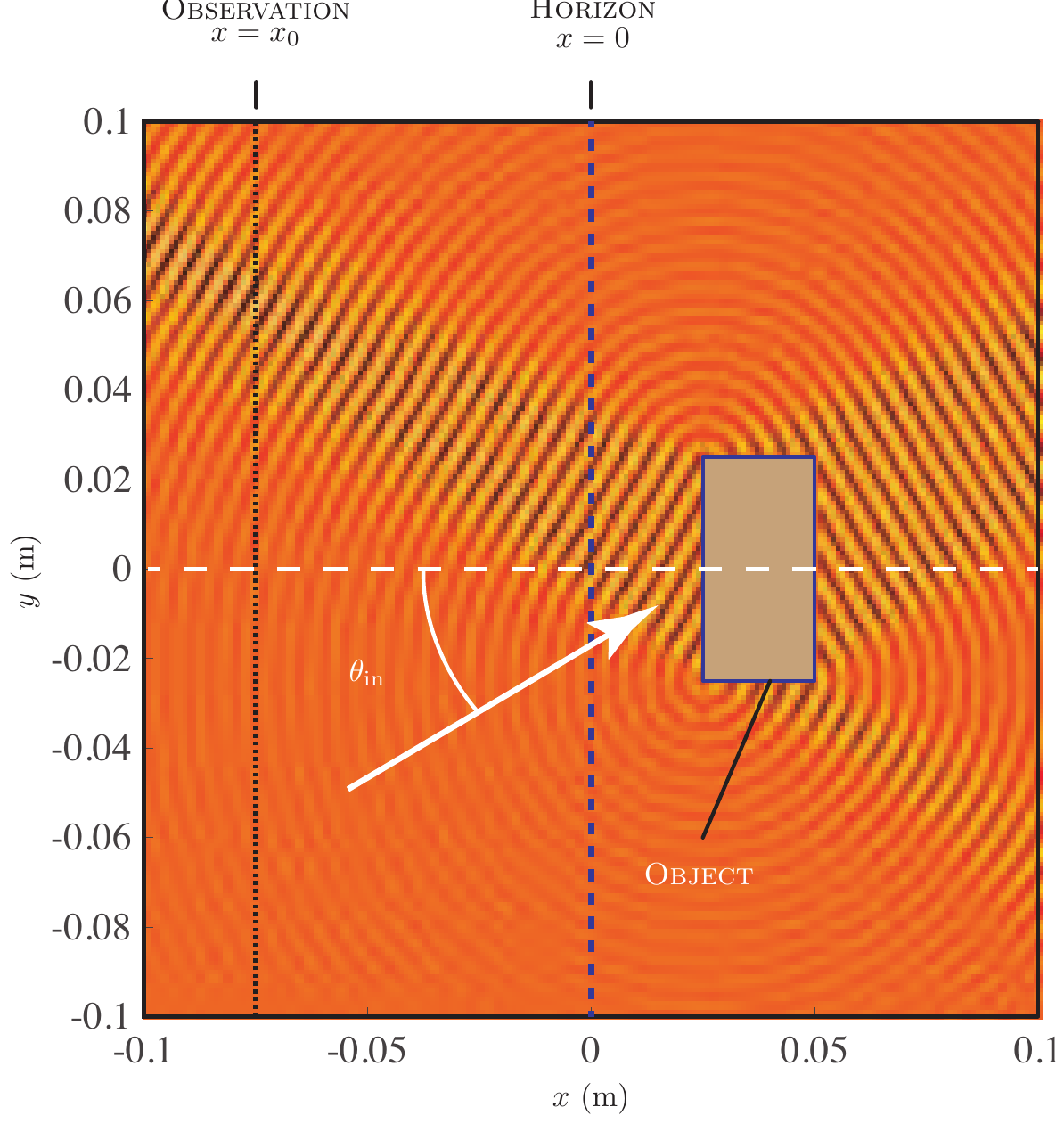}\caption{Re$\{E_{z,s}^\text{ref.}(x,y)\}$ - Object Only}
     \end{subfigure}
     \begin{subfigure}[b]{0.3\textwidth}
         \centering
         \psfrag{a}[c][c][0.7]{$y$~(m)}
         \psfrag{e}[l][c][0.6]{Re\{$\chi$\}}
         \psfrag{d}[l][c][0.6]{Im\{$\chi$\}}
         \psfrag{b}[c][c][0.7]{$\chi_\text{mm}(y)$}
         \psfrag{c}[c][c][0.7]{$\chi_\text{ee}(y)$}
         \psfrag{T}[c][c][0.7]{\color{white}$\theta_\text{in}$}
         	\includegraphics[width=1\columnwidth]{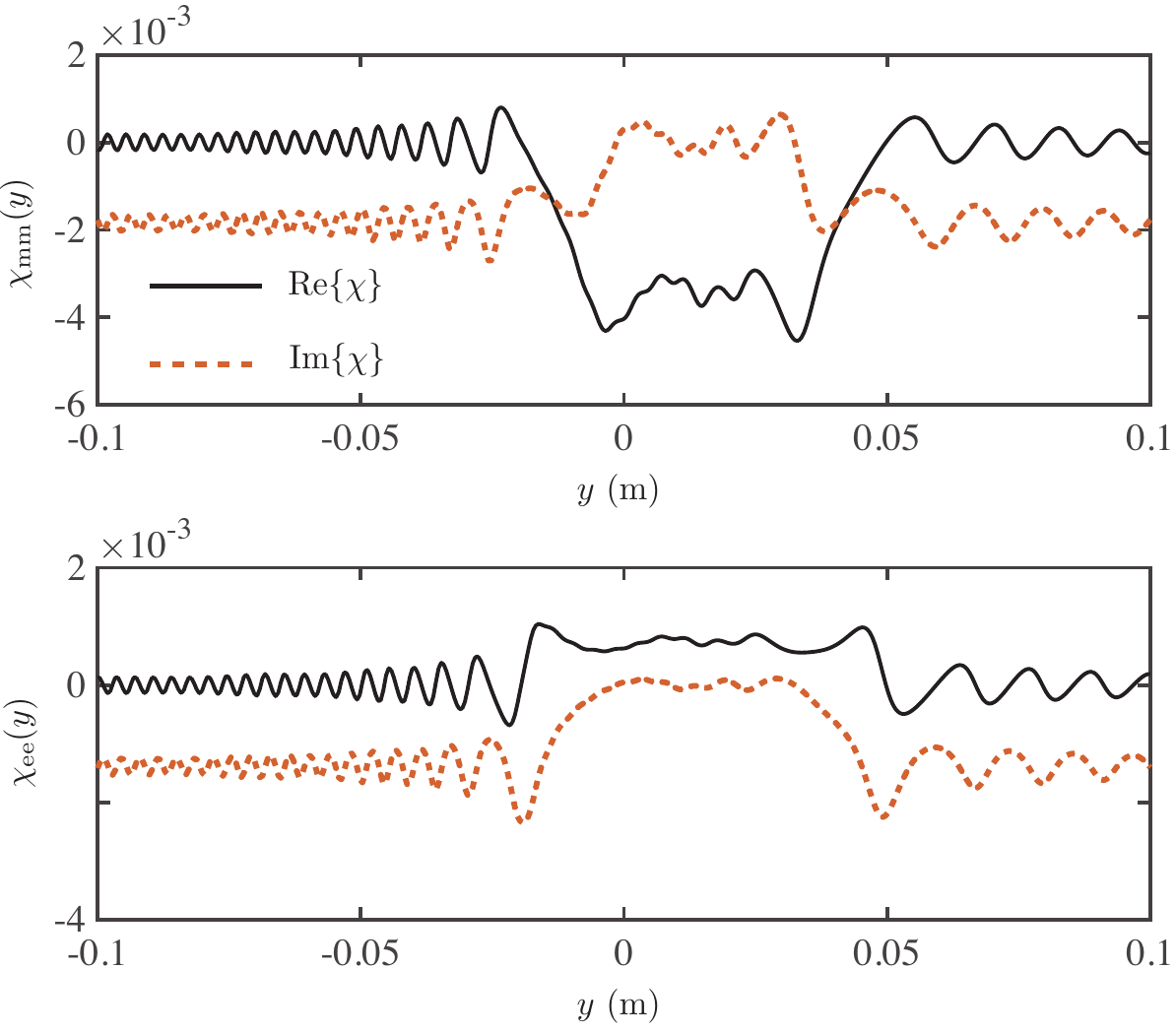}\caption{Synthesized Surface Susceptibilities}
     \end{subfigure}
     \par\bigskip
     \begin{subfigure}[b]{0.3\textwidth}
         \centering
         \psfrag{a}[c][c][0.7]{$y$~(m)}
         \psfrag{b}[c][c][0.7]{$x$~(m)}
         \psfrag{M}[c][c][0.7]{\color{white}\textsc{Metasurface}}
         \psfrag{O}[c][c][0.6]{\shortstack{\textsc{Observation}\\$x=x_0$}}
         \psfrag{V}[c][c][0.6]{\color{white}\textsc{\shortstack{Virtual \\Object}}}
         \psfrag{T}[c][c][0.7]{\color{white}$\theta_\text{in}$}
	\psfrag{T}[c][c][0.7]{\color{white}$\theta_\text{in}$}
	\includegraphics[width=\columnwidth]{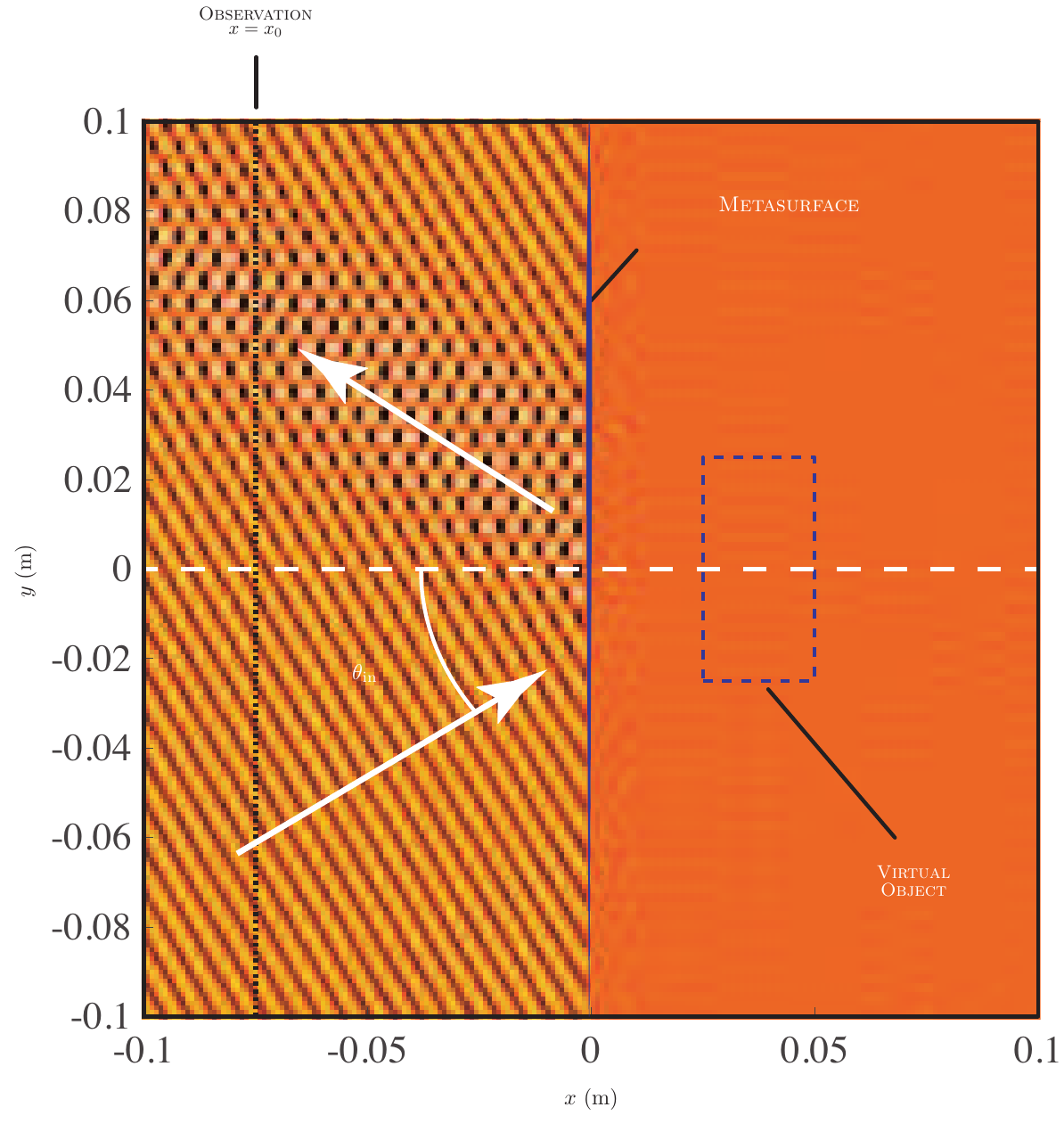}\caption{Re$\{E_{z,t}^\text{ms.}(x,y)\}$ - Metasurface only}
     \end{subfigure}
     \begin{subfigure}[b]{0.3\textwidth}
         \centering
         \psfrag{a}[c][c][0.7]{$y$~(m)}
         \psfrag{b}[c][c][0.7]{$x$~(m)}
         \psfrag{M}[c][c][0.7]{\color{white}\textsc{Metasurface}}
         \psfrag{I}[c][c][0.6]{\shortstack{\textsc{Horizon}\\ $x=0$}}
         \psfrag{O}[c][c][0.6]{\shortstack{\textsc{Observation}\\ $x=x_0$}}
         \psfrag{V}[c][c][0.6]{\color{white}\textsc{\shortstack{Virtual \\Object}}}
         \psfrag{T}[c][c][0.7]{\color{white}$\theta_\text{in}$}
         	\includegraphics[width=1\columnwidth]{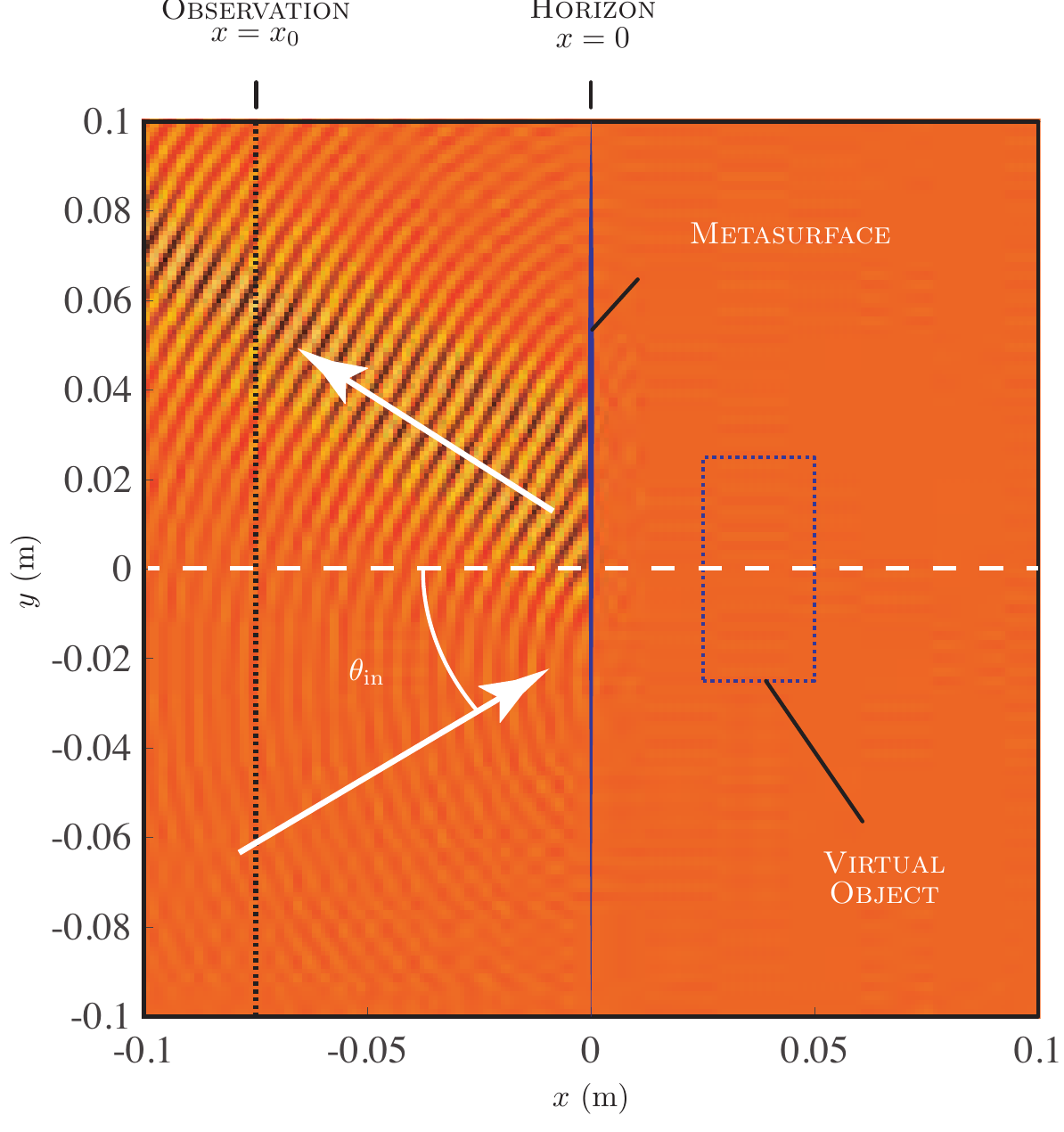}\caption{Re$\{E_{z,s}^\text{ms.}(x,y)\}$ - Metasurface only}
     \end{subfigure}
     \begin{subfigure}[b]{0.3\textwidth}
         \centering
         \psfrag{a}[c][c][0.7]{$y$~(m)}
         \psfrag{d}[c][c][0.6]{$|E_z(0, y)|$~(norm.)}   
	\psfrag{e}[c][c][0.6]{$|E_z(x_0, y)|$~(norm.)}     
	\psfrag{f}[l][c][0.6]{$|E_{z,s}^\text{ref.}(x_0,y)|$}    
	\psfrag{g}[l][c][0.6]{$|E_{z,s}^\text{ms.}(x_0,y)|$}   
	\psfrag{m}[l][c][0.6]{Re\{$E_{z,s}^\text{ref.}(0,y)$\}, Re\{$E_{z,s}^\text{ms.}(0_-,y)$\}}  
	\psfrag{n}[l][c][0.6]{Im\{$E_{z,s}^\text{ref.}(0,y)$\}, Im\{$E_{z,s}^\text{ms.}(0_-,y)$\}}       
	\includegraphics[width=1\columnwidth]{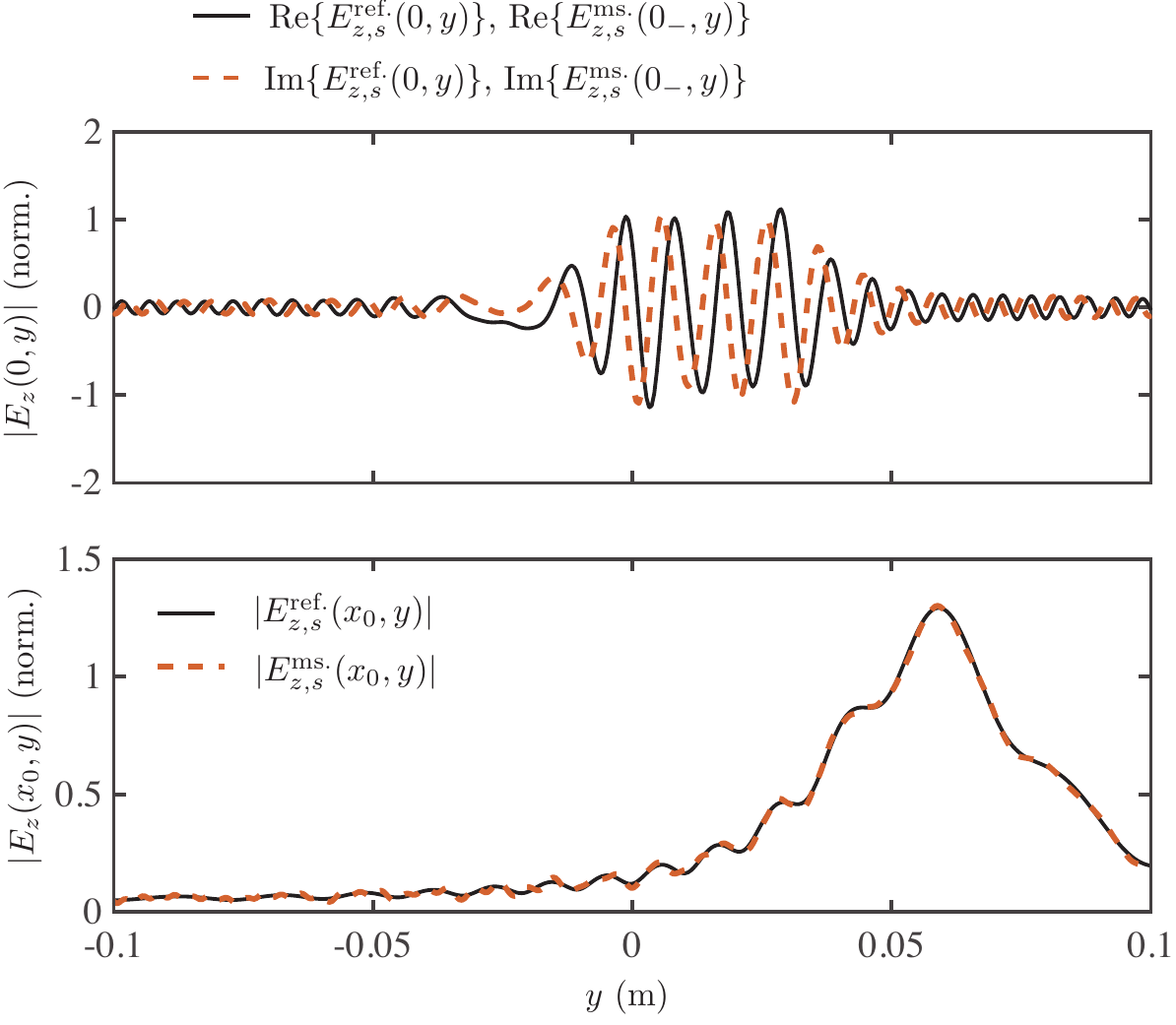}\caption{Reconstructed Scattered Fields}
     \end{subfigure}               
\par\bigskip
     \begin{subfigure}[b]{0.3\textwidth}
         \centering
         \psfrag{a}[c][c][0.7]{$y$~(m)}
         \psfrag{e}[l][c][0.6]{Re\{$\chi$\}}
         \psfrag{d}[l][c][0.6]{Im\{$\chi$\}}
         \psfrag{b}[c][c][0.7]{$\chi_\text{mm}(y)$}
         \psfrag{c}[c][c][0.7]{$\chi_\text{ee}(y)$}
	\includegraphics[width=1\columnwidth]{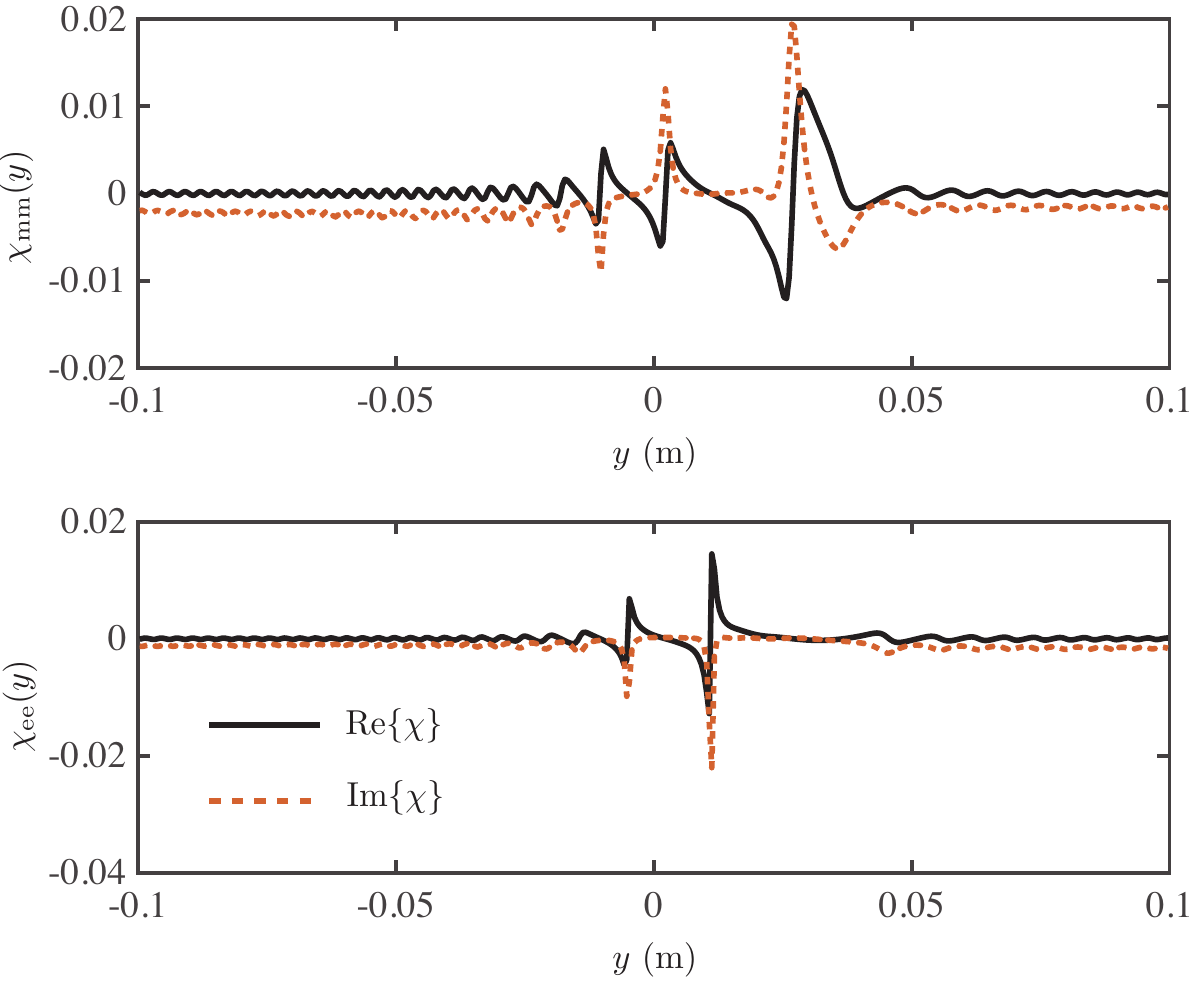}\caption{Susceptibilities -  Curvilinear Metasurface}
     \end{subfigure}
     \begin{subfigure}[b]{0.3\textwidth}
         \centering
         \psfrag{a}[c][c][0.7]{$y$~(m)}
         \psfrag{b}[c][c][0.7]{$x$~(m)}
          \psfrag{M}[c][c][0.7]{\color{white}\textsc{Metasurface}}
         \psfrag{O}[c][c][0.6]{\shortstack{\textsc{Observation}\\ $x=x_0$}}
         \psfrag{V}[c][c][0.6]{\color{white}\textsc{\shortstack{Virtual \\Object}}}
         \psfrag{T}[c][c][0.7]{\color{white}$\theta_\text{in}$}
         	\includegraphics[width=\columnwidth]{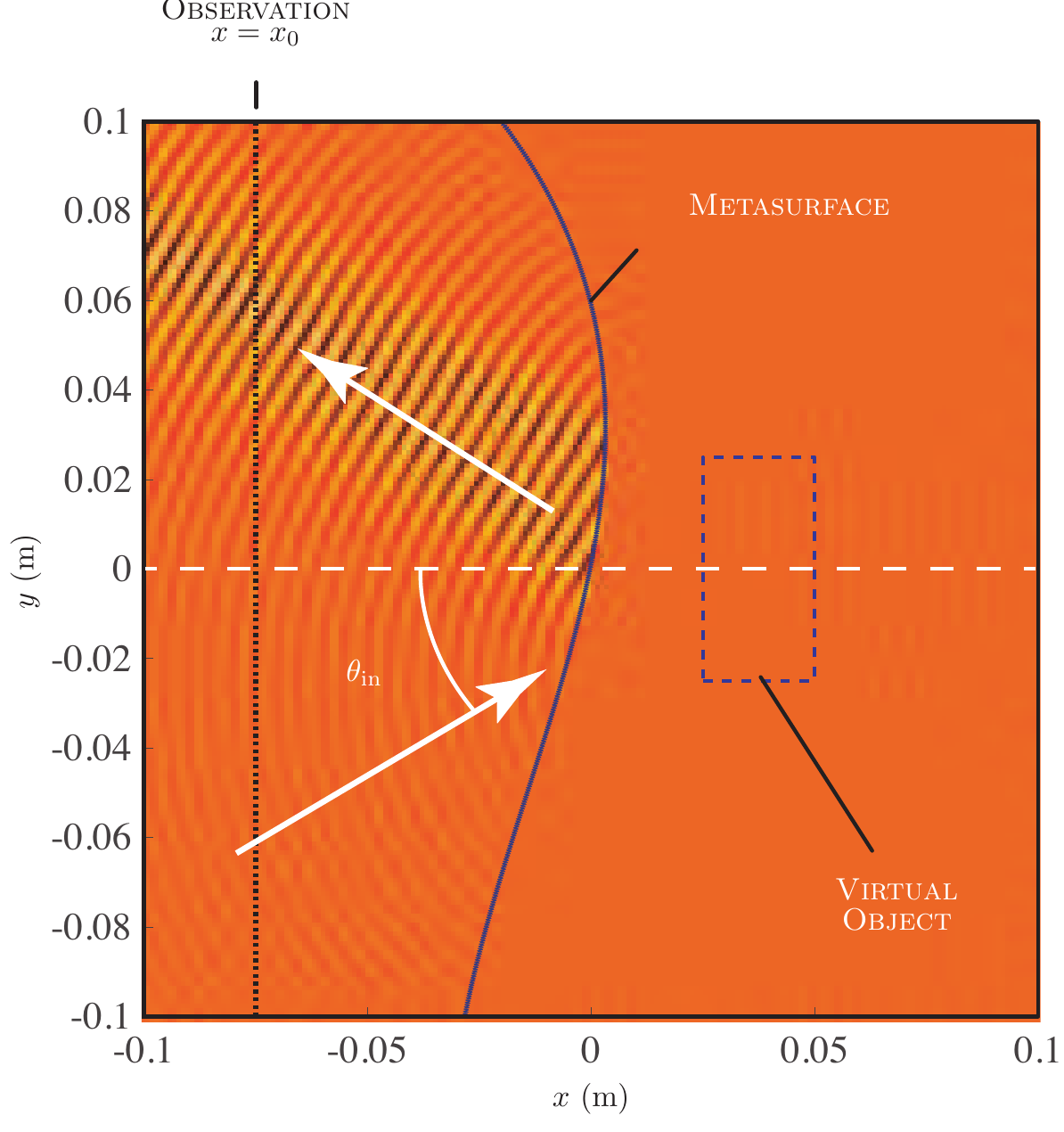}\caption{Re$\{E_{z,s}^\text{ms.}(x,y)\}$ -  Curvilinear Metasurface}
     \end{subfigure}
     \begin{subfigure}[b]{0.3\textwidth}
         \centering
         \psfrag{a}[c][c][0.7]{$y$~(m)}
         \psfrag{d}[c][c][0.6]{$|E_z(0, y)|$~(norm.)}   
	\psfrag{e}[c][c][0.6]{$|E_z(x', y)|$~(norm.)}     
	\psfrag{f}[l][c][0.6]{$|E_{z,s}^\text{ref.}(x_0,y)|$}    
	\psfrag{g}[l][c][0.6]{$|E_{z,s}^\text{ms.}(x_0,y)|$}     
	\psfrag{m}[l][c][0.6]{Re\{$E_{z,s}^\text{ref.}(0,y)$\}, Re\{$E_{z,s}^\text{ms.}(0_-,y)$\}}  
	\psfrag{n}[l][c][0.6]{Im\{$E_{z,s}^\text{ref.}(0,y)$\}, Im\{$E_{z,s}^\text{ms.}(0_-,y)$\}}   
	\includegraphics[width=1\columnwidth]{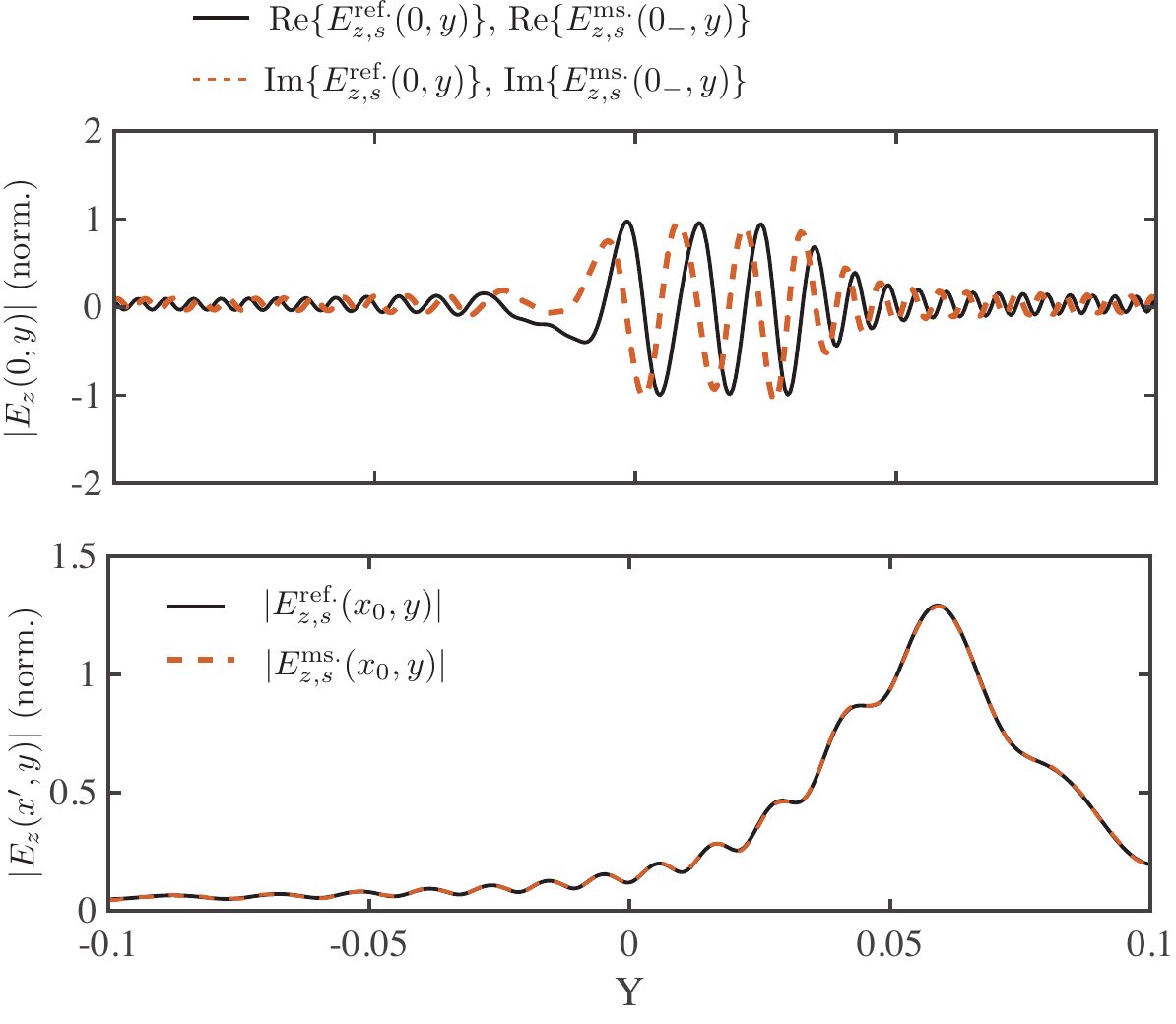}\caption{Reconstructed Scattered Fields}
     \end{subfigure}    
\caption{Front-lit posterior illusion with front illumination, where the illumination field is identical to the incident field, using a flat (a-f) and a curvilinear metasurface (g-i), respectively. The simulation parameters are:  $f = 60$~GHz ($\lambda = 5$~mm), $x_0 = -15\lambda$, $x_h = x_m = 0$. Rectangular PEC object $5\lambda \times 10\lambda$ centered at $x = 7.5\lambda$, metasurface length $\ell_\text{ms} = 120\lambda$. Both incident and illumination are oblique uniform plane-waves of $|E_z|=1.0$ with 30$^\circ$ tilt measured from the $x-$axis. Notation: $\psi^\text{ref.}_{z,s}$ represents the $z-$component of the scattered fields $\psi$ for the simulation with the reference object only; $\psi^\text{ms.}_{z,t}$ represents the $z-$component of the total fields $\psi$ for the illusion simulation with the metasurface and no object.}\label{Fig:FLPFI}
\end{center}
\end{figure*}

\begin{figure*}[htbp]
\begin{center}
     \begin{subfigure}[b]{0.3\textwidth}
         \centering
         \psfrag{a}[c][c][0.7]{$y$~(m)}
         \psfrag{e}[l][c][0.6]{Re\{$\chi$\}}
         \psfrag{d}[l][c][0.6]{Im\{$\chi$\}}
         \psfrag{b}[c][c][0.7]{$\chi_\text{mm}(y)$}
         \psfrag{c}[c][c][0.7]{$\chi_\text{ee}(y)$}
	\includegraphics[width=1\columnwidth]{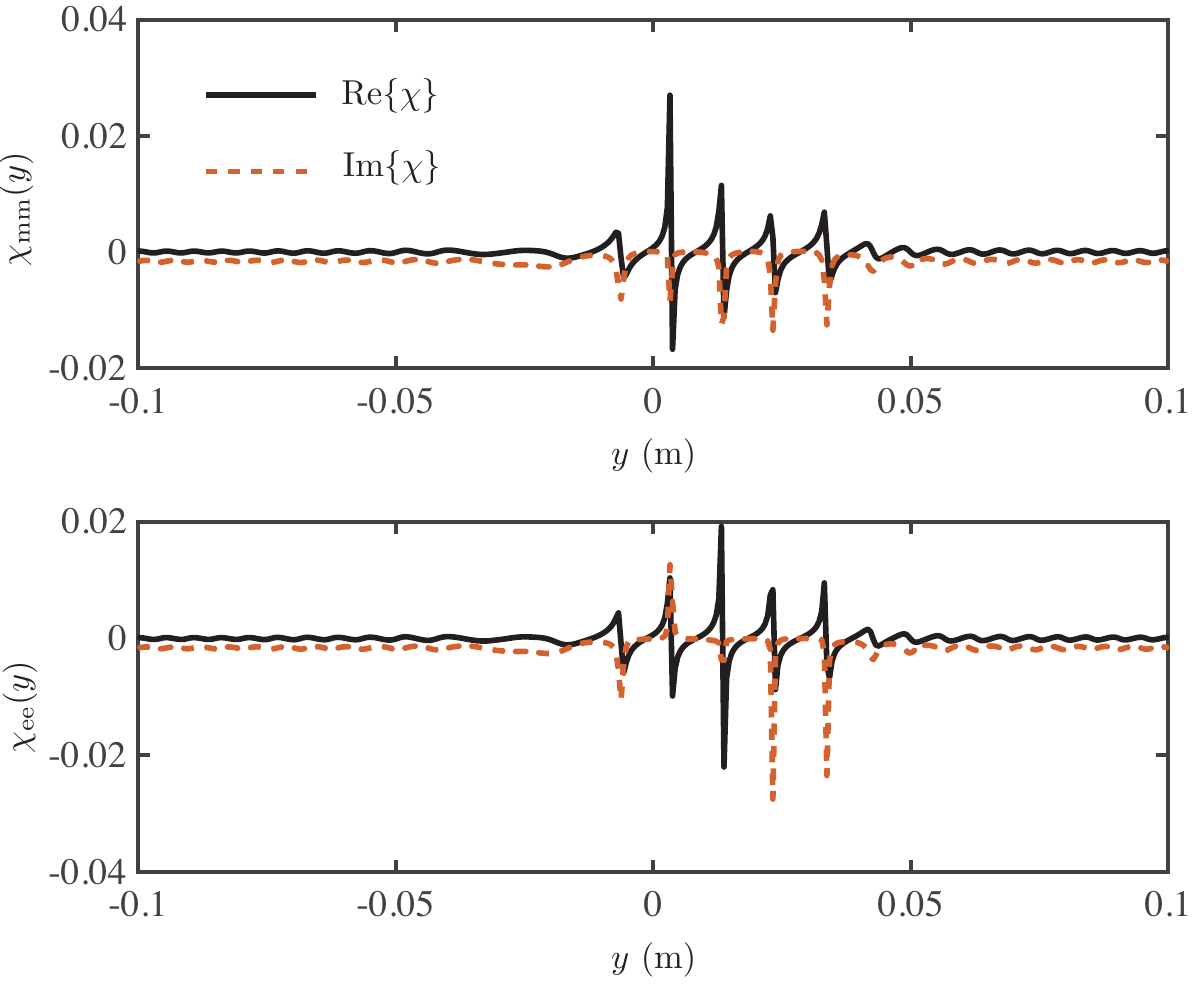}\caption{Synthesized Surface Susceptibilities}
     \end{subfigure}
     \begin{subfigure}[b]{0.3\textwidth}
         \centering
         \psfrag{a}[c][c][0.7]{$y$~(m)}
         \psfrag{b}[c][c][0.7]{$x$~(m)}
         \psfrag{M}[c][c][0.7]{\color{white}\textsc{Metasurface}}
         \psfrag{O}[c][c][0.6]{\shortstack{\textsc{Observation}\\ $x=x_0$}}
         \psfrag{V}[c][c][0.6]{\color{white}\textsc{\shortstack{Virtual \\Object}}} 
         \psfrag{I}[c][c][0.6]{\shortstack{\textsc{Horizon}\\ $x=0$}}        
	\includegraphics[width=\columnwidth]{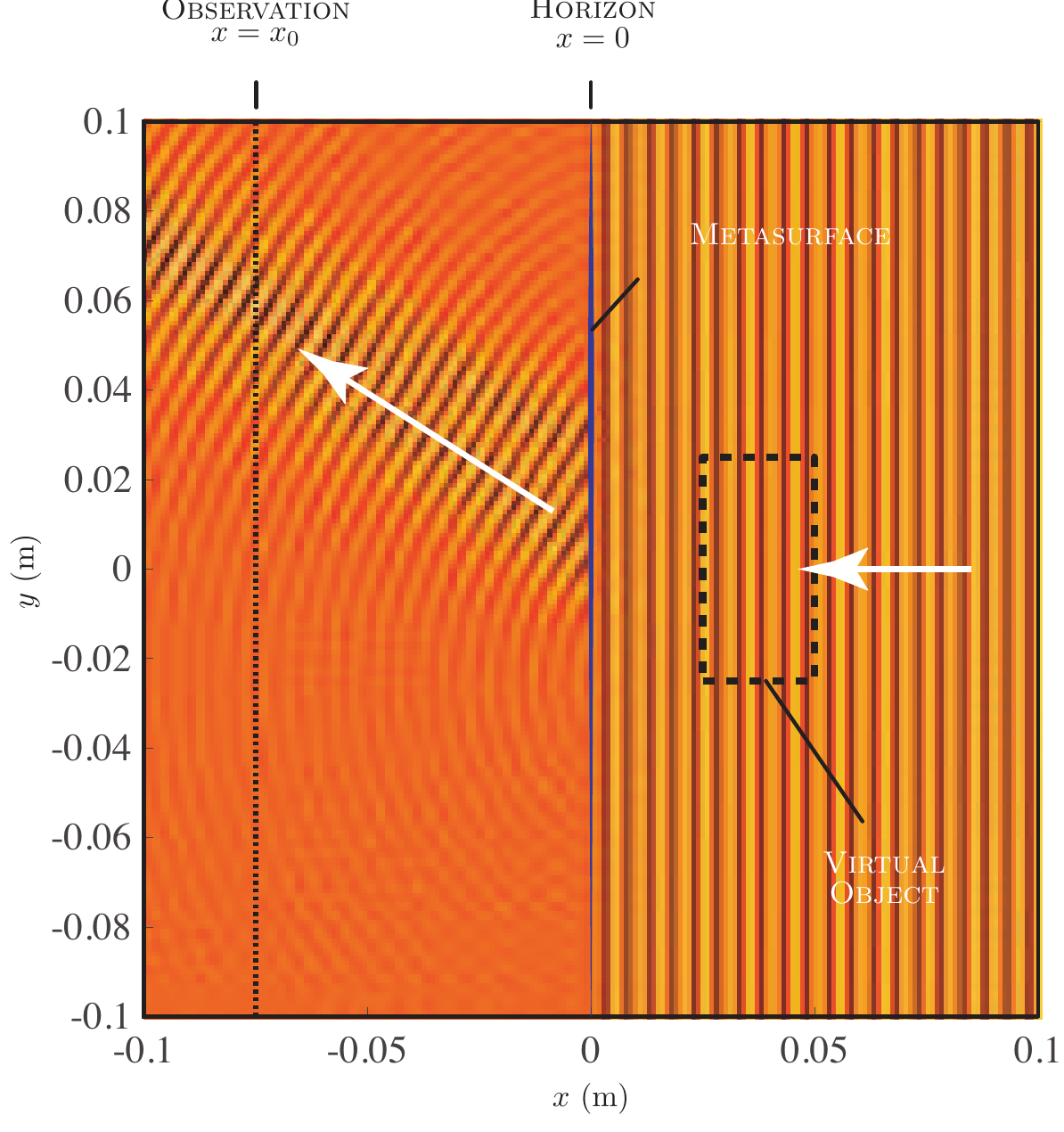}\caption{Re$\{E_{z,t}^\text{ms.}(x,y)\}$ -  Metasurface only}
     \end{subfigure}
     %
%     \begin{subfigure}[b]{0.23\textwidth}
%         \centering
%         \psfrag{a}[c][c][0.7]{$\J,~\K$}
%	\includegraphics[width=0.9\columnwidth]{Results/FrontLitPosterior/PosteriorPWFrontLitBackIllRectPEC/Scattered_Ez_synth.eps}\caption{$E_z$ Scattered fields - MS}
%         
%     \end{subfigure}
          %
     \begin{subfigure}[b]{0.3\textwidth}
         \centering
         \psfrag{a}[c][c][0.7]{$y$~(m)}
         \psfrag{d}[c][c][0.6]{$|E_z(0, y)|$~(norm.)}   
	\psfrag{e}[c][c][0.6]{$|E_z(x_0, y)|$~(norm.)}     
	\psfrag{f}[l][c][0.6]{$|E_{z,s}^\text{ref.}(x_0,y)|$}    
	\psfrag{g}[l][c][0.6]{$|E_{z,s}^\text{ms.}(x_0,y)|$}   
	\psfrag{m}[l][c][0.6]{Re\{$E_{z,s}^\text{ref.}(0,y)$\}, Re\{$E_{z,s}^\text{ms.}(0_-,y)$\}}  
	\psfrag{n}[l][c][0.6]{Im\{$E_{z,s}^\text{ref.}(0,y)$\}, Im\{$E_{z,s}^\text{ms.}(0_-,y)$\}}   
	\includegraphics[width=1\columnwidth]{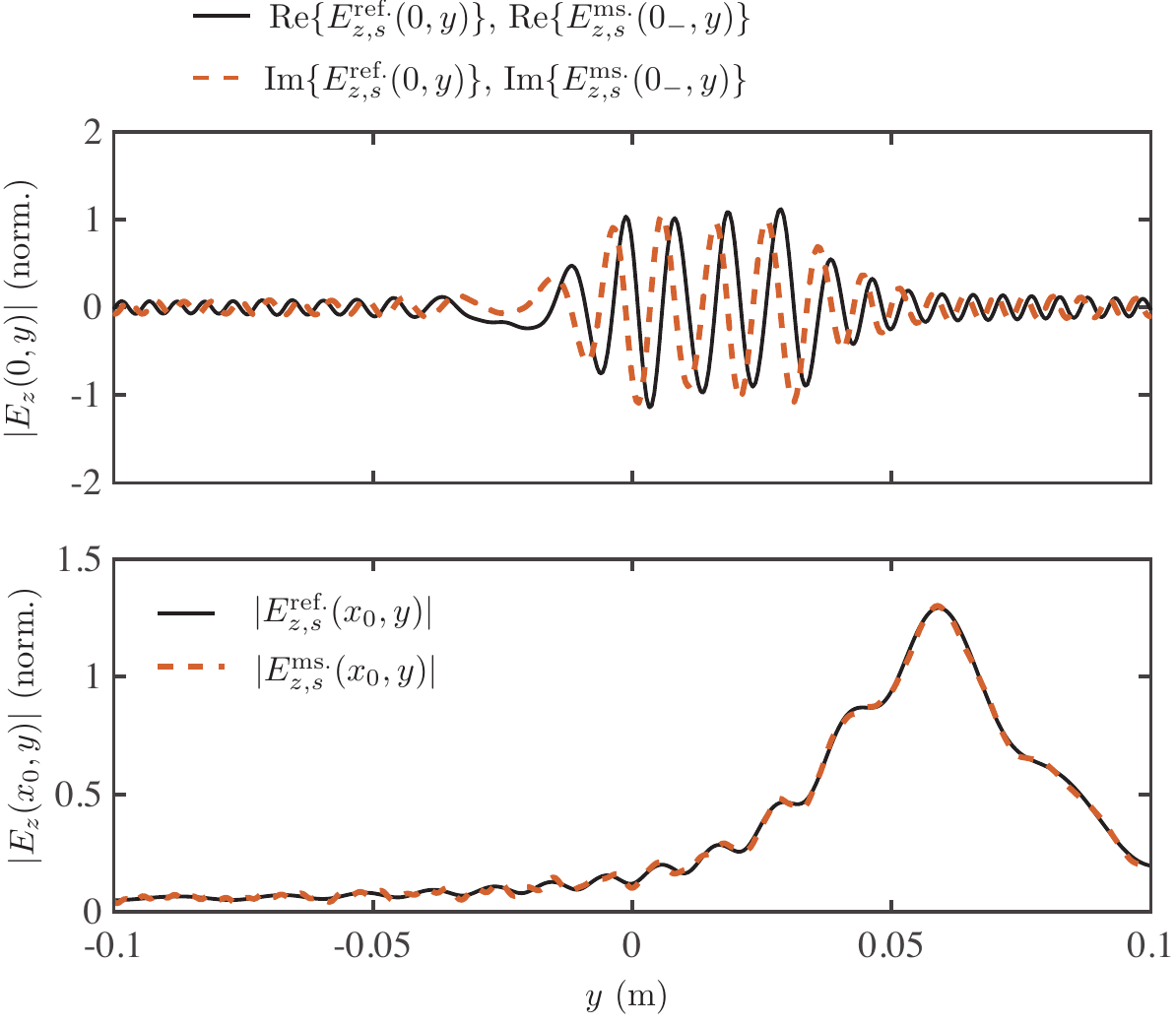}\caption{Reconstructed Scattered Fields}
     \end{subfigure}     
     \par\bigskip
 \caption{Front-lit posterior illusion with back illumination, where the illumination field is a plane-wave incidenting from the right-half region. Simulation parameters are: Rectangular PEC object $5\lambda \times 10\lambda$ centered at $x= 7.5\lambda$; the illumination is a uniform plane wave traveling along $-x$ direction with $|E_z| = 1.0$. Rest of the parameters are same as that mentioned in caption of Fig.~\ref{Fig:FLPFI}.}\label{Fig:FLPBI}
\end{center}
\end{figure*}

\section{Results and Discussion}\label{Sec:Res}

To illustrate the synthesis procedure for metasurface holograms and the ability to reconstruct desired fields, let us consider a few examples. Given that a large number of illusion configurations are possible, however, only few will be presented for better readability and tractability, and to highlight various aspects of the proposed synthesis procedures. 

The simulation setup follows the illustration of Fig.~\ref{Fig:Illusion_Setup}, in the $x-y$ plane, and the frequency of operation is chosen to be $60$~GHz ($\lambda = 0.005$~m) where the observer is fixed at $x_0 = -15\lambda$. The surface meshing is set to $\lambda/10$ based on proper convergence and the metasurface has a physical extent of $120\lambda$ unless otherwise noted.

\subsection{Front-lit Posterior Illusion with Front/Back-illumination}

We first synthesize a metasurface hologram that creates an illusion of a rectangular PEC (Perfect Electric Conductor) object. The first step is to determine the scattered fields from a real PEC object when excited with a reference field - a uniform plane-wave - incidenting at an angle of $\theta_\text{in}$ as shown in Fig.~\ref{Fig:FLPFI}. The reference total and scattered fields are shown in Fig.~\ref{Fig:FLPFI}(a) and (b), respectively. The ideal PEC object creates a shadow region behind it, and produces strong scattered fields through its flat faces and sharp corner diffraction. Next, the Horizon is defined at $x=0$, which is an arbitrary choice as long as it is lying to the left of the object. The scattered fields (both amplitude and phase) are now recorded on the horizon, which our metasurface hologram must recreate.

We then remove the object and a metasurface hologram is introduced. For posterior illusion, the metasurface is placed directly at the horizon at $x=0$. For this first test, the illumination fields are kept identical to the reference fields. The metasurface susceptibilities are next computed using the Horizon fields computed earlier assuming a planar configuration, and it is found that it is sufficient to use only the electric and magnetic surface susceptibilities, $\chi_\text{ee}$ and $\chi_\text{mm}$, to recreate the fields. They are shown in Fig.~\ref{Fig:FLPFI}(c) as a function of space. Their complex nature suggests that the metasurface must produce a spatially varying transformation of both amplitude and phase. 

Then using the standard BEM-GSTC technique of \cite{stewart2019scattering}, the response of this synthesized metasurface is simulated when excited with an illumination field. The resulting total and scattered fields are shown in Fig.~\ref{Fig:FLPFI}(d-e). The scattered fields of Fig.~\ref{Fig:FLPFI}(e) produced by the metasurface hologram must be compared with that of the object only in Fig.~\ref{Fig:FLPFI}(b). The metasurface hologram successfully recreates the scattered fields on the left of the horizon, while producing zero fields on the other side as imposed in the synthesis steps. The 1D recreated scattered fields at the metasurface location are compared with the desired fields at the Horizon and shown to be perfectly superimposed in Fig.~\ref{Fig:FLPFI}(f), as expected. Furthermore, at the observer located at $x=x_0$, there is an excellent match between the recreated scattered fields and the original fields of the object. Therefore, from the perspective of the observer, the metasurface hologram is perceived as the original PEC rectangular object -- a virtual object image is formed behind the metasurface.

\begin{figure*}[htbp]
\begin{center}
     \begin{subfigure}[b]{0.3\textwidth}
         \centering
         \psfrag{a}[c][c][0.7]{$y$~(m)}
         \psfrag{b}[c][c][0.7]{$x$~(m)}
         \psfrag{I}[c][c][0.6]{\shortstack{\textsc{Metasurface}\\ $x=x_m$}}
         \psfrag{O}[c][c][0.6]{\shortstack{\textsc{Observation}\\ $x=x_0$}}
         \psfrag{C}[c][c][0.6]{\shortstack{\textsc{Horizon}\\ $x=0$}}
         \psfrag{R}[c][c][0.6]{\color{white}\textsc{\shortstack{Object}}}       
         \psfrag{T}[c][c][0.7]{\color{white}$\theta_\text{in}$}        
	\includegraphics[width=1\columnwidth]{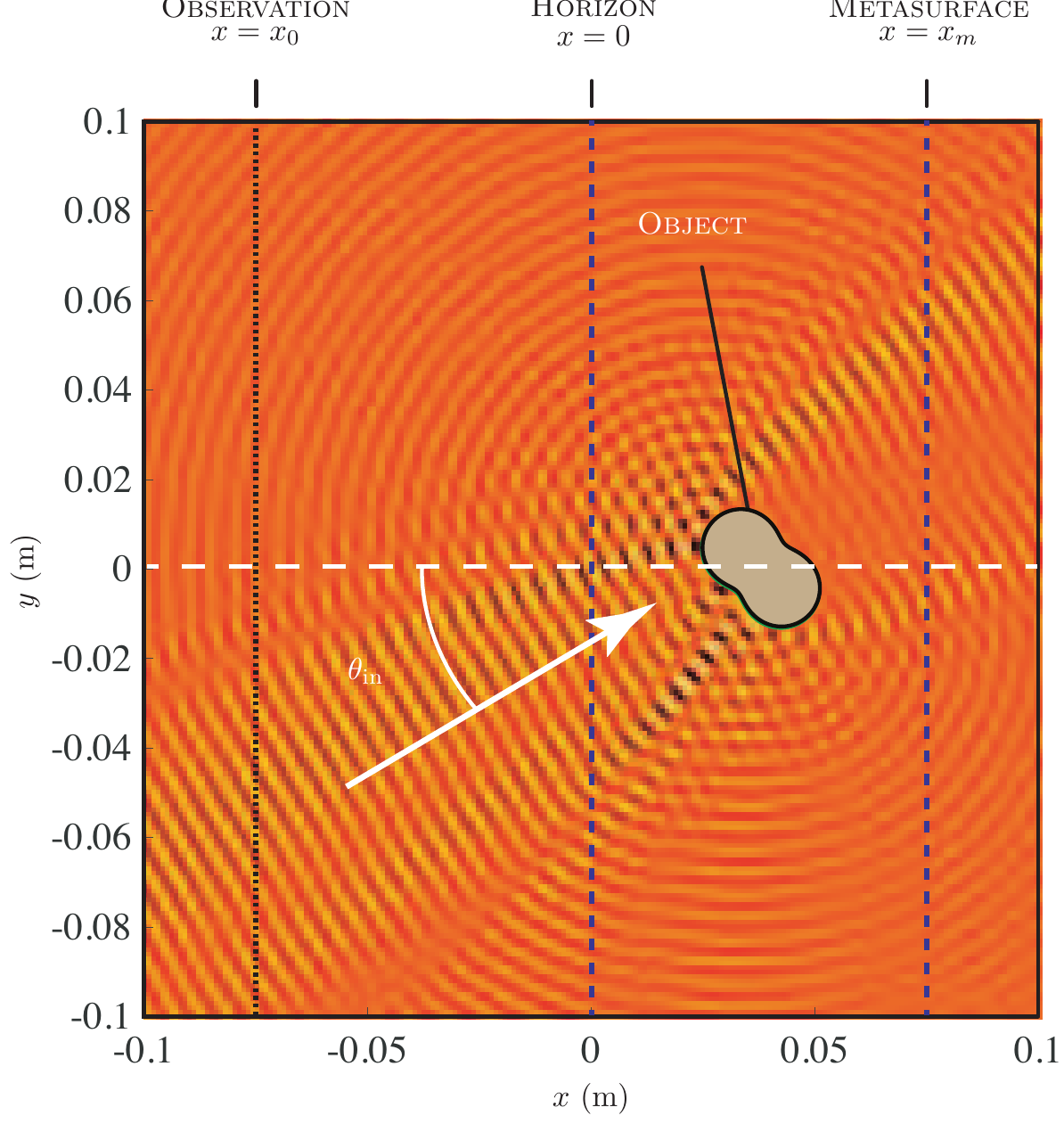}\caption{Re$\{E_{z,t}^\text{ref.}(x,y)\}$ - Object only}
     \end{subfigure}
     \begin{subfigure}[b]{0.3\textwidth}
         \centering
         \psfrag{a}[c][c][0.7]{$y$~(m)}
         \psfrag{b}[c][c][0.7]{$x$~(m)}
         \psfrag{I}[c][c][0.6]{\shortstack{\textsc{Metasurface}\\ $x=x_m$}}
         \psfrag{O}[c][c][0.6]{\shortstack{\textsc{Observation}\\ $x=x_0$}}
         \psfrag{C}[c][c][0.6]{\shortstack{\textsc{Horizon}\\ $x=0$}}
         \psfrag{R}[c][c][0.6]{\color{white}\textsc{\shortstack{Object}}}   
         \psfrag{T}[c][c][0.7]{\color{white}$\theta_\text{in}$}
	\includegraphics[width=1\columnwidth]{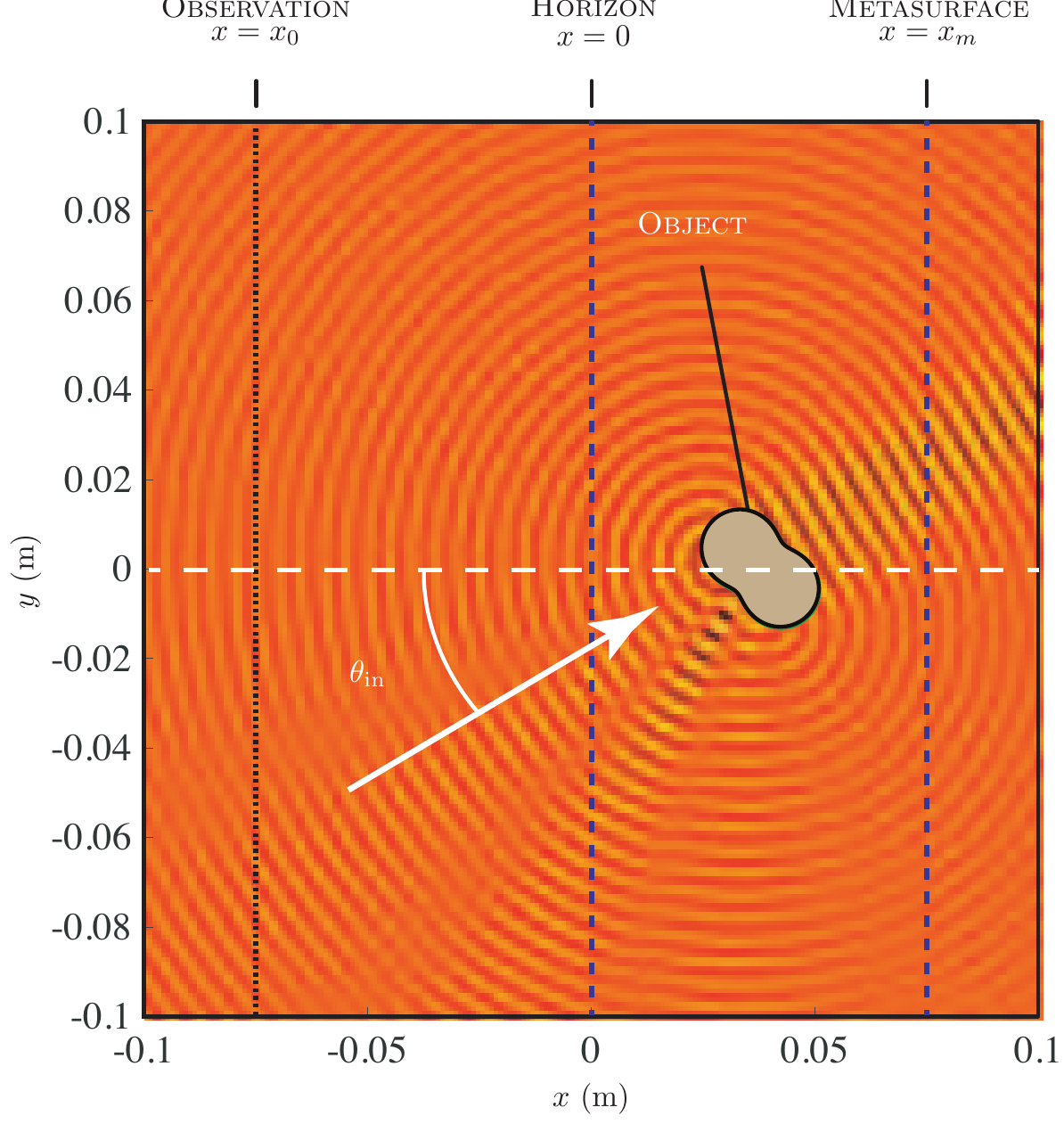}\caption{Re$\{E_{z,s}^\text{ref.}(x,y)\}$ - Object only}
     \end{subfigure}
     \begin{subfigure}[b]{0.3\textwidth}
         \centering
         \psfrag{a}[c][c][0.7]{$y$~(m)}
         \psfrag{b}[c][c][0.7]{$x$~(m)}
         \psfrag{I}[c][c][0.6]{\shortstack{\textsc{Metasurface}\\ $x=x_m$}}
         \psfrag{O}[c][c][0.6]{\shortstack{\textsc{Observation}\\ $x=x_0$}}
         \psfrag{C}[c][c][0.6]{\shortstack{\textsc{Horizon}\\ $x=0$}}
         \psfrag{R}[c][c][0.6]{\color{white}\textsc{\shortstack{Object}}}   
         \psfrag{B}[c][c][0.8]{\color{white}\textsc{\shortstack{Back-propagated}}}  
	\includegraphics[width=1\columnwidth]{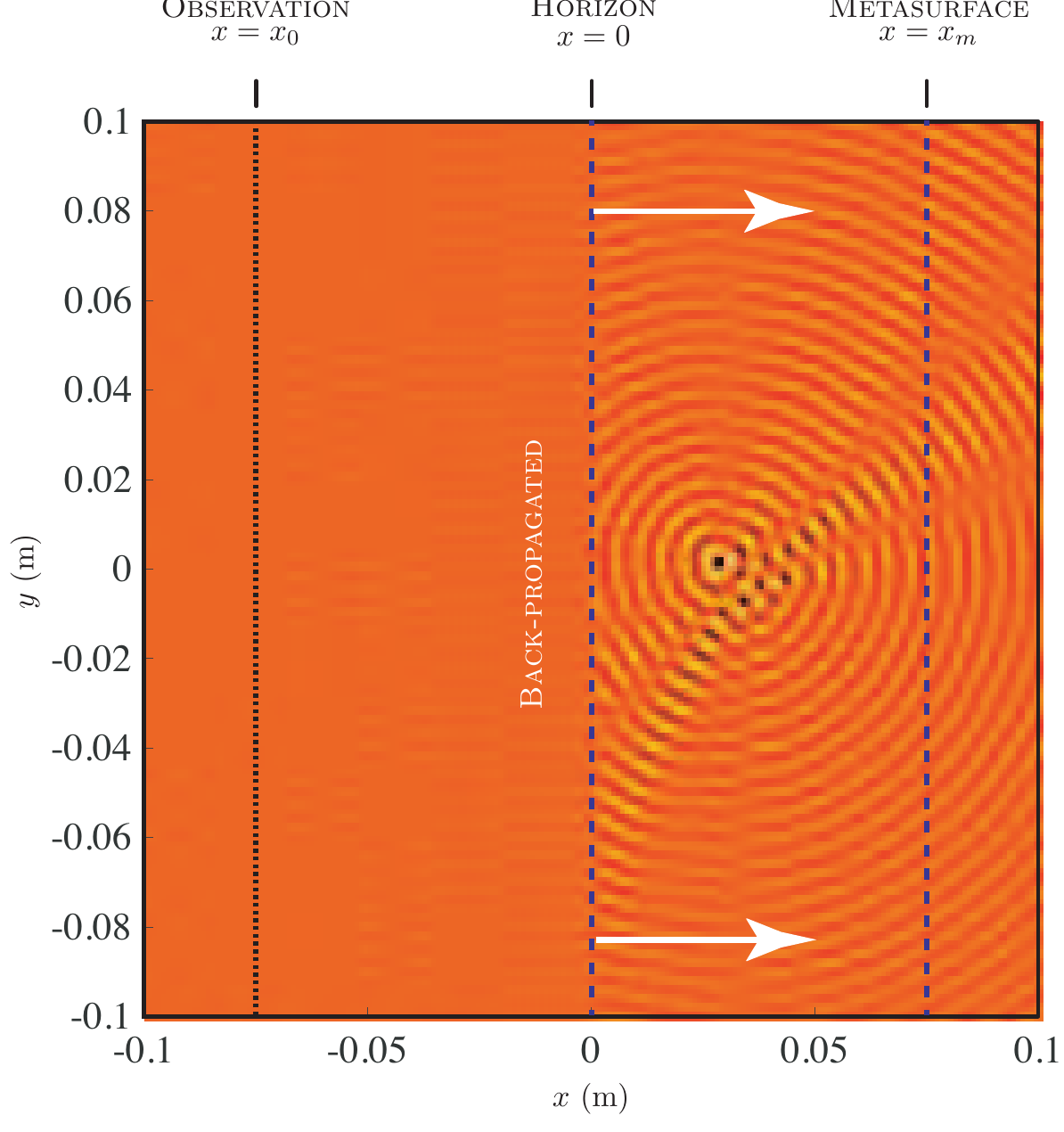}\caption{Re$\{E_{z,s}^\text{bck}(x > 0,y)\}$ - Reverse-propagation}
     \end{subfigure}
     \par\bigskip
     \begin{subfigure}[b]{0.3\textwidth}
         \centering
         \psfrag{a}[c][c][0.7]{$y$~(m)}
         \psfrag{e}[l][c][0.6]{Re\{$\chi$\}}
         \psfrag{d}[l][c][0.6]{Im\{$\chi$\}}
         \psfrag{b}[c][c][0.7]{$\chi_\text{mm}(y)$}
         \psfrag{c}[c][c][0.7]{$\chi_\text{ee}(y)$}
	\includegraphics[width=1\columnwidth]{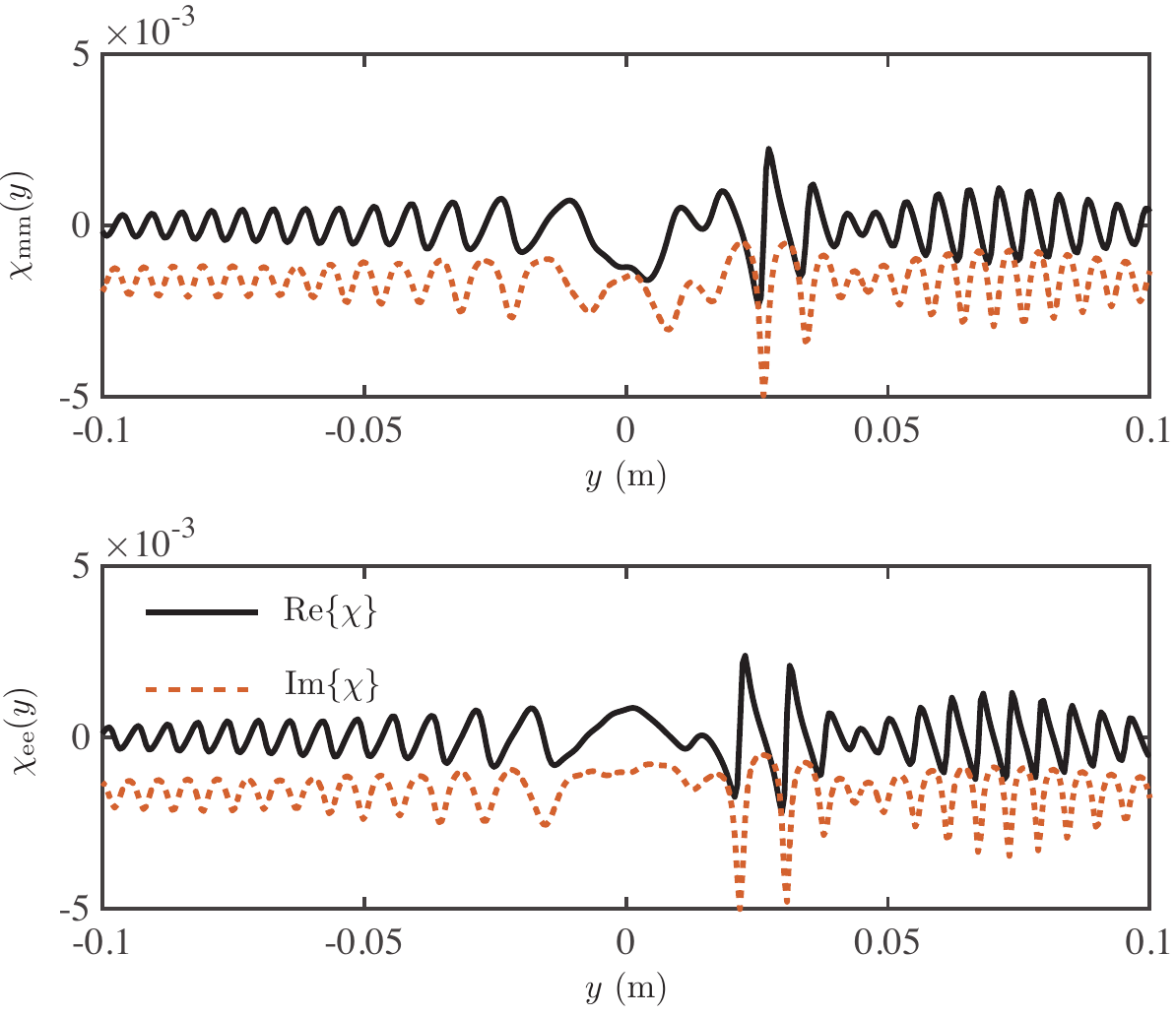}\caption{Synthesized Surface Susceptibilities}
     \end{subfigure}
     \begin{subfigure}[b]{0.3\textwidth}
         \centering
         \psfrag{a}[c][c][0.7]{$y$~(m)}
         \psfrag{b}[c][c][0.7]{$x$~(m)}
         \psfrag{I}[c][c][0.6]{\shortstack{\textsc{Metasurface}\\ $x=x_m$}}
         \psfrag{O}[c][c][0.6]{\shortstack{\textsc{Observation}\\ $x=x_0$}}
         \psfrag{C}[c][c][0.6]{\shortstack{\textsc{Horizon}\\ $x=0$}}
         \psfrag{R}[c][c][0.6]{\color{white}\textsc{\shortstack{Virtual \\Object}}}  
         \psfrag{M}[c][c][0.6]{\color{white}\textsc{Metasurface}}   
         \psfrag{T}[c][c][0.7]{\color{white}$\theta_\text{in}$}       
	\includegraphics[width=1\columnwidth]{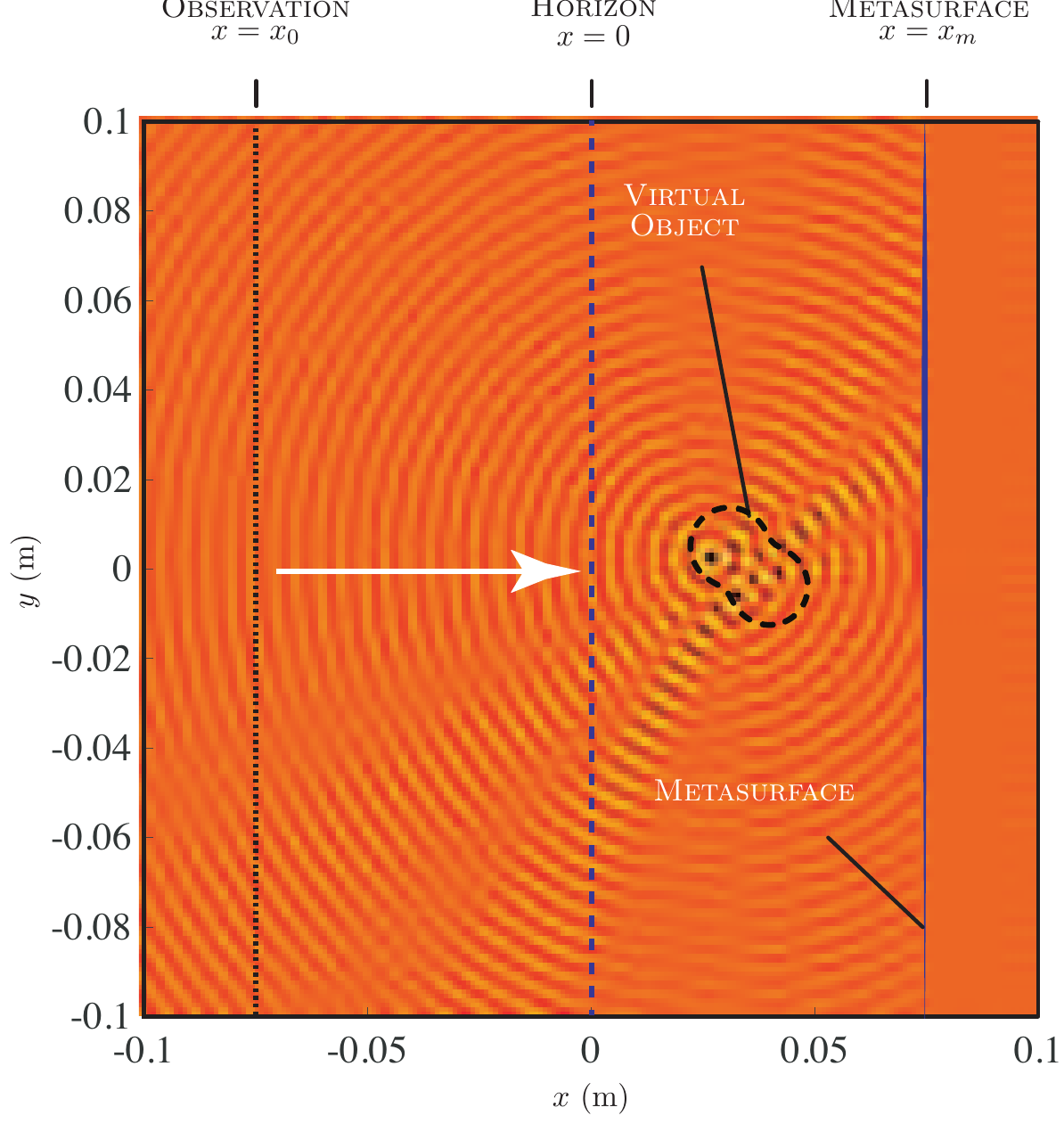}\caption{Re$\{E_{z,s}^\text{ms.}(x,y)\}$ - Metasurface only}
     \end{subfigure}
     \begin{subfigure}[b]{0.3\textwidth}
         \centering
         \psfrag{a}[c][c][0.7]{$y$~(m)}
         \psfrag{d}[c][c][0.6]{$|E_z(0, y)|$~(norm.)}   
	\psfrag{e}[c][c][0.6]{$|E_z(x_0, y)|$~(norm.)}     
	\psfrag{f}[l][c][0.6]{$|E_{z,s}^\text{ref.}(x_0,y)|$}    
	\psfrag{g}[l][c][0.6]{$|E_{z,s}^\text{ms.}(x_0,y)|$}   
	\psfrag{m}[l][c][0.6]{Re\{$E_{z,s}^\text{bck.}(x_m,y)$\}, Re\{$E_{z,s}^\text{ms.}(x_{m-},y)$\}}  
	\psfrag{n}[l][c][0.6]{Im\{$E_{z,s}^\text{bck.}(x_m,y)$\}, Im\{$E_{z,s}^\text{ms.}(x_{m-},y)$\}}   
	\includegraphics[width=1\columnwidth]{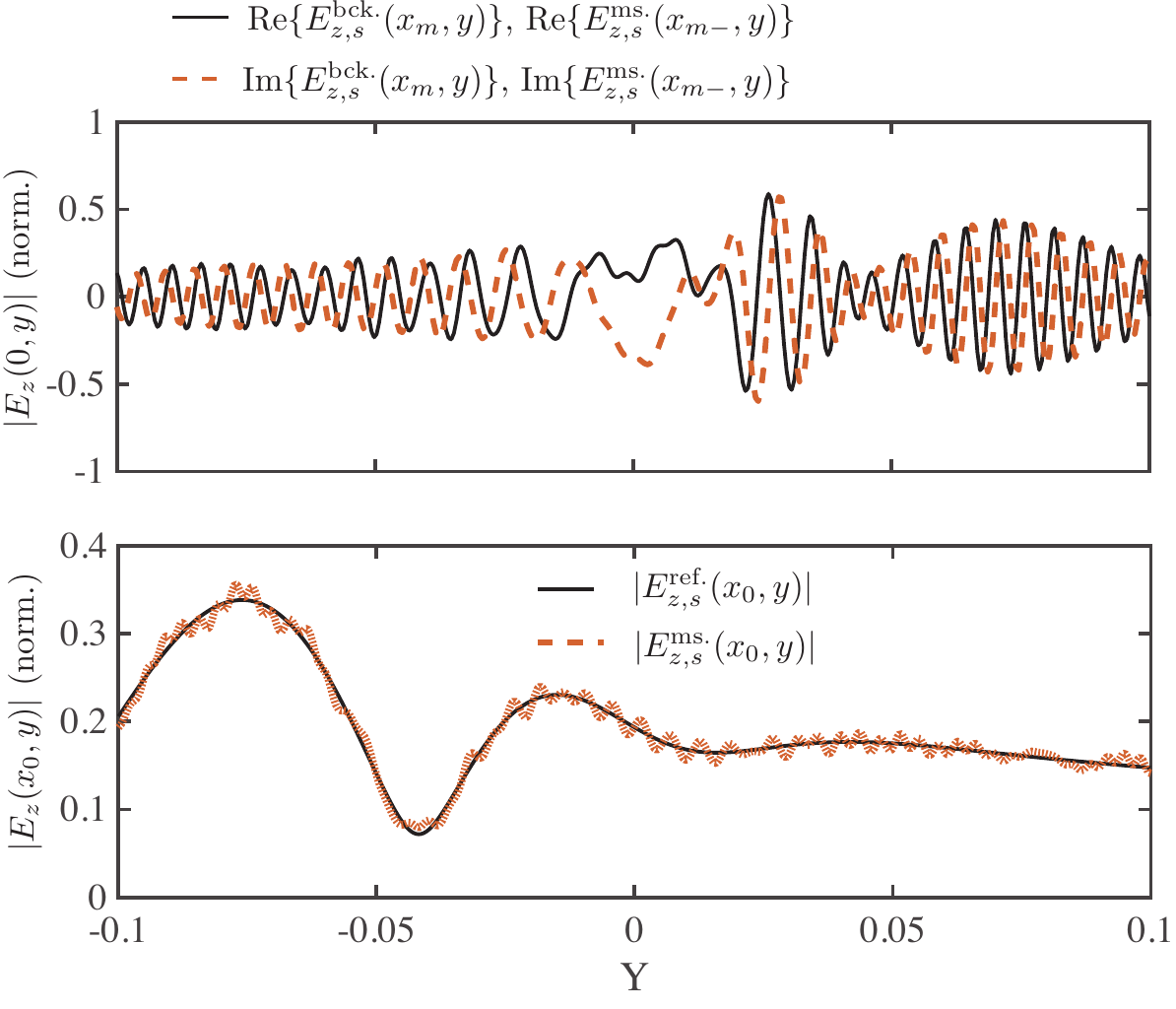}\caption{Reconstructed Scattered Fields}
     \end{subfigure}               
\caption{Front-lit anterior illusion with front illumination using the reverse-propagation technique, with a Gaussian incident field. The simulation parameters are: Parametrized PEC object centered at $x = 7.5\lambda$, $x_h=0,~x_m= 15\lambda$, incident field is a Gaussian-wave at $30^\circ$ from $x-$axis, with $|E_z| = 1.0$ and width of $6\lambda$, illumination is a uniform plane-wave propagating along $+x$-axis with $|E_z|=1.0$ and $\ell_\text{ms} = 120\lambda$.
}\label{Fig:FLAFI}
\end{center}
\end{figure*}

\begin{figure*}[htbp]
\begin{center}
     \begin{subfigure}[b]{0.3\textwidth}
         \centering
         \psfrag{a}[c][c][0.7]{$y$~(m)}
         \psfrag{e}[l][c][0.6]{Re\{$\chi$\}}
         \psfrag{d}[l][c][0.6]{Im\{$\chi$\}}
         \psfrag{b}[c][c][0.7]{$\chi_\text{mm}(y)$}
         \psfrag{c}[c][c][0.7]{$\chi_\text{ee}(y)$}
	\includegraphics[width=\columnwidth]{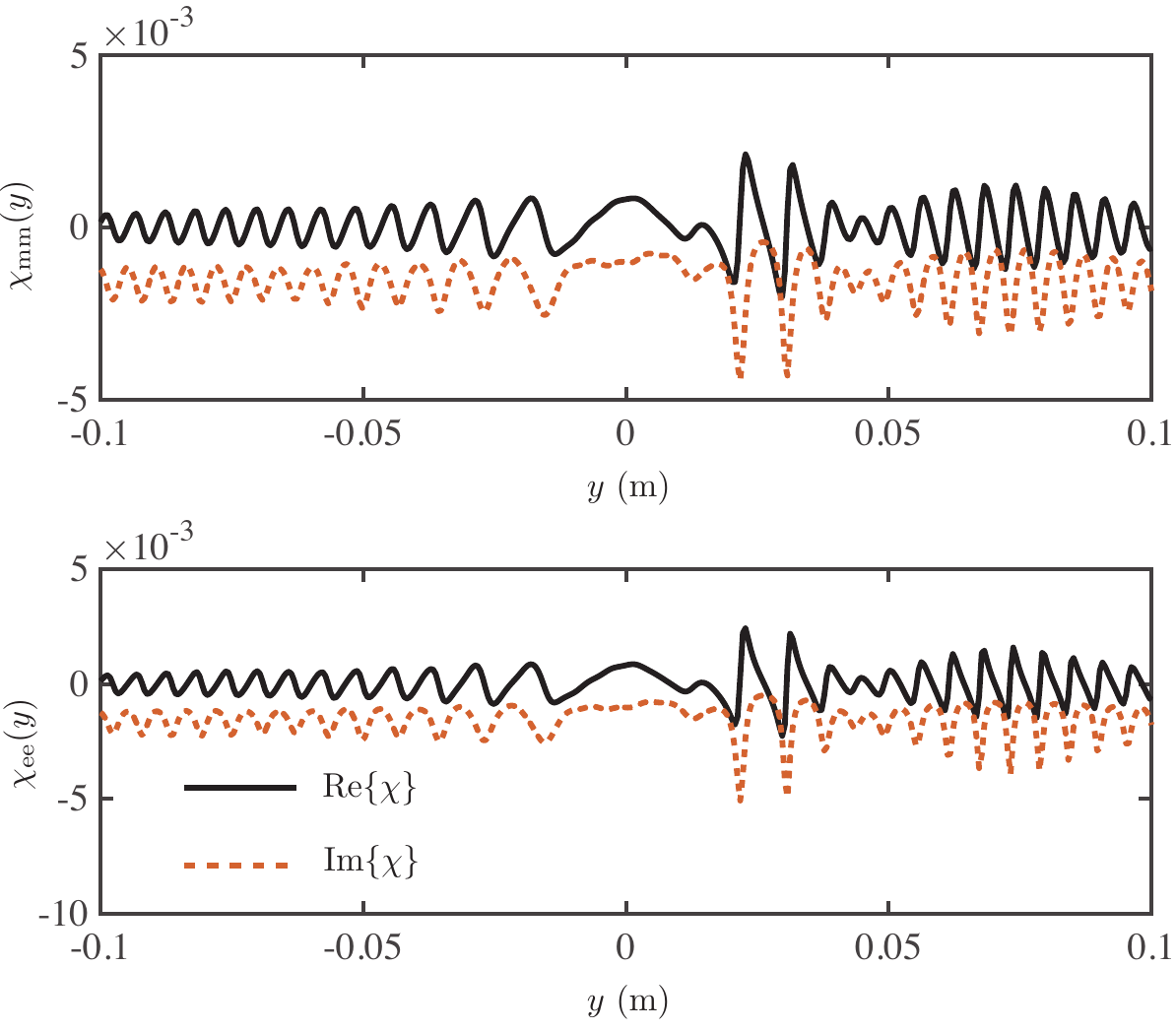}\caption{Synthesized Surface Susceptibilities}       
     \end{subfigure}
     \begin{subfigure}[b]{0.3\textwidth}
         \centering
	 \psfrag{a}[c][c][0.7]{$y$~(m)}
         \psfrag{b}[c][c][0.7]{$x$~(m)}
         \psfrag{I}[c][c][0.6]{\shortstack{\textsc{Metasurface}\\ $x=x_m$}}
         \psfrag{O}[c][c][0.6]{\shortstack{\textsc{Observation}\\ $x=x_0$}}
         \psfrag{C}[c][c][0.6]{\shortstack{\textsc{Horizon}\\ $x=0$}}
         \psfrag{R}[c][c][0.6]{\color{white}\textsc{\shortstack{Virtual \\Object}}}  
         \psfrag{M}[c][c][0.6]{\color{white}\textsc{Metasurface}}   
        	\includegraphics[width=\columnwidth]{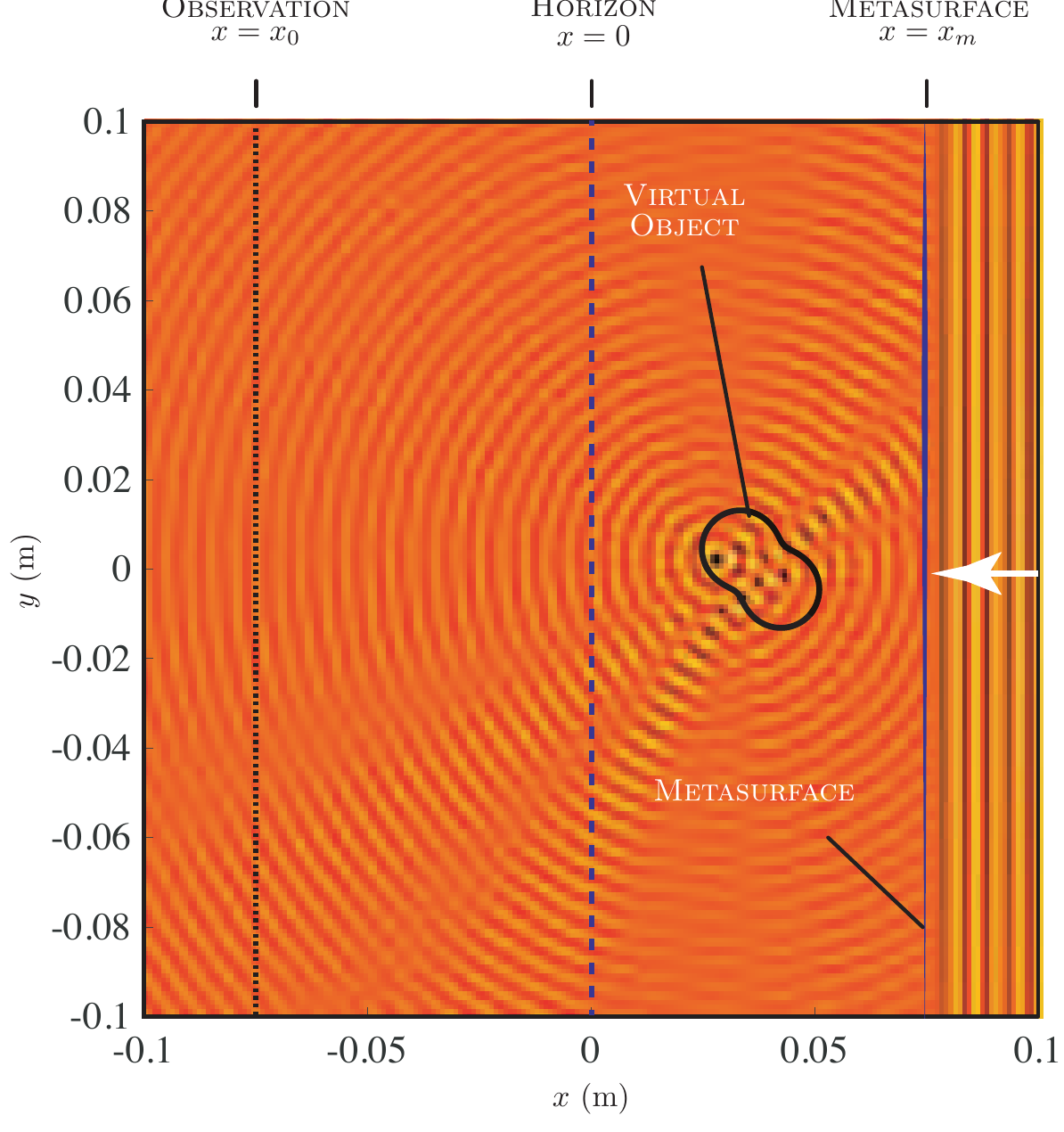}\caption{Re$\{E_{z,t}^\text{ms.}(x,y)\}$ - Metasurface only}       
     \end{subfigure}
     %
%     \begin{subfigure}[b]{0.23\textwidth}
%         \centering
%         \psfrag{a}[c][c][0.7]{$\J,~\K$}
%	\includegraphics[width=0.9\columnwidth]{Results/FrontLitAnterior/AnteriorPECFrontLitBackIll_MIll_1_wgIll_0.19986/Scattered_Ez_synth.eps}\caption{$E_z$ Scattered fields - MS}
%         
%     \end{subfigure}
%
     \begin{subfigure}[b]{0.3\textwidth}
         \centering
         \psfrag{a}[c][c][0.7]{$y$~(m)}
         \psfrag{d}[c][c][0.6]{$|E_z(0, y)|$~(norm.)}   
	\psfrag{e}[c][c][0.6]{$|E_z(x_0, y)|$~(norm.)}     
	\psfrag{f}[l][c][0.6]{$|E_{z,s}^\text{ref.}(x_0,y)|$}    
	\psfrag{g}[l][c][0.6]{$|E_{z,s}^\text{ms.}(x_0,y)|$}   
	\psfrag{m}[l][c][0.6]{Re\{$E_{z,s}^\text{bck.}(x_m,y)$\}, Re\{$E_{z,s}^\text{ms.}(x_{m-},y)$\}}  
	\psfrag{n}[l][c][0.6]{Im\{$E_{z,s}^\text{bck.}(x_m,y)$\}, Im\{$E_{z,s}^\text{ms.}(x_{m-},y)$\}}   
	\includegraphics[width=1\columnwidth]{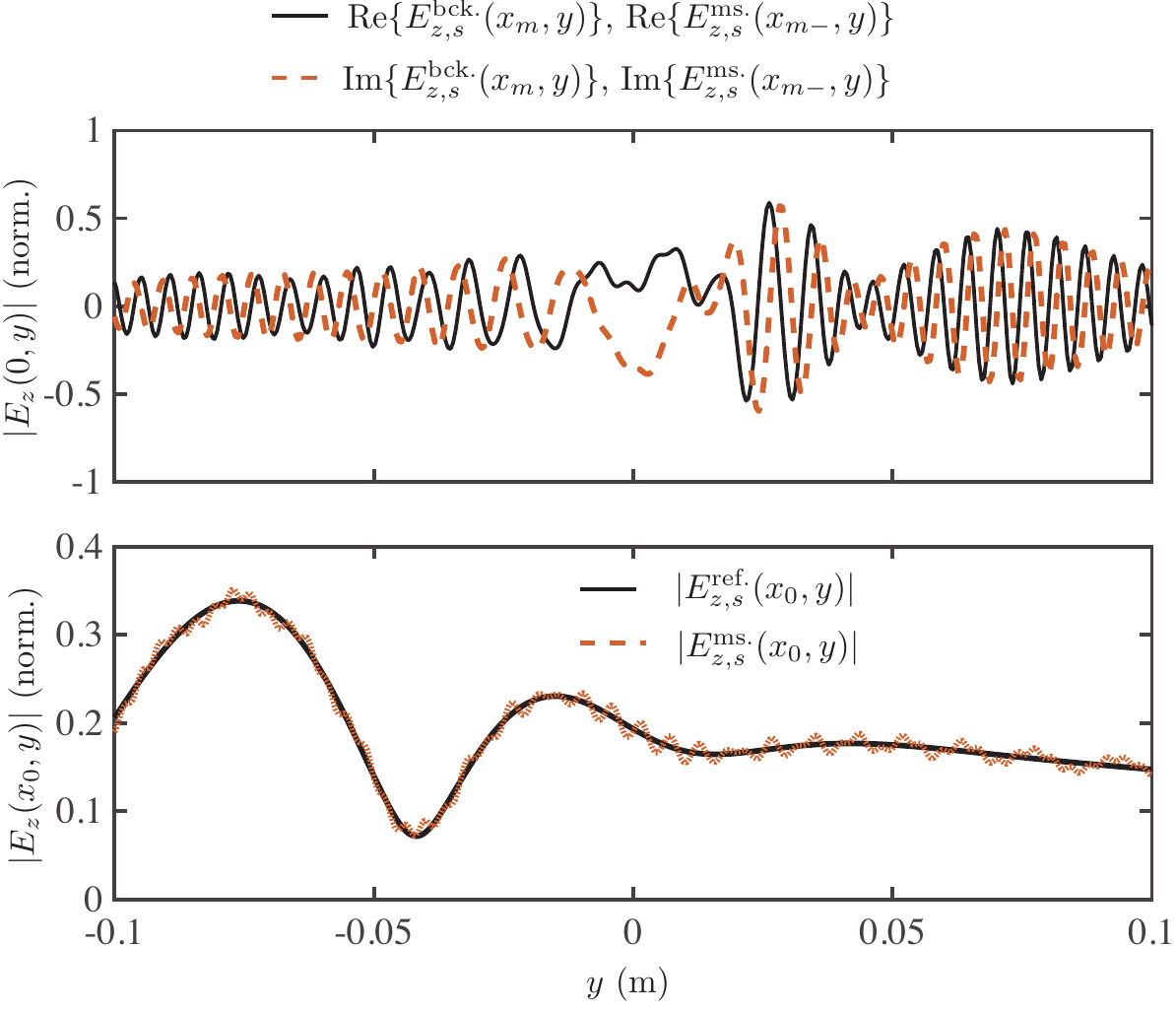}\caption{Reconstructed Scattered Fields}        
     \end{subfigure}     
 \caption{Front-lit anterior illusion with back illumination using the reverse-propagation technique, where the illumination is a Gaussian wave from the right half-space and different from the incident field. The simulation parameters are: Parametrized PEC object centered at $x = 7.5\lambda$, incident field is a Gaussian-wave at $30^\circ$ from $x-$axis, with $|E_z| = 1.0$ and width of $6\lambda$, illumination field is a Gaussian-wave at $0^\circ$ from $x-$axis, with $|E_z| = 1.0$ and width of $40\lambda$.}\label{Fig:FLABI}
\end{center}
\end{figure*}

The synthesized surface susceptibilities strongly depend on the illumination fields and the shape of the metasurface. The topography of the metasurface is not limited to planar configurations. For instance, if a curvilinear metasurface is preferred, the scattered fields from the object can also be recreated, however, a different set of surface characteristics are needed. Fig.~\ref{Fig:FLPFI}(g) shows one example of the synthesized susceptibilities with its corresponding scattered fields shown in Fig.~\ref{Fig:FLPFI}(h). As before, this curvilinear metasurface recreates the desired scattered fields perfectly everywhere to the left of the horizon as shown in Fig.~\ref{Fig:FLPFI}(i). However, this time the surface susceptibilities of Fig.~\ref{Fig:FLPFI}(g) are significantly different from the ones of a planar metasurface of Fig.~\ref{Fig:FLPFI}(c). In particular, the curvilinear metasurface exhibits alternating loss-gain regions, compared to the purely passive susceptibilities of the planar metasurface. This illustrates that the choice of metasurface shape is an important design parameter to be considered in practical hologram designs.

Now let us consider back-illumination of the metasurface, in this case, the metasurface hologram is excited from the right side of the surface -  a uniform plane wave of normal incidence (this is, of course, a case of illumination field being different from the reference field). Fig.~\ref{Fig:FLPBI}(a) shows the synthesized susceptibilities with its corresponding scattered fields shown in Fig.~\ref{Fig:FLPBI}(b). The reference object's scattered fields are perfectly recreated left of the horizon, with a virtual object image formed behind the metasurface. The 1D plot comparison of Fig.~\ref{Fig:FLPBI}(c) confirms the perfect field creation everywhere including that at the observer. Compared to the front illumination, where the metasurface was transforming the illumination fields in reflection, the back illumination requires a transmission-type metasurface, with zero reflection, i.e. a perfectly matched metasurface. Such a metasurface is typically realized using a \emph{Huygens'} source configuration, and is generally difficult to implement compared to reflection-type surfaces.

It should be noted that in all these examples, once the metasurface hologram is synthesized for a given illumination - front or back - the illusion is produced perfectly in the entire region of space left of the horizon, irrespective of the location and extent of the observer. However, we should be aware that the complexity of the metasurface design rests on the spatial variation of the surface susceptibilities, which is controlled by the choice of the reference object, metasurface shape and the illumination conditions, thereby producing virtually an infinite number of configuration possibilities. The hologram designer will need to make judicious choices to achieve practically realizable illusion configurations. For instance, in practice a purely passive metasurface may be desired compared to a loss-gain metasurface which requires an actively powered metasurface, as some of these above results demand.

\subsection{Front-lit Anterior Illusion with Front/Back-illumination}

\begin{figure*}[htbp]
\begin{center}
     \begin{subfigure}[b]{0.3\textwidth}
         \centering
          \psfrag{a}[c][c][0.7]{$y$~(m)}
         \psfrag{b}[c][c][0.7]{$x$~(m)}
         \psfrag{I}[c][c][0.6]{\shortstack{\textsc{Metasurface}\\ $x=x_m$}}
         \psfrag{O}[c][c][0.6]{\shortstack{\textsc{Observation}\\ $x=x_0$}}
         \psfrag{C}[c][c][0.6]{\shortstack{\textsc{Horizon}\\ $x=0$}}
         \psfrag{R}[c][c][0.6]{\color{white}\textsc{\shortstack{Object}}}  
	\psfrag{T}[c][c][0.7]{\color{white}$\theta_\text{in}$}
	\includegraphics[width=\columnwidth]{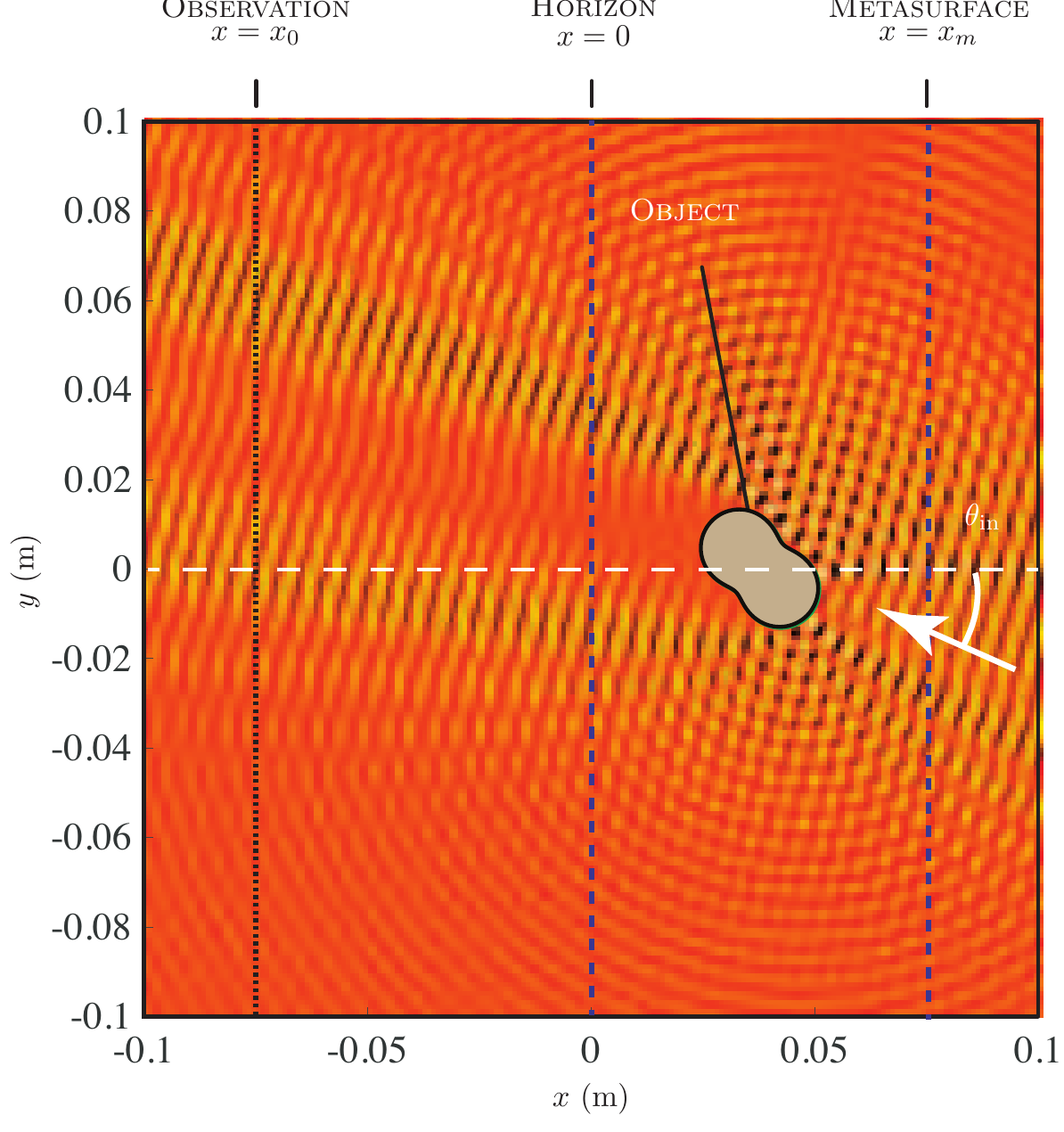}\caption{Re$\{E_{z,t}^\text{ref.}(x,y)\}$ - Object only}
         
     \end{subfigure}
     \begin{subfigure}[b]{0.3\textwidth}
         \centering
          \psfrag{a}[c][c][0.7]{$y$~(m)}
         \psfrag{b}[c][c][0.7]{$x$~(m)}
         \psfrag{I}[c][c][0.6]{\shortstack{\textsc{Metasurface}\\ $x=x_m$}}
         \psfrag{O}[c][c][0.6]{\shortstack{\textsc{Observation}\\ $x=x_0$}}
         \psfrag{C}[c][c][0.6]{\shortstack{\textsc{Horizon}\\ $x=0$}}
         \psfrag{R}[c][c][0.6]{\color{white}\textsc{\shortstack{Object}}}  
         \psfrag{T}[c][c][0.7]{\color{white}$\theta_\text{in}$}
	\includegraphics[width=\columnwidth]{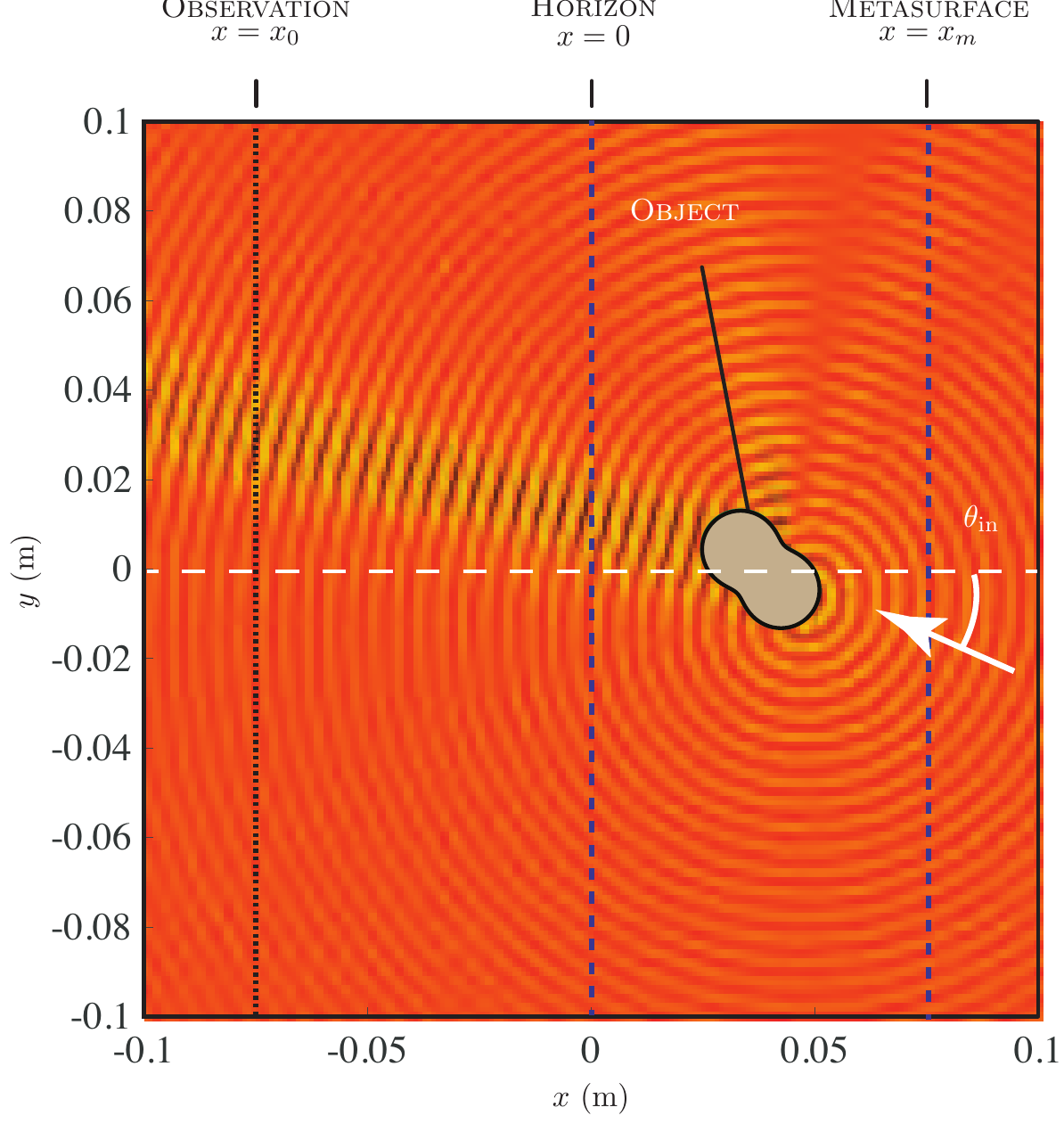}\caption{Re$\{E_{z,s}^\text{ref.}(x,y)\}$ - Object only}
         
     \end{subfigure}
     \begin{subfigure}[b]{0.3\textwidth}
         \centering
         \psfrag{a}[c][c][0.7]{$y$~(m)}
         \psfrag{b}[c][c][0.7]{$x$~(m)}
         \psfrag{I}[c][c][0.6]{\shortstack{\textsc{Metasurface}\\ $x=x_m$}}
         \psfrag{O}[c][c][0.6]{\shortstack{\textsc{Observation}\\ $x=x_0$}}
         \psfrag{C}[c][c][0.6]{\shortstack{\textsc{Horizon}\\ $x=0$}}
         \psfrag{R}[c][c][0.6]{\color{white}\textsc{\shortstack{Object}}}   
         \psfrag{B}[c][c][0.8]{\color{white}\textsc{\shortstack{Back-propagated}}}  
	\includegraphics[width=\columnwidth]{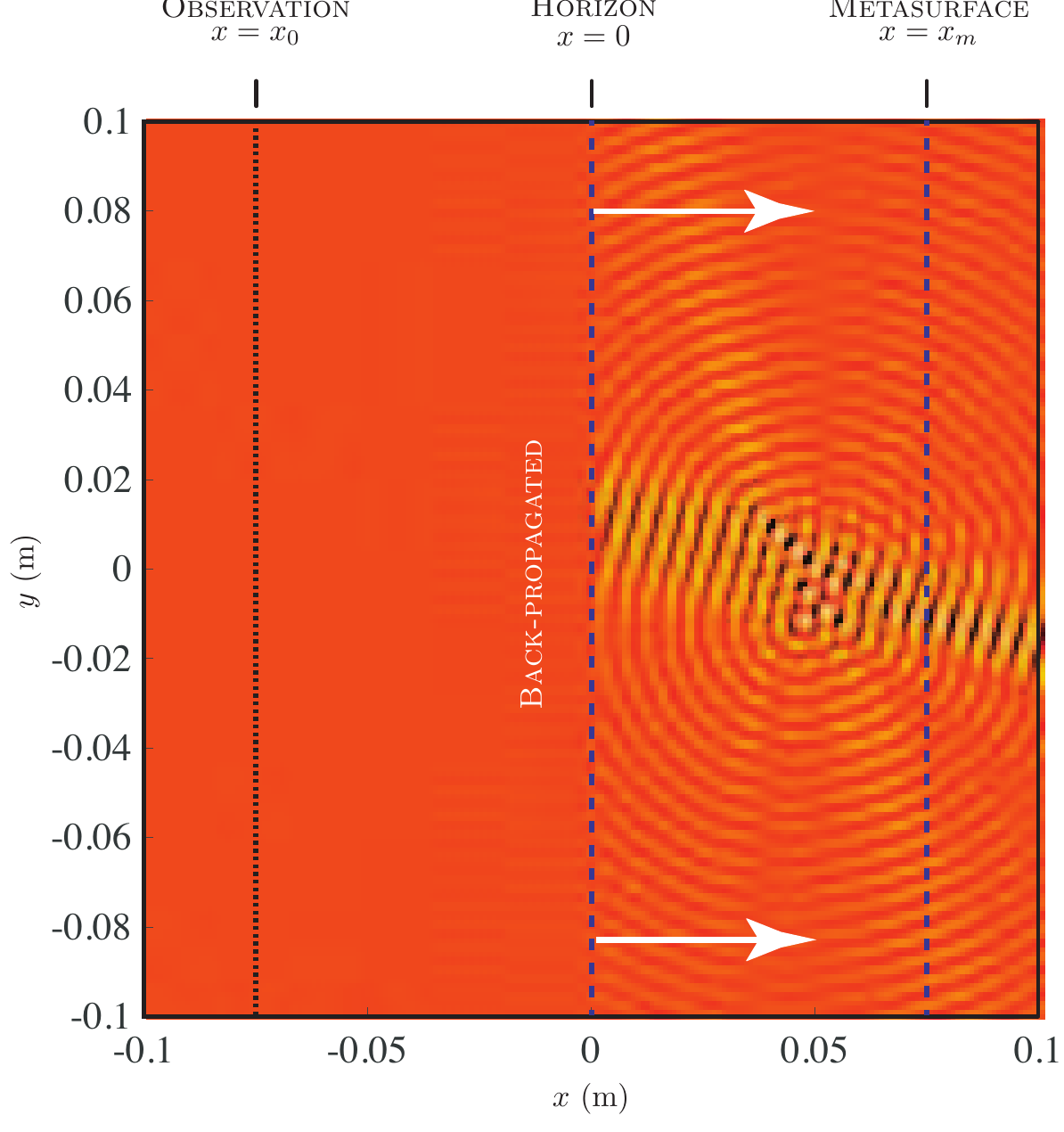}\caption{Re$\{E_{z,s}^\text{bck}(x < 0,y)\}$ - Reverse-propagation}
         
     \end{subfigure}
     \par\bigskip
     \begin{subfigure}[b]{0.3\textwidth}
         \centering
         \psfrag{a}[c][c][0.7]{$y$~(m)}
         \psfrag{e}[l][c][0.6]{Re\{$\chi$\}}
         \psfrag{d}[l][c][0.6]{Im\{$\chi$\}}
         \psfrag{b}[c][c][0.7]{$\chi_\text{mm}(y)$}
         \psfrag{c}[c][c][0.7]{$\chi_\text{ee}(y)$}
	\includegraphics[width=1\columnwidth]{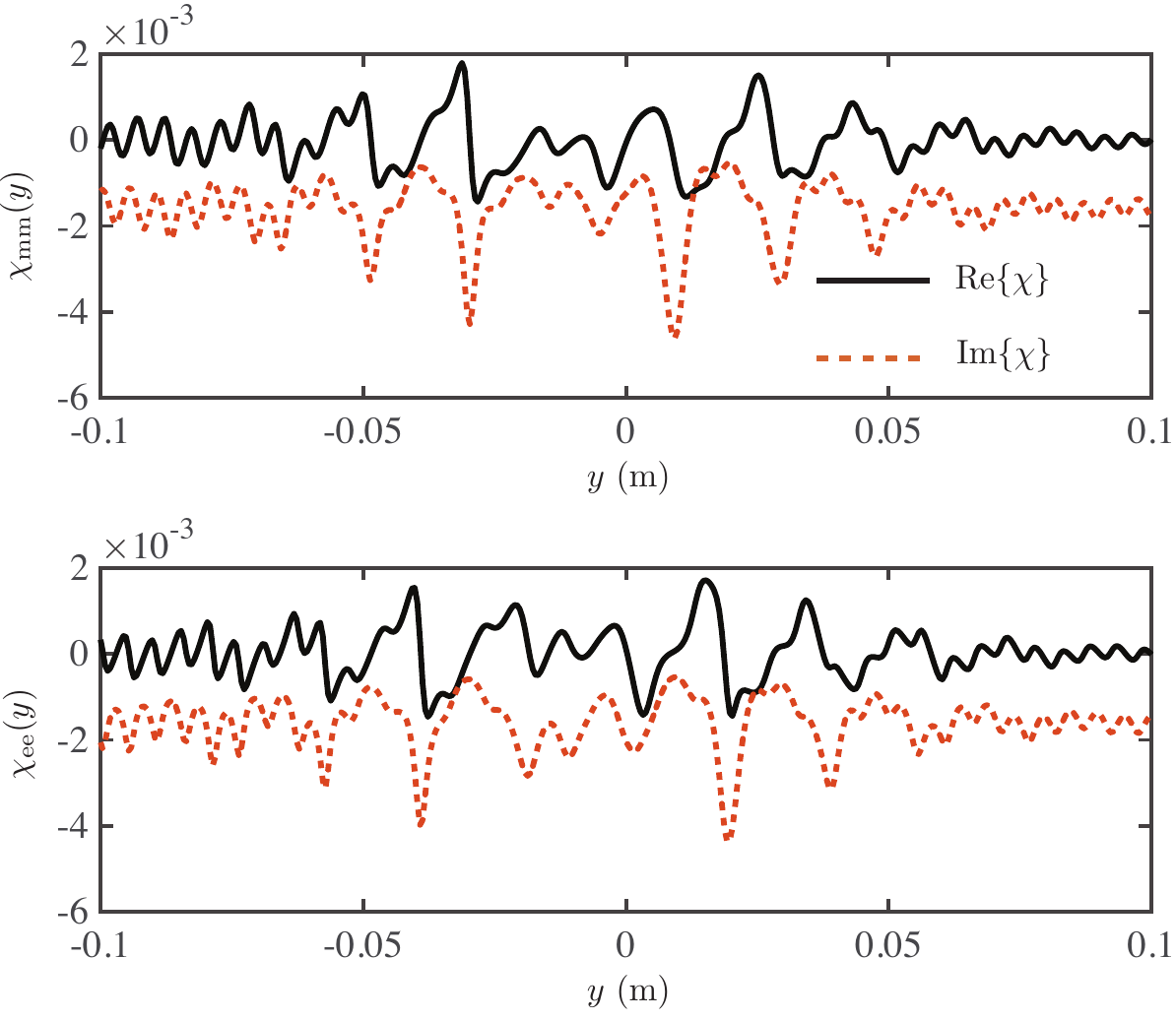}\caption{Synthesized Surface Suceptibilities}
         
     \end{subfigure}
     \begin{subfigure}[b]{0.3\textwidth}
         \centering
	 \psfrag{a}[c][c][0.7]{$y$~(m)}
         \psfrag{b}[c][c][0.7]{$x$~(m)}
         \psfrag{I}[c][c][0.6]{\shortstack{\textsc{Metasurface}\\ $x=x_m$}}
         \psfrag{O}[c][c][0.6]{\shortstack{\textsc{Observation}\\ $x=x_0$}}
         \psfrag{C}[c][c][0.6]{\shortstack{\textsc{Horizon}\\ $x=0$}}
         \psfrag{R}[c][c][0.6]{\color{white}\textsc{\shortstack{Virtual \\Object}}}  
         \psfrag{M}[c][c][0.6]{\color{white}\textsc{Metasurface}}   
         \psfrag{T}[c][c][0.7]{\color{white}$\theta_\text{in}$}
	\includegraphics[width=\columnwidth]{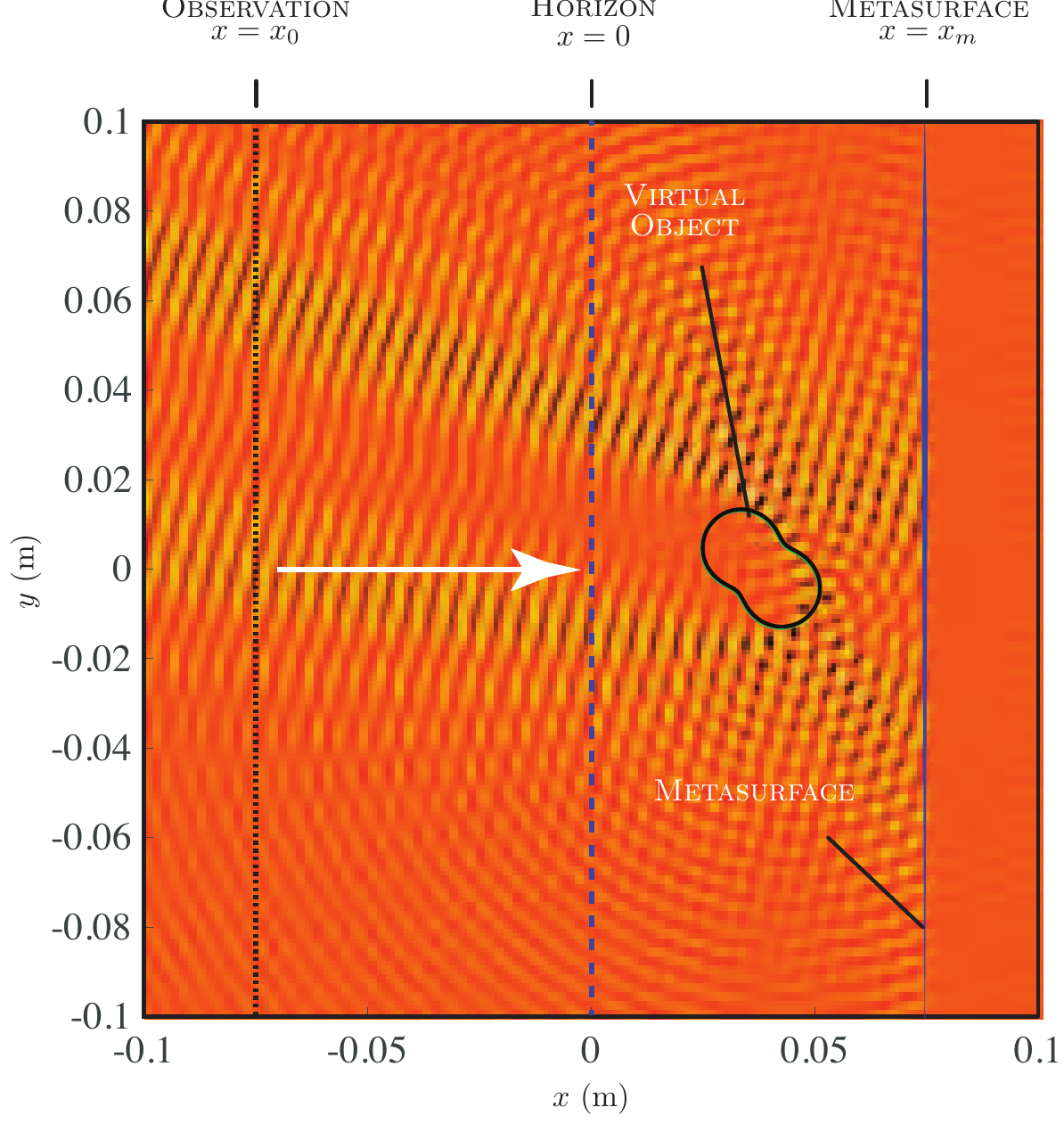}\caption{Re$\{E_{z,s}^\text{ms.}(x,y)\}$ - Metasurface Only}
         
     \end{subfigure}
     \begin{subfigure}[b]{0.3\textwidth}
         \centering
         \psfrag{a}[c][c][0.7]{$y$~(m)}
         \psfrag{d}[c][c][0.6]{$|E_z(0, y)|$~(norm.)}   
	\psfrag{e}[c][c][0.6]{$|E_z(x_0, y)|$~(norm.)}     
	\psfrag{f}[l][c][0.6]{$|E_{z,s}^\text{ref.}(x_0,y)|$}    
	\psfrag{g}[l][c][0.6]{$|E_{z,s}^\text{ms.}(x_0,y)|$}   
	\psfrag{m}[l][c][0.6]{Re\{$E_{z,s}^\text{bck.}(x_m,y)$\}, Re\{$E_{z,s}^\text{ms.}(x_{m-},y)$\}}  
	\psfrag{n}[l][c][0.6]{Im\{$E_{z,s}^\text{bck.}(x_m,y)$\}, Im\{$E_{z,s}^\text{ms.}(x_{m-},y)$\}}   
	\includegraphics[width=1\columnwidth]{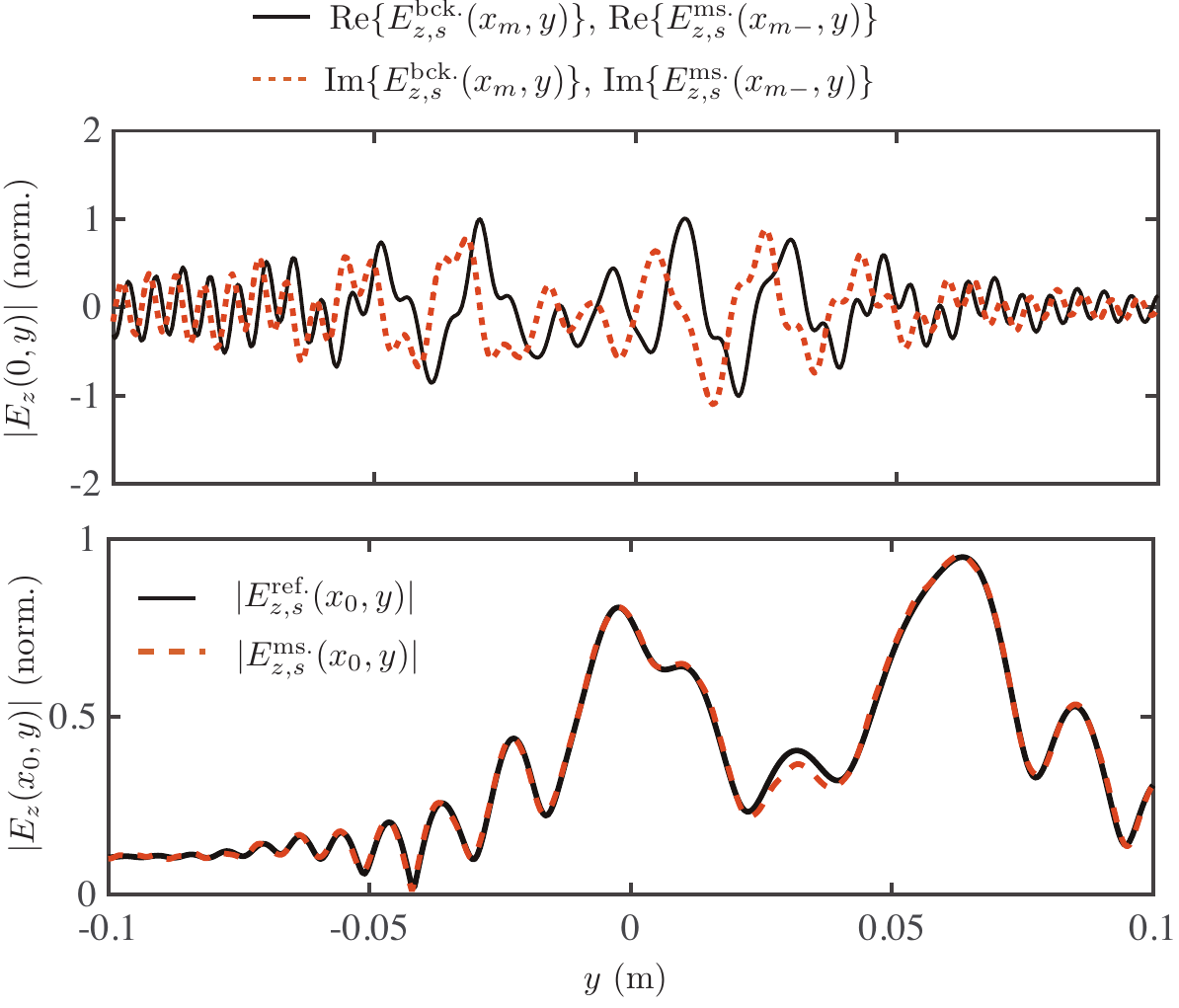}\caption{Reconstructed Scattered Fields}
         
     \end{subfigure}               
\caption{Back-lit anterior illusion with front illumination using the reverse-propagation technique, where the illumination is a Gaussian-wave from the left half-space. The simulation parameters are: Parametrized PEC object centered at $x  = 7.5\lambda$, incident Gaussian-wave at $-15^\circ$ from $x-$axis, with $|E_z| = 1.0$ and width of $10\lambda$ propagating along $-x$, illuminating Gaussian-wave at $0^\circ$ from $x-$axis, with $|E_z| = 1.0$ and width of $40\lambda$ propagating along $+x$.}\label{Fig:BLAFI}
\end{center}
\end{figure*}

We shall now deal with the case of an Anterior illusion, where the metasurface hologram is placed behind the object at $x_m$. Fig,~\ref{Fig:FLAFI}(a-b) shows the total and scattered fields generated by a parameterized curvi-linear reference PEC object, which is excited by a Gaussian beam from the bottom left of the object producing a front-lit object. Complex scattered fields are produced along with a shadow behind the object. Since, the object is in between the horizon and the metasurface, the horizon fields at $x_h$ must be mathmatically reverse-propagated to the metasurface location, before the susceptibility synthesis can be performed. 

This reverse-propagation of the horizon fields is shown in Fig.~\ref{Fig:FLAFI}(c). In this step, the object is removed, and following the procedure of Sec.~\ref{Sec:III-B}, the fields are reverse propagated to the desired metasurface location at $x_m$ towards the right. The fields are naturally zero on the left of the horizon. The reverse propagation creates an artificial field distribution which appears to focus the horizon fields before diverging again at the metasurface location. The resulting fields $E^\text{bck.}_{z,s}$ (and other associated fields and components) are now the desired scattered fields that the metasurface hologram must recreate in the absence of the reference object under specified illumination fields.

Next, for specified illumination fields (front illuminated normally incident uniform plane-wave) and the reverse-propagated horizon fields, the metasurface surface susceptibilities are synthesized, as shown in Fig.~\ref{Fig:FLAFI}(d). The resulting scattered fields are further shown in Fig.~\ref{Fig:FLAFI}(e) along with the fields comparisons at the metasurface and observer location with the reference fields. The scattered fields from the metasurface at the metasurface are perfectly recreated, while an excellent reconstruction of the fields at the observer location is seen with slight ripples.

A similar field reconstruction is observed when the metasurface is re-synthesized for back illumination, where a uniform plane-wave is normally incident from the right, as shown in Fig.~\ref{Fig:FLABI}. As expected, for a matched metasurface, the electric and magnetic surface susceptibilities are balanced emulating a Huygens' source configuration. Moreover, in both the front and back illumination case, the synthesized susceptibilities are found to be purely lossy. Finally, in both cases, the virtual image of the object is formed in front of the metasurface. It should be noted that, while the horizon location is arbitrary, it is limited to the left-most extent of the object, beyond which the illusion is perfectly created.

\begin{figure*}[htbp]
\begin{center}
     \begin{subfigure}[b]{0.3\textwidth}
         \centering
         \psfrag{a}[c][c][0.7]{$y$~(m)}
         \psfrag{b}[c][c][0.7]{$x$~(m)}
         \psfrag{I}[c][c][0.6]{\shortstack{\textsc{Metasurface}\\ $x=x_m$}}
         \psfrag{O}[c][c][0.6]{\shortstack{\textsc{Observation}\\ $x=x_0$}}
         \psfrag{C}[c][c][0.6]{\shortstack{\textsc{Horizon}\\ $x=0$}}
         \psfrag{R}[c][c][0.6]{\color{white}\textsc{\shortstack{Virtual \\Object}}}  
         \psfrag{M}[c][c][0.6]{\color{white}\textsc{Metasurface}}   
	\includegraphics[width=\columnwidth]{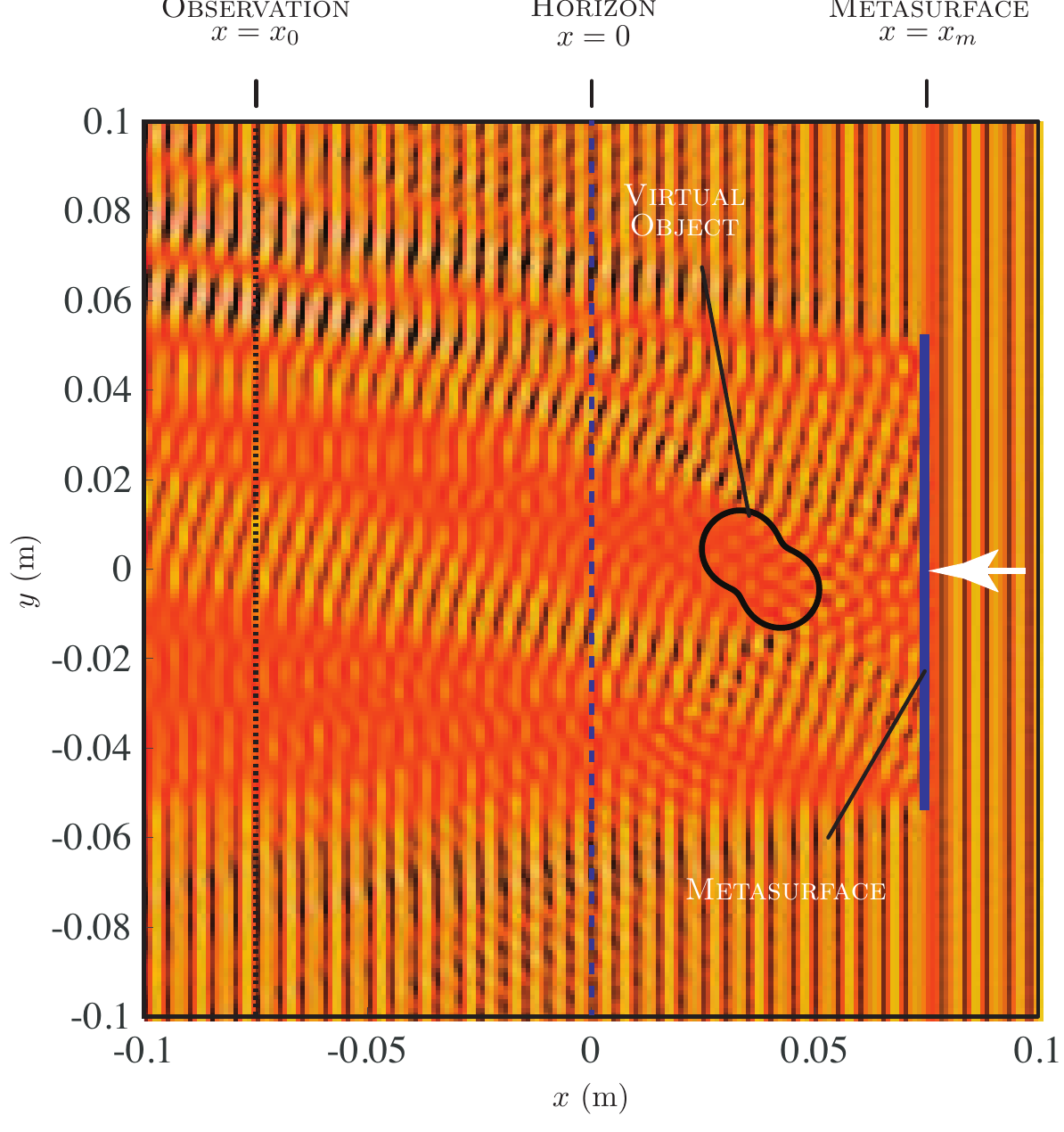}
         
     \end{subfigure}
     \begin{subfigure}[b]{0.3\textwidth}
         \centering
         \psfrag{a}[c][c][0.7]{$y$~(m)}
         \psfrag{b}[c][c][0.7]{$x$~(m)}
         \psfrag{I}[c][c][0.6]{\shortstack{\textsc{Metasurface}\\ $x=x_m$}}
         \psfrag{O}[c][c][0.6]{\shortstack{\textsc{Observation}\\ $x=x_0$}}
         \psfrag{C}[c][c][0.6]{\shortstack{\textsc{Horizon}\\ $x=0$}}
         \psfrag{R}[c][c][0.6]{\color{white}\textsc{\shortstack{Virtual \\Object}}}  
         \psfrag{M}[c][c][0.6]{\color{white}\textsc{Metasurface}}   
	\includegraphics[width=\columnwidth]{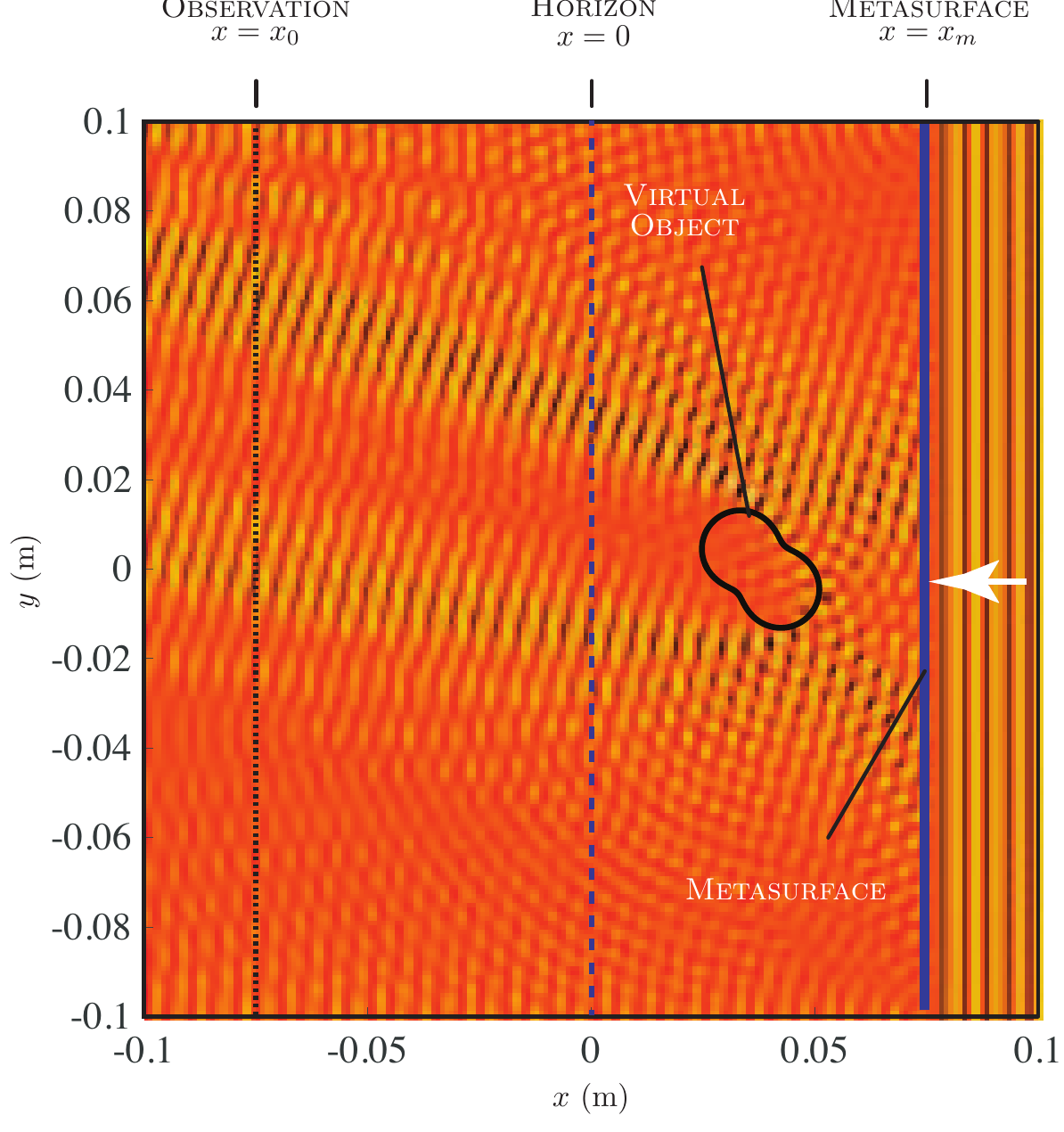}
         
     \end{subfigure}
     \begin{subfigure}[b]{0.3\textwidth}
         \centering
         \psfrag{a}[c][c][0.7]{$y$~(m)}
         \psfrag{b}[c][c][0.7]{$x$~(m)}
         \psfrag{I}[c][c][0.6]{\shortstack{\textsc{Metasurface}\\ $x=x_m$}}
         \psfrag{O}[c][c][0.6]{\shortstack{\textsc{Observation}\\ $x=x_0$}}
         \psfrag{C}[c][c][0.6]{\shortstack{\textsc{Horizon}\\ $x=0$}}
         \psfrag{R}[c][c][0.6]{\color{white}\textsc{\shortstack{Virtual \\Object}}}  
         \psfrag{M}[c][c][0.6]{\color{white}\textsc{Metasurface}}   
	\includegraphics[width=\columnwidth]{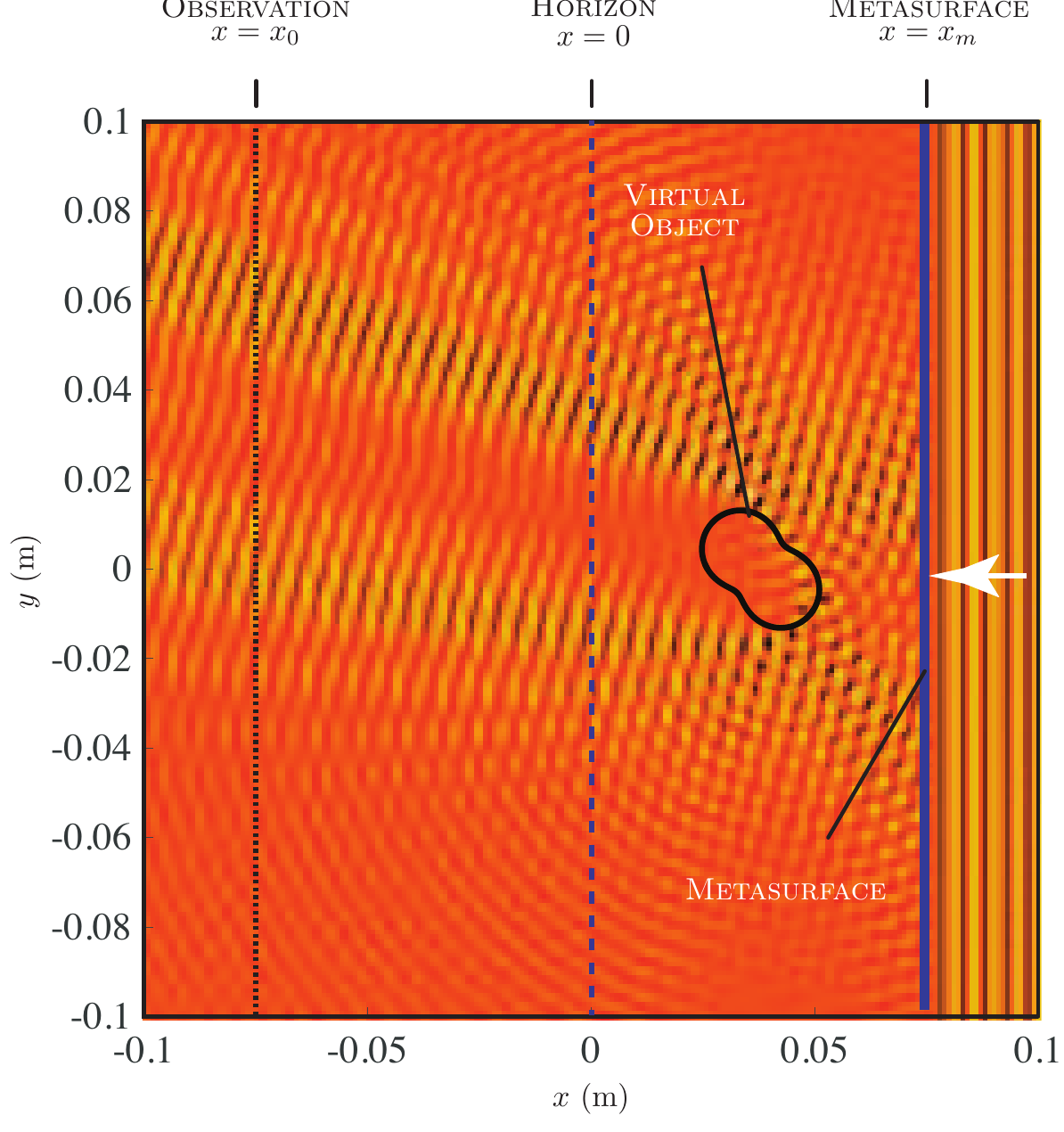}
         
     \end{subfigure}
     \par\bigskip
     \begin{subfigure}[b]{0.3\textwidth}
         \centering
         \psfrag{a}[c][c][0.7]{$y$~(m)}
         \psfrag{d}[c][c][0.6]{$|E_z(0, y)|$~(norm.)}   
	\psfrag{e}[c][c][0.6]{$|E_z(x_0, y)|$~(norm.)}     
	\psfrag{f}[l][c][0.6]{$|E_{z,s}^\text{ref.}(x_0,y)|$}    
	\psfrag{g}[l][c][0.6]{$|E_{z,s}^\text{ms.}(x_0,y)|$}   
	\psfrag{m}[l][c][0.6]{Re\{$E_{z,s}^\text{bck.}(x_m,y)$\}, Re\{$E_{z,s}^\text{ms.}(x_{m-},y)$\}}  
	\psfrag{n}[l][c][0.6]{Im\{$E_{z,s}^\text{bck.}(x_m,y)$\}, Im\{$E_{z,s}^\text{ms.}(x_{m-},y)$\}}   
	\includegraphics[width=\columnwidth]{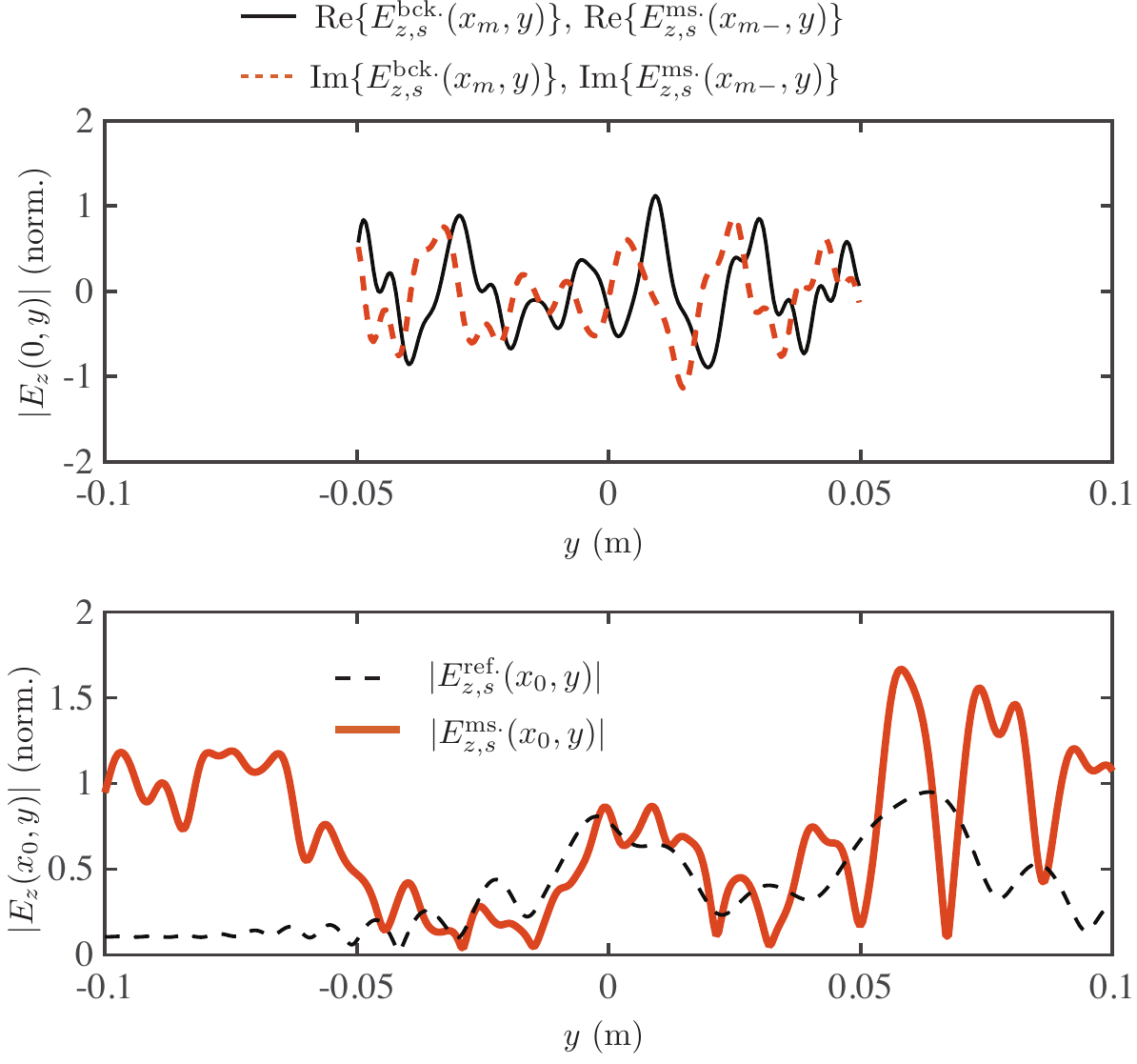}\caption{$\ell_\text{ms} = 20\lambda$}
         
     \end{subfigure}
     \begin{subfigure}[b]{0.3\textwidth}
         \centering
         \psfrag{a}[c][c][0.7]{$y$~(m)}
         \psfrag{d}[c][c][0.6]{$|E_z(0, y)|$~(norm.)}   
	\psfrag{e}[c][c][0.6]{$|E_z(x_0, y)|$~(norm.)}     
	\psfrag{f}[l][c][0.6]{$|E_{z,s}^\text{ref.}(x_0,y)|$}    
	\psfrag{g}[l][c][0.6]{$|E_{z,s}^\text{ms.}(x_0,y)|$}   
	\psfrag{m}[l][c][0.6]{Re\{$E_{z,s}^\text{bck.}(x_m,y)$\}, Re\{$E_{z,s}^\text{ms.}(x_{m-},y)$\}}  
	\psfrag{n}[l][c][0.6]{Im\{$E_{z,s}^\text{bck.}(x_m,y)$\}, Im\{$E_{z,s}^\text{ms.}(x_{m-},y)$\}}   
	\includegraphics[width=\columnwidth]{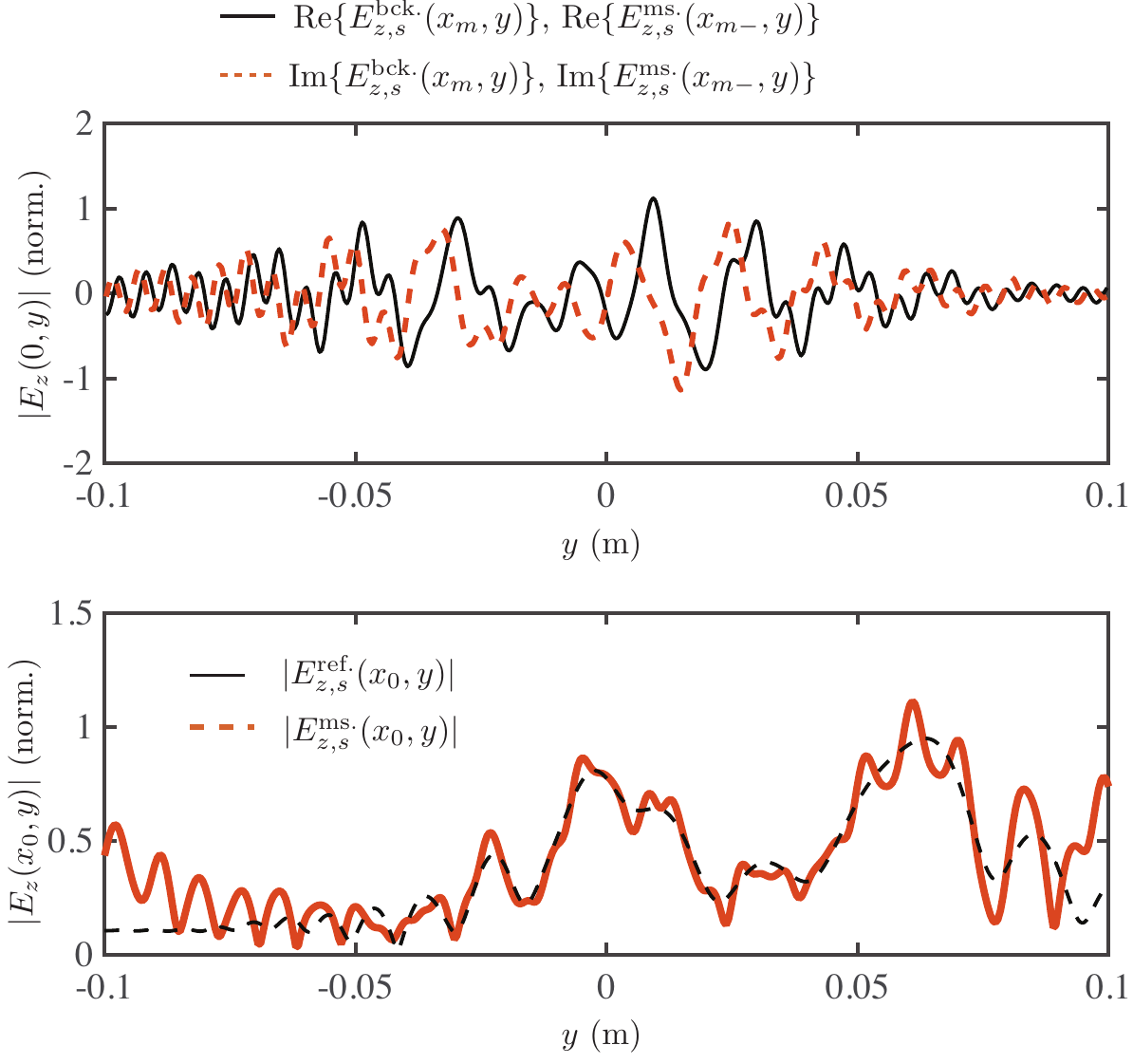}\caption{$\ell_\text{ms} = 40\lambda$}
         
     \end{subfigure}
     \begin{subfigure}[b]{0.3\textwidth}
         \centering
         \psfrag{a}[c][c][0.7]{$y$~(m)}
         \psfrag{d}[c][c][0.6]{$|E_z(0, y)|$~(norm.)}   
	\psfrag{e}[c][c][0.6]{$|E_z(x_0, y)|$~(norm.)}     
	\psfrag{f}[l][c][0.6]{$|E_{z,s}^\text{ref.}(x_0,y)|$}    
	\psfrag{g}[l][c][0.6]{$|E_{z,s}^\text{ms.}(x_0,y)|$}   
	\psfrag{m}[l][c][0.6]{Re\{$E_{z,s}^\text{bck.}(x_m,y)$\}, Re\{$E_{z,s}^\text{ms.}(x_{m-},y)$\}}  
	\psfrag{n}[l][c][0.6]{Im\{$E_{z,s}^\text{bck.}(x_m,y)$\}, Im\{$E_{z,s}^\text{ms.}(x_{m-},y)$\}}   
	\includegraphics[width=\columnwidth]{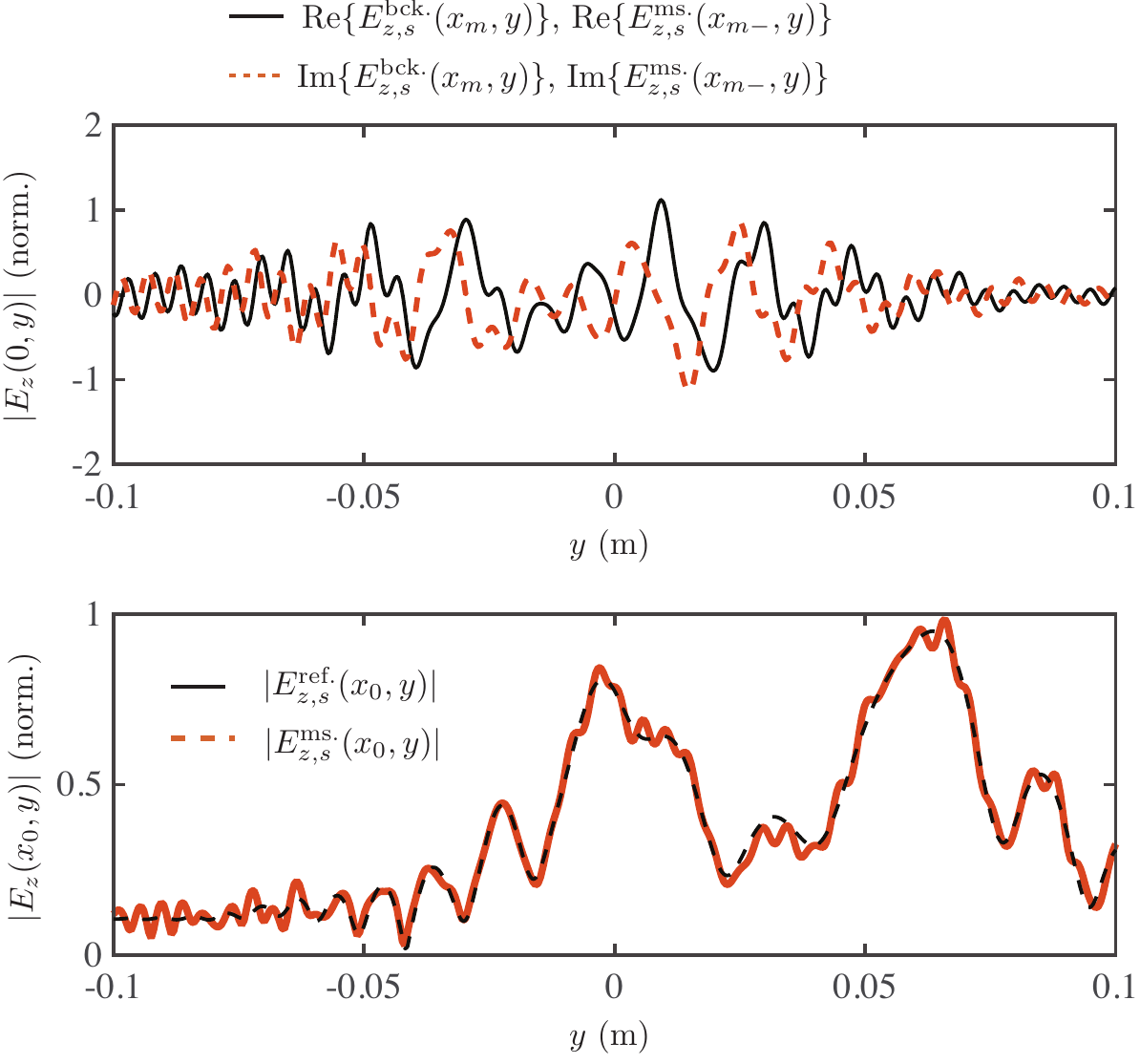}\caption{$\ell_\text{ms} = 80\lambda$}
         
     \end{subfigure}               
\caption{Effect of metasurface length on the fidelity of the reconstructed scattered fields, using the back-lit anterior front illumination case of Fig.~\ref{Fig:BLAFI}. Simulation parameters are: Parametrized PEC object centered at $x = 7.5\lambda$, incident Gaussian-wave at $-15^\circ$ from $x-$axis, with $|E_z| = 1.0$ and width of $10\lambda$ propagating along $-x$, illuminating plane-wave at $0^\circ$ from $x-$axis, with $|E_z| = 1.0$ propagating along $+x$.}\label{Fig:FLPFI_LV}
\end{center}
\end{figure*}

\subsection{Back-lit Anterior Illusion with front-illumination}

The last example of this section is a case of back-lit anterior illusion with front illumination. In this case, the same parametrized object of Fig.~\ref{Fig:FLAFI} and \ref{Fig:FLABI}, is excited with a reference Gaussian beam from the back of the object. The computed total and scattered fields from the object are shown in Fig.~\ref{Fig:BLAFI}(a) and (b). Compared to all the previous front-lit cases, the observer this time measures not only the scattered fields, but also the reference fields. Consequently, the total fields are captured at the horizon. These total fields are then reverse propagated to the desired metasurface location behind the object (anterior illusion), as shown in Fig.~\ref{Fig:BLAFI}(c).

Next for the specified illumination fields - a front illuminating Gaussian beam - the metasurface is synthesized, and the resulting susceptibilities are shown in Fig.~\ref{Fig:BLAFI}(d). The final fields scattered from the metasurface are shown in Fig.~\ref{Fig:BLAFI}(e) along with the field comparisons at the metasurface and the observer location in Fig.~\ref{Fig:BLAFI}(f). In this case, the scattered fields from the metasurface must be compared to the total fields of the reference object, i.e. Fig.~\ref{Fig:BLAFI}(a), in the region left of the horizon. As expected, a near-perfect reconstruction of the reference fields from the metasurface hologram is observed throughout the observation region.

These near-perfect results may be compared to the previous case of a front-lit object (Fig.~\ref{Fig:FLAFI} and \ref{Fig:FLABI}), where slight ripples were observed in the recreated fields. While the object is the same in both cases excited with a Gaussian beam, the two reference fields are quite different due to front vs back lit conditions. This further suggests that the nature of the reference fields to be recreated by the hologram impacts the accuracy of the reconstructed fields.

\section{Practical Aspects of Metasurface Holograms}\label{Sec:Prac}

\subsection{Effect of Finite-sized Metasurfaces} 

So far, the metasurface has been treated as an essentially ideally large surface, separating the two half regions. However, practical metasurfaces are finite-sized in nature, and this finite extent of the metasurface may have important consequences on the quality of scattered field recreation using our holograms. The key distortions are the presence of secondary diffraction from the surface edges and spilling over of the illumination fields around the metasurface, and reducing the field-of-view of the illusion, in some cases.

Let us take the previous example of back-lit anterior illusion with front illumination case, where the length of the metasurface hologram is changed. Fig.~\ref{Fig:FLPFI_LV}(a) shows the case of a metasurface of length $20\lambda$, modeled with a dielectric (non-scattering) boundary on each side. The 2D total fields clearly show the strong penetration of the back illumination towards the left of the horizon. Fig.~\ref{Fig:FLPFI_LV}(a) also shows the 2D field comparison at the metasurface and the observer compared to the ideal case of a large metasurface. It is clear that the recreated fields significantly differ from the desired fields, with some  resemblance near the center of the observer only. Large field oscillations are present on either side of the metasurface due to the illumination field spilling around the metasurface. This could also be attributed to any surface waves reflected off the edge discontinuities of the surface and forming standing-wave type fields on the surface. As the metasurface is made larger to $40\lambda$, the field reconstruction improves as shown in Fig.~\ref{Fig:FLPFI_LV}(b), which eventually becomes near-identical to the desired ones for a length of $80\lambda$, as shown in Fig.~\ref{Fig:FLPFI_LV}(c). A further improvement is shown in Fig.~\ref{Fig:BLAFI}(f) in which a surface of length $120\lambda$ was used.

\subsection{Effect of Illumination Field Strength} 

\begin{figure*}[htbp]
\begin{center}
     \begin{subfigure}[b]{0.3\textwidth}
         \centering
         \psfrag{a}[c][c][0.7]{$y$~(m)}
         \psfrag{e}[l][c][0.6]{Re\{$\chi$\}}
         \psfrag{d}[l][c][0.6]{Im\{$\chi$\}}
         \psfrag{b}[c][c][0.7]{$\chi_\text{mm}(y)$}
         \psfrag{c}[c][c][0.7]{$\chi_\text{ee}(y)$}
	\includegraphics[width=1\columnwidth]{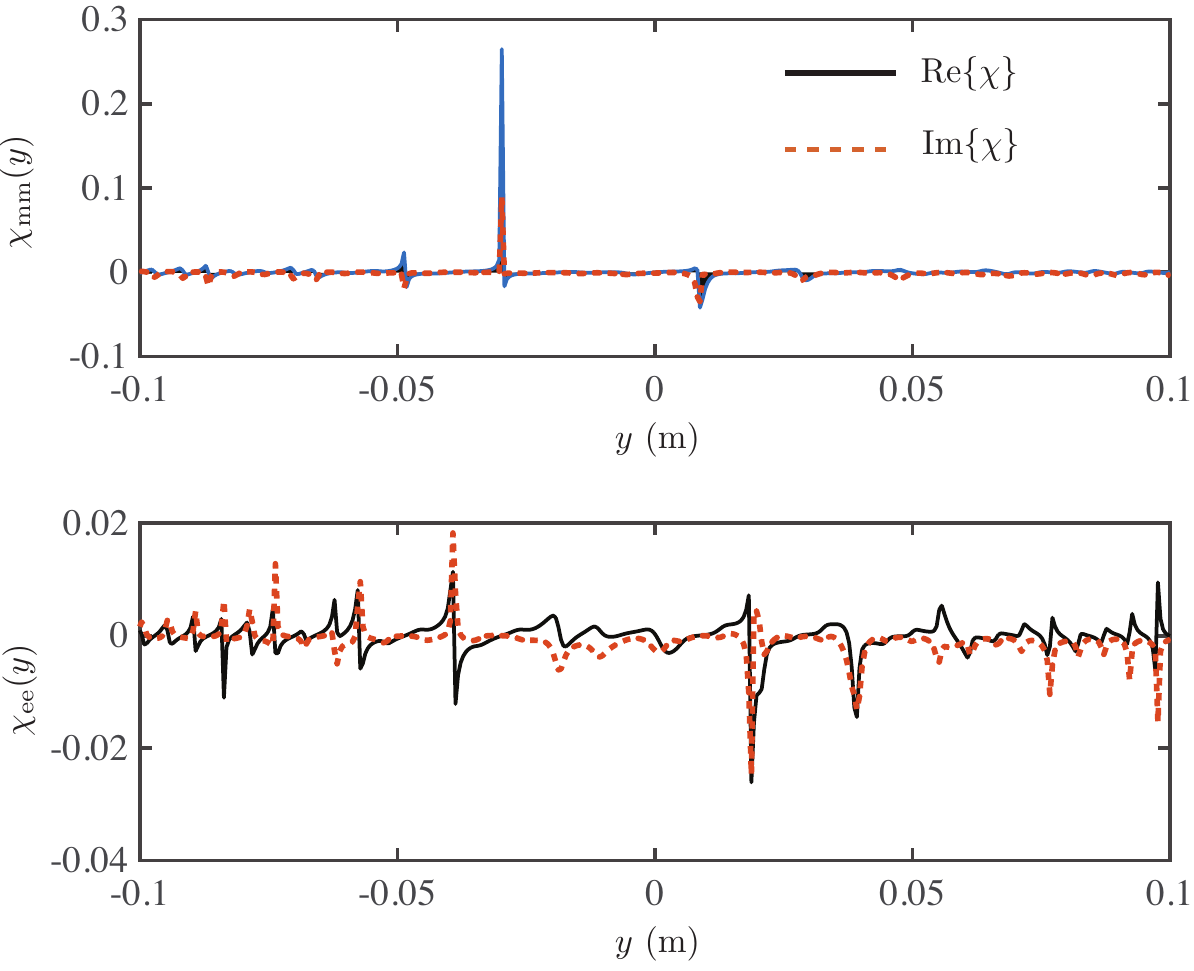}
         
     \end{subfigure}
     \begin{subfigure}[b]{0.3\textwidth}
         \centering
         \psfrag{a}[c][c][0.7]{$y$~(m)}
         \psfrag{e}[l][c][0.6]{Re\{$\chi$\}}
         \psfrag{d}[l][c][0.6]{Im\{$\chi$\}}
         \psfrag{b}[c][c][0.7]{$\chi_\text{mm}(y)$}
         \psfrag{c}[c][c][0.7]{$\chi_\text{ee}(y)$}
	\includegraphics[width=1\columnwidth]{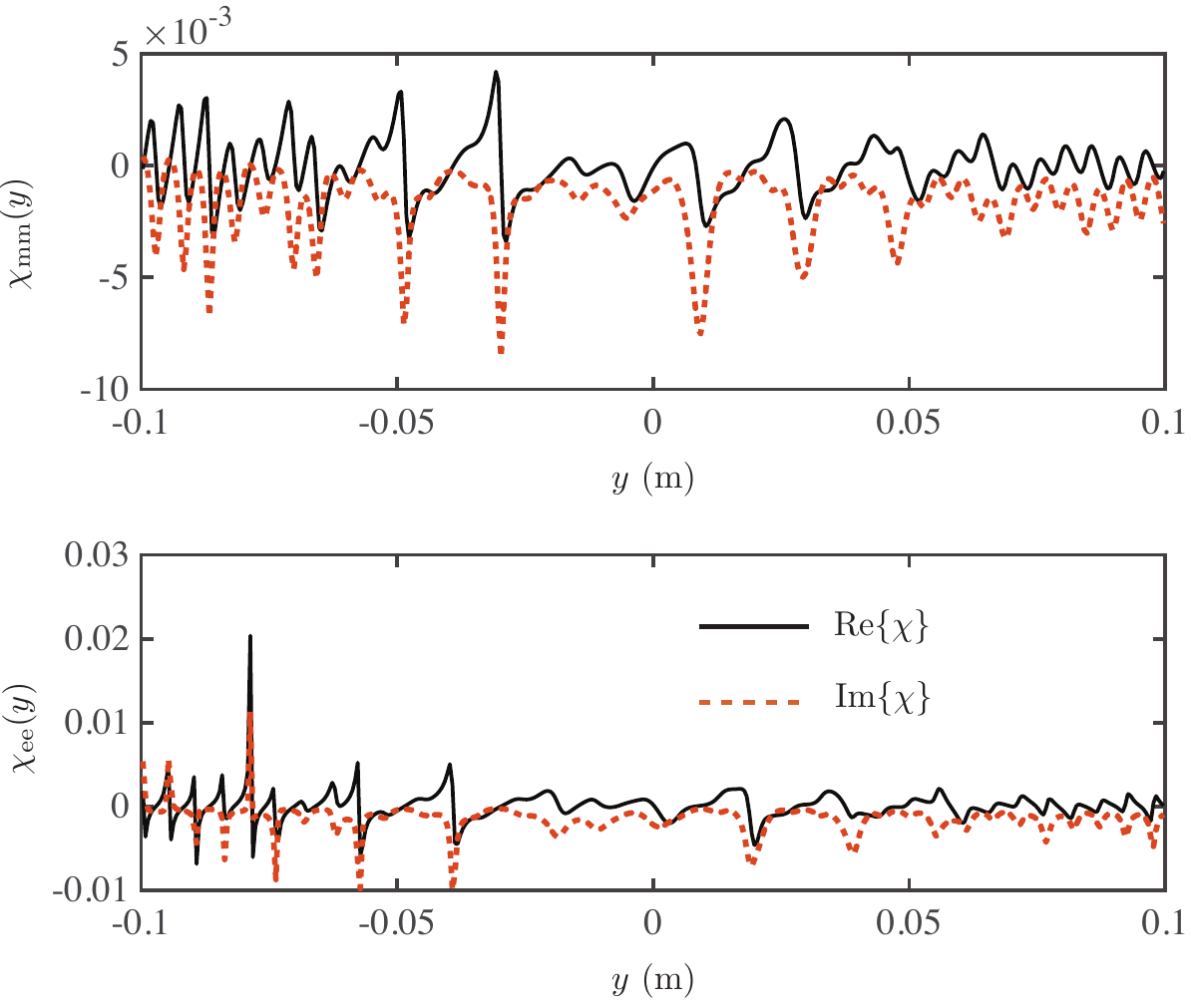}
         
     \end{subfigure}
     \begin{subfigure}[b]{0.3\textwidth}
         \centering
         \psfrag{a}[c][c][0.7]{$y$~(m)}
         \psfrag{e}[l][c][0.6]{Re\{$\chi$\}}
         \psfrag{d}[l][c][0.6]{Im\{$\chi$\}}
         \psfrag{b}[c][c][0.7]{$\chi_\text{mm}(y)$}
         \psfrag{c}[c][c][0.7]{$\chi_\text{ee}(y)$}
	\includegraphics[width=1\columnwidth]{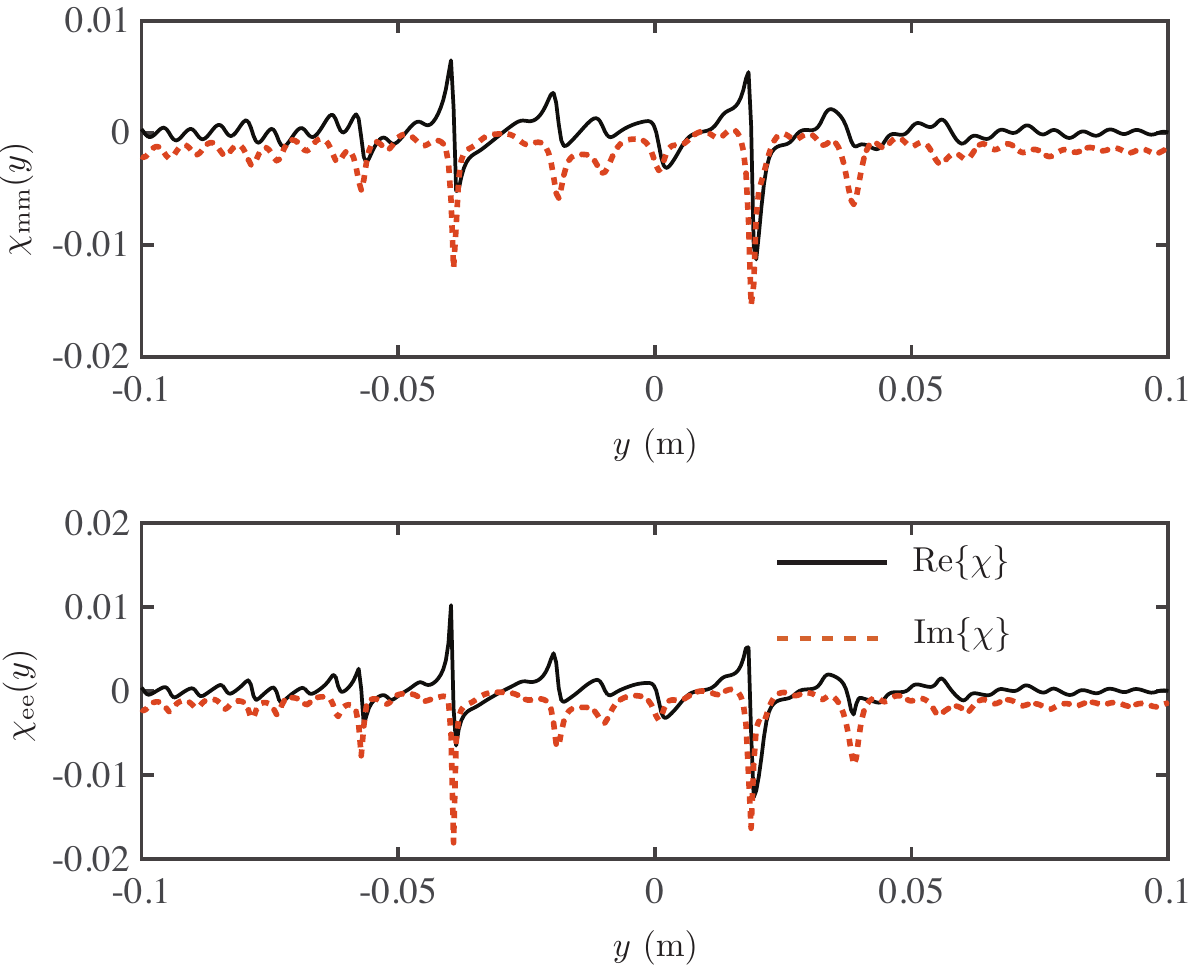}
	
     \end{subfigure}
     \par\bigskip
     \begin{subfigure}[b]{0.3\textwidth}
         \centering
         \psfrag{a}[c][c][0.7]{$y$~(m)}
         \psfrag{d}[c][c][0.6]{$|E_z(0, y)|$~(norm.)}   
	\psfrag{e}[c][c][0.6]{$|E_z(x_0, y)|$~(norm.)}     
	\psfrag{f}[l][c][0.6]{$|E_{z,s}^\text{ref.}(x_0,y)|$}    
	\psfrag{g}[l][c][0.6]{$|E_{z,s}^\text{ms.}(x_0,y)|$}   
	\psfrag{m}[l][c][0.6]{Re\{$E_{z,s}^\text{bck.}(x_m,y)$\}, Re\{$E_{z,s}^\text{ms.}(x_{m-},y)$\}}  
	\psfrag{n}[l][c][0.6]{Im\{$E_{z,s}^\text{bck.}(x_m,y)$\}, Im\{$E_{z,s}^\text{ms.}(x_{m-},y)$\}}    
	\includegraphics[width=1\columnwidth]{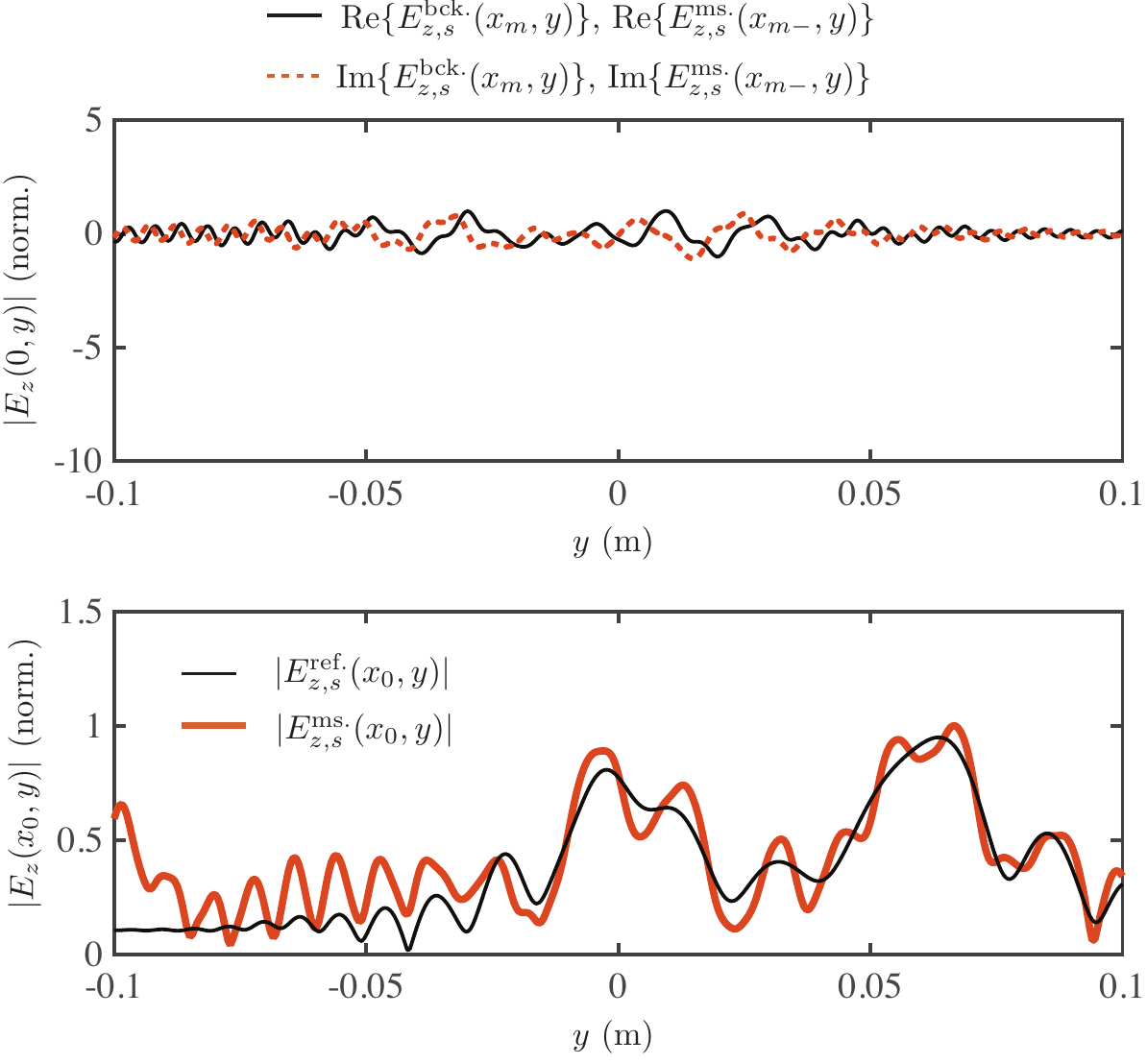}\caption{$|E_{z}^\text{ill.}| = 1.0$}
         
     \end{subfigure}
     \begin{subfigure}[b]{0.3\textwidth}
         \centering
         \psfrag{a}[c][c][0.7]{$y$~(m)}
         \psfrag{d}[c][c][0.6]{$|E_z(0, y)|$~(norm.)}   
	\psfrag{e}[c][c][0.6]{$|E_z(x_0, y)|$~(norm.)}     
	\psfrag{f}[l][c][0.6]{$|E_{z,s}^\text{ref.}(x_0,y)|$}    
	\psfrag{g}[l][c][0.6]{$|E_{z,s}^\text{ms.}(x_0,y)|$}   
	\psfrag{m}[l][c][0.6]{Re\{$E_{z,s}^\text{bck.}(x_m,y)$\}, Re\{$E_{z,s}^\text{ms.}(x_{m-},y)$\}}  
	\psfrag{n}[l][c][0.6]{Im\{$E_{z,s}^\text{bck.}(x_m,y)$\}, Im\{$E_{z,s}^\text{ms.}(x_{m-},y)$\}}   
	\includegraphics[width=1\columnwidth]{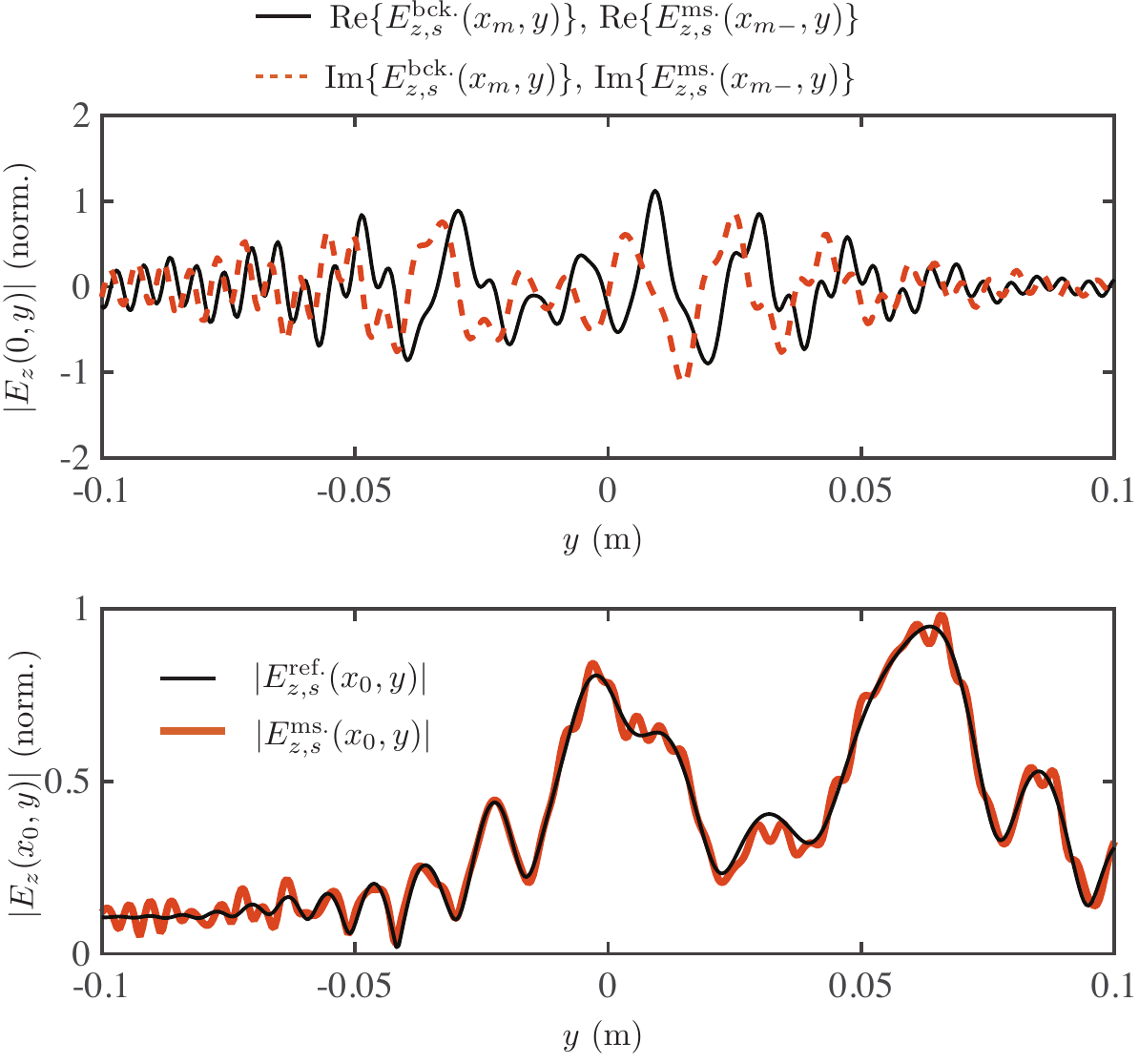}\caption{$|E_{z}^\text{ill.}| = 1.25$}
         
     \end{subfigure}
     \begin{subfigure}[b]{0.3\textwidth}
         \centering
         \psfrag{a}[c][c][0.7]{$y$~(m)}
         \psfrag{d}[c][c][0.6]{$|E_z(0, y)|$~(norm.)}   
	\psfrag{e}[c][c][0.6]{$|E_z(x_0, y)|$~(norm.)}     
	\psfrag{f}[l][c][0.6]{$|E_{z,s}^\text{ref.}(x_0,y)|$}    
	\psfrag{g}[l][c][0.6]{$|E_{z,s}^\text{ms.}(x_0,y)|$}   
	\psfrag{m}[l][c][0.6]{Re\{$E_{z,s}^\text{bck.}(x_m,y)$\}, Re\{$E_{z,s}^\text{ms.}(x_{m-},y)$\}}  
	\psfrag{n}[l][c][0.6]{Im\{$E_{z,s}^\text{bck.}(x_m,y)$\}, Im\{$E_{z,s}^\text{ms.}(x_{m-},y)$\}}   
	\includegraphics[width=1\columnwidth]{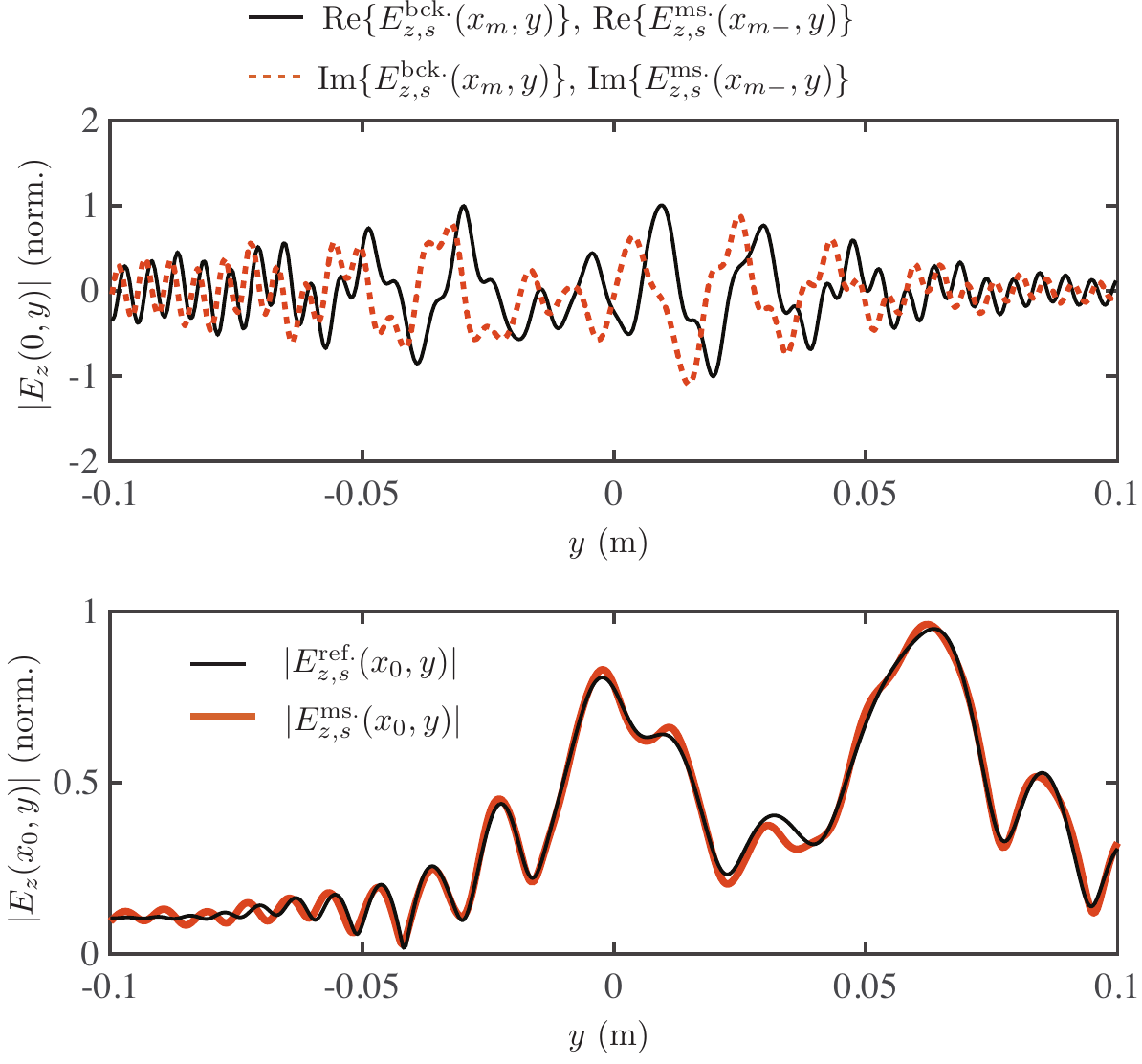}\caption{$|E_{z}^\text{ill.}| = 1.5$}
         
     \end{subfigure}               
\caption{Effect of illumination strength on the synthesized surface susceptibilities, demonstrated using the case of a back-lit anterior illusion with front illumination case. Simulation parameters are: Parametrized PEC object centered at $x = 7.5\lambda$, incident Gaussian-wave at $-15^\circ$ from $x-$axis, and width of $10\lambda$ propagating along $-x$, illuminating Gaussian-wave at $0^\circ$ from $x-$axis with width of $15\lambda$ propagating along $+x$.}\label{Fig:BLAFI_AV}
\end{center}
\end{figure*}

It is clearly evident throughout all these examples, that the synthesized metasurface susceptibilities strongly depend on the nature and configuration of the illumination fields -- how the metasurface hologram is illuminated. In general, the synthesized surfaces exhibit varied regions of loss-gain characteristics to enable the recreation of the desired fields along with a large variation in the susceptibility values. Such surfaces are practically challenging to implement due to design complexity requiring active surfaces and  purely passive surfaces maybe desired. One simple way to influence the passivity of the surface, is to increase the illumination field strength. This could be seen as a practical way to enforce a passive surface using an external control. 

In all the previous examples, the illumination field strength was kept the same as the reference field. Fig.~\ref{Fig:BLAFI_AV} shows the effect of the illumination strength on the synthesized surface susceptibilities of the metasurface hologram using the back-lit anterior illusion with front illumination case of Fig.~\ref{Fig:BLAFI}. Fig.~\ref{Fig:BLAFI_AV}(a) shows the nominal case of unity amplitude, where the synthesized susceptibilities show large peaking values of $\chi_\text{ee}$ and $\chi_\text{mm}$ in several local regions, along with several active regions on the surface where the fields are amplified. When the illumination strength is increased to 1.25, the susceptibility values significantly improve to lower values, with the surface becoming near passive and also producing a better reconstruction of the fields, as shown in Fig.~\ref{Fig:BLAFI_AV}(b). With a further increase in the illumination strength to 1.5, the susceptibility values drop to small values, although still exhibiting some local regions of gain. However, this time the desired fields are near-perfectly reconstructed, as seen in Fig.~\ref{Fig:BLAFI_AV}(c).

This example illustrates the sensitivity of the metasurface hologram design to illumination field strengths and demonstrates that it is an important parameter to take into account. It should be noted that the increased illumination strength is equivalent to using a lower reference field to compute the desired scattered fields. This disparity in the fields will simply manifest as an illusion of lower brightness under normal illumination conditions.

%\begin{itemize}
%    \item Parametrized PEC object centered at $x_{off} = 7.5\lambda$
%    \item Horizon at 0 length of $120\lambda$
%    \item Incident -- rightside gaussian at -15$\deg$ $E_z = 1.0$ width of $10\lambda$
%    \item Illumination -- leftside gaussian 0$\deg$ $E_z = 1.0, 1.25, 1.6$ width of $15\lambda$
%    \item MS length $120\lambda$ with dielectric padding $25\lambda$ on either side
%\end{itemize}

\section{Conclusions}\label{Sec:Con}

A systematic and structured approach has been presented to exploit the rich field transformation capabilities of EM metasurfaces for creating a variety of EM illusions using the concept of metasurface holograms. A holistic approach of metasurface hologram synthesis has been undertaken here from a system point of view, where the desired fields detected by the observer are first recreated and then fed back into the overall metasurface synthesis problem. Considering the complexity of the problem and large number of configuration possibilities, the approach of classifying them in terms of front/back-lit posterior/anterior illusions using front/back illumination has been adopted for better organization and tractability of the overall synthesis problem. These classifications are based on specific relationships between the reference object to be recreated, the observer measuring the object, the orientation and placement of the reference and illumination field, and the desired placement of the metasurface hologram creating a virtual image. Consequently, a general design procedure to synthesize metasurface holograms has been proposed based on the IE-GSTC method and its EM illusion creation capabilities has been demonstrated using several examples. The proposed synthesis framework results in the determination of the spatially varying surface susceptibilities describing the EM properties of the metasurface. The synthesis technique combines the integral form of the Maxwell's equations, and the corresponding field propagators with the GSTCs describing the field interaction with zero thickness metasurfaces. It has next been implemented using the BEM approach using a rigorous and compact matrix formulation which is capable of synthesizing planar as well as curvilinear metasurface holograms and for arbitrary specifications of the reference object. Finally, the impact of the metasurface size and the illumination field strength on the quality of the reconstructed scattered fields has been discussed 
with respect to the feasibility of practical metasurface holograms.

The number of possible configurations and situations for creating EM illusions using metasurface holograms are virtually unlimited. While the handful of examples presented here have been strategically chosen to highlight and illustrate several of the salient features of the hologram synthesis, the presented framework represents a flexible test-bed to explore a wider variety of illusion scenarios before undertaking practical demonstrations. It further highlights the unprecedented capabilities of EM metasurfaces in achieving very complex wave transformations. While only scalar surface susceptibilities have been employed in this work, the BEM-GSTC framework is easily extendable to fully tensorial descriptions of the surface as was done in \cite{stewart2019scattering}. Furthermore, the usage of GSTCs combined with surface susceptibility description of zero-thickness surfaces is also a very efficient tool for modeling complex objects (and not limited to PEC objects as used here). This makes the proposed numerical framework complete for handling arbitrarily complex problems using a common IE-GSTC infrastructure. It should also be finally remarked that, while a vast number of techniques have been proposed for designing optical holograms, the proposed technique represents a rigorous full-vectorial field-based approach for metasurface synthesis, as opposed to methods typically based on paraxial approximations at optical frequencies \cite{Goodman_Fourier_Optics}. Furthermore, compared to the existing works on metasurface synthesis, the system level approach undertaken here where the desired fields are first computed from physical considerations, results in a metasurface synthesis problem that is likely to be well-posed, as opposed to the possibility of an otherwise arbitrary and physically disconnected problem. Therefore, this work provides a set of important tools to metasurface hologram designers for creating a myriad of complex EM illusions throughout the EM spectrum.

%\section{Acknowledgement}
%
%The authors acknowledge funding from the Department of National Defence's Innovation for Defence Excellence and Security (IDEaS) Program in support of this work.

\bibliographystyle{IEEEtran}
\bibliography{2020_Metasurface_Illusions_TAP_Smy}

\end{document}